\documentclass[english]{article}
\usepackage[T1]{fontenc}
\usepackage[latin9]{inputenc}
\usepackage{amsmath}
\usepackage{amssymb}
\usepackage{esint}
\usepackage{amsthm}
\usepackage{babel}

\begin{document}

\title{Affine Bodies Revisited. \\
Constraints, Symmetry, Analytical Methods \\
and Some Perspectives}

\author{J.~J.~S\l{}awianowski, B.~Go\l{}ubowska, E.~E.~Ro\.{z}ko,\\ V. Kovalchuk, A.~Martens, E.~Gobcewicz
\\
\\
Institute of Fundamental Technological Research,\\
Polish Academy of Sciences,\\
5B, Pawinskiego str., 02--106 Warsaw, Poland\\
e-mails: jslawian@ippt.gov.pl, bgolub@ippt.gov.pl,\\
erozko@ippt.gov.pl, vkoval@ippt.gov.pl,\\
amartens@ippt.gov.pl, gobicz@o2.pl}

\date{ }
\maketitle
\begin{abstract}
\noindent The purpose of this publication is to derive and discuss equations of motion of affinely rigid (homogeneously deformable) body moving in Euclidean space of general dimension $n$. Our aim is to present some analytical methods and to discuss geometric structure and invariance properties of the theory. Some perspectives of further developments are also discussed. 
\end{abstract}

\section{Introduction}

The theory of affinely-rigid bodies (shortly -- affine bodies) was formulated and treated in a systematic way by J. J. S\l{}awianowski \cite{JJS1975} \cite{JJS1982} in terms of the modern differential-geometric and algebraic methods. Earlier, the concept was implicitly present in Eringen's theory of micromorphic continua. Those continua, roughly speaking, consist no longer of structureless material points, but rather of infinitesimal affine bodies, i.e., material points with attached linear bases \cite{Erin1964} \cite{Erin1968}. This was an obvious generalization of Cosserat's theory; the structure elements (``grains'') not only rotate but also undergo homogeneous deformations. We are aware also of the existence of some much earlier Russian papers (private information), however, having no explicit references we were unable to find them. Later on the subject developed rapidly both on the level of theory and applications and attracted the attention of many researchers \cite{Bur2008} \cite{Bur1996} \cite{Cap2003} \cite{Cas1995} \cite{Che2004} \cite{Gol2004} \cite{Lew1990} \cite{Mig2003} \cite{5Mig2003} \cite{Papad2001} \cite{Rei1998} \cite{Rei1996} \cite{Roz2005} \cite{Rub1985} \cite{2Rub1985} \cite{Simo1991} \cite{IJJSco2004} \cite{IIJJSco2005} \cite{Sousa1994} \cite{Sol2000} \cite{Sol1999} \cite{Vill2005}. What concerns theory and mathematical foundations, deep studies by Burov and Chevalier \cite{Bur2008} \cite{Bur1996} are recommended. 

Affinely rigid body is continuous or discrete or even finite system of material points constrained in such a way that all affine relationships (but not necessarily metrical ones like distances and angles) are preserved during any admissible motion. Such a body performs rigid rotations, translations and homogeneous deformations. It has a finite number of degrees of freedom, however, deformations being admitted, it is something between (metrically) rigid body (gyroscope) and general deformable continuum. Let us mention that some important general problems concerning constrained continua were studied systematically by Wo\'{z}niak and Kleiber \cite{Kleiber1991} \cite{Woz1992} \cite{Woz1988}.

This is an interesting model from the purely mathematical point of view of rational analytical mechanics, but there are also remarkable applications in a wide variety of problems. Let us quote some of them: 
\begin{itemize}
\item problems of macroscopic elasticity in which the length of excited waves is comparable with the linear size of the body 
\item molecular dynamics, molecular crystals, fullerens 
\item dynamics of inclusions, suspensions and bubbles 
\item structured continua 
\item nuclear dynamics 
\item astrophysics, geophysics 
\item robotics 
\item modified finite elements method 
\item dynamics of one-dimensional chains/lattices. 
\end{itemize}

Let us also stress some important theoretical point. Namely, in sufficiently small regions every smooth deformation is homogeneous; this is just the very idea of differential calculus. Analysis of affine bodies contributes much to the general theory of non-constrained continua. Any deformation gives rise to some field of homogeneous deformations. And fields of homogeneous deformations, if not necessarily holonomic, are useful when analyzing internal stresses \cite{Gold1960}.

Below we discuss certain dynamical models of affine bodies including problems of the partial separability and integrability. Presented are ideas and models which are rather new, up to our knowledge never touched in literature. Namely, in all papers we know (many of them quoted in the references) it is only kinematics that is based on affine geometry, but on the dynamical level affine symmetry is broken and restricted to the Euclidean group of motions or to some its subgroups. Unlike this we discuss also models the dynamics of which is affinely-invariant. This is quite a new program, very interesting at least from the purely academic point of view. But there are also some reasons to expect that the suggested models are also dynamically viable and that on the fundamental level of physical phenomena the ``large'' affine symmetry of dynamical laws is more justified and desirable than the restricted invariance under isometries. Moreover, we show that instead using some explicit potential energy expression, one can encode the dynamics of elastic vibrations in an appropriate from of affinely-invariant kinetic energy. The resulting geodetic models in a sense resemble the procedure of Maupertuis principle, where the orbits of motion are geodetics of an appropriate metric tensor build as some conformal modification of the ``true'' geometric fundamental tensor \cite{Arn1978}. There is also some similarity to the concept of effective mass known from solid state physics \cite{Kittel2005}.

In a sense of our affinely-invariant geodetic models may be interpreted as a discretization of the Arnold description of ideal fluids in terms of geodetic Hamiltonian systems on the group of volume-preserving diffeomorphisms \cite{Arn1978}. This is a very drastic discretization, reducing the continuum cardinality of degrees of freedom to the finite one, namely $n(n+1)$, $n$ denoting the dimension of the physical space ($3$ in realistic models).

It is interesting to mention about certain interesting procedures based on modified finite elements methods following the ideas of M. Rubin and his co-workers \cite{Rub1985} \cite{2Rub1985}. In its most traditional version the idea of finite elements was to cover the bulk of deformable body with a mesh of tetrahedrons or parallelepipeds (or sometimes-other simple figures) of sufficiently small size; similarly, the surface was replaced by a mesh of triangles or parallelepipeds (or other simple two-dimensional figures). Those small parts were assumed to be (in a good approximation) homogeneously deformable. Basing on this assumption one replaced a system of partial differential equations by a discrete system of difference equations. And this approximation enabled one to perform directly the computer-aided numerical calculations. The time variable is then also discretized and its continuity is replaced by some mesh. But there is also another procedure, when instead of the spatio-temporal finite elements one uses also spatial ones. This hybrid method consist in an approximate representation of continuum as a system of mutually interacting affine (homogeneously deformable) bodies. But the time variable is continuous and for the mentioned system the usual methods of analytical mechanics and qualitative theory of dynamical systems are used. In various problems this hybrid discretized-analytical approach seems to be more appropriate. In any case the powerful methods of dynamical systems theory may be used.

\section{Some basic concepts. Review, notation, geometry.}

Although, as mentioned, we admit discrete or even finite systems of material points, it is more convenient to use the standard terms of continua. As is well-known, configurations of continuous media are described by diffeomorphisms $\Phi$ of the material space $N$ (``Lagrange variables'') onto the physical space $M$ (``Euler variables'') \cite{Erin1964} \cite{Erin1962} \cite{Gold1960} \cite{Mars1994} \cite{Truesd1965} \cite{Truesd1972} \cite{Yav2008}. Those are assumed to be affine spaces of the same dimension $n$. Obviously, in applications $n=3$, sometimes $2$ or $1$, but for many reasons it is convenient to admit the artificially general dimension. What concerns fundamental concepts and notations in affine geometry we follow mainly the standard treatments like \cite{Berger1977} \cite{Berger1978}, cf. also \cite{Che2004} \cite{Helwig} \cite{JJS1975} \cite{JJS1982} \cite{IJJSco2004} \cite{IIJJSco2005}. There are problems where instead of the total $N$ we take some subset of $N$, e.g., in problems of shells, roads, membranes and strings. Sometimes $N$ from the very beginning has the lower dimension than $M$, however, here we do not consider such situations; for some examples see, e.g. \cite{Roz2005}. Motion is described by the time dependence of $\Phi$. Usually (not always) $M$ and $N$ are endowed with flat metric tensors $g$, $\eta$ in $M$, $N$; $g\in V^{*}\otimes V^{*}$, $\eta\in U^{*}\otimes U^{*}$ where $V$, $U$ are respectively linear spaces of translations (free vectors) in $M$, $N$, and $V^{*}$, $U^{*}$ are their duals (spaces of linear functions respectively on $V$, $U$). It is not necessary \cite{Erin1962} \cite{Erin1964} \cite{Erin1968}, but in this paper sufficient and convenient to use rectangular coordinates $a^{K}$, $y^{i}$ respectively in $N$ and $M$. They are what is often referred to as Lagrange and Euler coordinates. The components of $g$, $\eta$ are denoted by $g_{ij}$, $\eta_{AB}$; obviously in rectilinear rectangular coordinates we have: 
\begin{equation}
g_{ij}\underset{*}{=}\delta_{ij},\qquad\eta_{AB}\underset{*}{=}\delta_{AB}.\label{eq:1}
\end{equation}

The contravariant invariant metrics $g^{-1}\in V\otimes V$, $\eta^{-1}\in U\otimes U$ have components traditionally denoted by $g^{ij}$, $\eta^{AB}$ (upper case indices), where 
\begin{equation}
g^{ik}g_{kj}=\delta^{i}\!_{j},\qquad\eta^{AC}\eta_{CB}=\delta^{A}\,_{B}.\label{eq:2}
\end{equation}

Analytically $\Phi$ is represented by the dependence of $y^{i}$-variables on $a^{K}$-variables, 
\begin{equation}
y^{i}=y^{i}(a^{K}).\label{eq:3}
\end{equation}

When dealing with motions, i.e., time-dependent $\Phi\left(t\right)$, we introduce in addition the time-dependence to (\ref{eq:3}), 
\begin{equation}
y^{i}=y^{i}(t,\, a^{K}).\label{eq:4}
\end{equation}

The placement, i.e., deformation gradient, is analytically represented by the matrix $\chi$ with components 
\begin{equation}
\chi^{i}\!_{K}=\frac{\partial y^{i}}{\partial a^{K}}.\label{eq:5}
\end{equation}

This is a ``doubled'' quantity in terminology of Schouten and Veblen; for the fixed $a\in N$, $y\left(a\right)\in M$, the tensor identification of $\chi(a)$ is as follows: 
\begin{equation}
\chi(a)\in V\otimes U^{*}.\label{eq:6}
\end{equation}

Therefore, nothing like the symmetry or antisymmetry concept does apply to the object $\chi$.

Solving (\ref{eq:3}) or (\ref{eq:4}) with respect to Lagrange variables,
\begin{equation}
a^{K}=a^{K}(y^{i}),\qquad a^{K}=a^{K}(t,\, y^{i}),\label{eq:7}
\end{equation}
we obtain the analytical description of the inverse mappings $\Phi^{-1}$, $\Phi(t)^{-1}$ of $M$ onto $N$.

Lagrange velocity field is defined on $N$ and takes values in $V$,
\begin{equation}
\mathcal{V}^{i}(t,a)=\frac{\partial y^{i}}{\partial t}\left(t,a\right).\label{eq:8}
\end{equation}

Euler velocity field is defined on $M$ and takes values in $V$,
\begin{equation}
v^{i}(t,y)=\frac{\partial y^{i}}{\partial t}\left(t,a(t,y)\right)=\mathcal{V}^{i}\left(t,a(t,y)\right).\label{eq:9}
\end{equation}
where the partial differentiation $\partial / \partial t$ acts only on the first argument, not on the $t$-variable in $a(t,y)$. One can easily show that: 
\begin{equation}
v^{i}(t,y)=-\frac{\partial y^{i}}{\partial a^{K}}\left(t,a\right)\frac{\partial a^{K}(t,y)}{\partial t}=-\chi^{i}\!_{K}\left(t,a(t,y)\right)\frac{\partial a^{K}(t,y)}{\partial t}.\label{eq:10}
\end{equation}

In certain formulas we need locally co-moving, i.e., material expressions of those (and other) quantities. They are given respectively by: 
\begin{eqnarray}
\widehat{\mathcal{V}}^{K}(t,a) & := & \frac{\partial a^{K}}{\partial y^{i}}\left(t,a\right)\mathcal{V}^{i}(t,a)=\chi^{-1}\left(t,a\right)^{K}\!_{i} \: \mathcal{V}^{i}(t,a),\label{eq:11}\\
\hat{v}^{K}(t,y) & := & \frac{\partial a^{K}}{\partial y^{i}}\left(t,a(t,y)\right)v^{i}\left(t,y\right)=\chi^{-1}\left(t,a(t,y)\right)^{K}\!_{i}\ v^{i}(t,y).\label{eq:12}
\end{eqnarray}

We follow the conventions and terms used in the classical treatises by Eringen \cite{Erin1968} \cite{Erin1964}, so e.g., the Green and Cauchy deformation tensors $G$, $C$ are meant in the convention:
\begin{eqnarray}
G_{KL} & = & g_{ij}\frac{\partial y^{i}}{\partial a^{K}}\frac{\partial y^{j}}{\partial a^{L}}=g_{ij} \: \chi^{i}\!_{K} \: \chi^{j}\!_{L},\label{eq:13}\\
C_{ij} & = & \eta_{KL}\frac{\partial a^{K}}{\partial y^{i}}\frac{\partial a^{L}}{\partial y^{j}}=\eta_{KL} \: \chi^{-1}\!^{K}\!_{i} \: \chi^{-1}\!^{L}\!_{j};\label{eq:14}
\end{eqnarray}
to avoid the crowd of characters we do not write arguments. This does not seem to generate confusion. Using the modern notation of differential geometry, namely the pull-back symbols, we write \cite{Kob1963}:
\begin{equation}
G=\Phi^{\star}g,\qquad C=\left.\Phi^{-1}\right.^{\star}\eta.\label{eq:15}
\end{equation}

We use also the contravariant tensors $G^{-1}$, $C^{-1}$, where, by definition: 
\begin{equation}
\left.G^{-1}\right.^{KA}G_{AL}=\delta^{K}\!_{L},\qquad\left.C^{-1}\right.^{ia}C_{aj}=\delta^{i}\!_{j}.\label{eq:16}
\end{equation}
 Obviously 
\begin{equation}
\left.G^{-1}\right.^{KA}=\frac{\partial a^{K}}{\partial y^{i}}\frac{\partial a^{L}}{\partial y^{j}} \: g^{ij},\qquad\left.C^{-1}\right.^{ij}=\frac{\partial y^{i}}{\partial a^{K}}\frac{\partial y^{j}}{\partial a^{L}} \: \eta^{KL}.\label{eq:17}
\end{equation}

In terms of the push-forward notation: 
\begin{equation}
G^{-1}=\Phi_{\star}^{-1}g^{-1},\qquad C^{-1}=\Phi_{\star}\eta^{-1}.\label{eq:18}
\end{equation}

\textbf{Remark}; a very important warning: There is a danger of confusion when the Schouten kernel-index convention is used very automatically, without sufficient caution. Namely, it is strictly forbidden to confuse (\ref{eq:17}) with the metrically-raised contravariant versions of $G$, $C$, 
\begin{equation}
G^{KL}=\eta^{KA}\eta^{LB}G_{AB},\qquad C^{ij}=g^{ia}g^{jb}C_{ab}.\label{eq:19}
\end{equation}
One uses also the following mixed tensors: 
\begin{equation}
\widehat{G}^{K}\!_{L}:=\eta^{KA}G_{AL},\qquad\widehat{C}^{k}\!_{l}:=g^{ka}C_{al}. \label{eq:20}
\end{equation}

The tensorial ``membership'' of those objects (when taken at a fixed argument value) is as follows: 
\begin{equation}
G\in U^{*}\otimes U^{*},\quad G^{-1}\in U\otimes U,\quad C\in V^{*}\otimes V^{*},\quad C^{-1}\in V\otimes V,\label{eq:21}
\end{equation}
\begin{equation}
\widehat{G}\in U\otimes U^{*}\simeq L\left(U\right),\qquad\widehat{C}\in V\otimes V^{*}\simeq L\left(V\right).\label{eq:22}
\end{equation}
Obviously (\ref{eq:21}) are symmetric and positively definite.

The pairs of tensors $\left(G,\eta\right)$, $\left(C,g\right)$ give rise to deformation invariants. They may be chosen in infinity of ways, for example we can take: 
\begin{equation}
\mathcal{K}_{a}\left[\phi\right]=Tr\left(\widehat{G}^{a}\right)=Tr\left(\widehat{C}^{-a}\right),\qquad a=1,\ldots,n.\label{eq:23}
\end{equation}

In the non-deformed situations we have $G=\eta$, $C=g$. Sometimes it is convenient to use the deformation measures vanishing in the non-deformed state; the most important of them, denoted by $E$, $e$ are in \cite{Erin1968} referred to as Lagrange and Euler deformation tensors, 
\begin{equation}
E:=\frac{1}{2}\left(G-\eta\right),\qquad e:=\frac{1}{2}\left(g-C\right).\label{eq:24}
\end{equation}

Obviously, their tensorial ``membership'' is $E\in U^{\star}\otimes U^{\star}$, $e\in V^{\star}\otimes V^{\star}$ and they are symmetric. Following (\ref{eq:20}) we can construct the mixed tensors $\widehat{E}\in U\otimes U^{\star}$, $\hat{e}\in V\otimes V^{\star}$, namely 
\begin{equation}
\widehat{E}^{A}\!_{B}=\eta^{AC}E_{CB},\qquad\hat{e}^{i}\!_{j}=g^{ik}\:e_{kj},\label{eq:25}
\end{equation}
and the corresponding invariants 
\begin{equation}
Tr\left(\widehat{E}^{a}\right),\qquad Tr\left(\hat{e}^{a}\right);\label{eq:26}
\end{equation}
obviously, those invariants are functions of (\ref{eq:23}).

Spatial gradient of the Euler velocity field (\ref{eq:9}), 
\begin{equation}
\Omega^{i}\!_{j}=\frac{\partial v^{i}}{\partial y^{j}}\label{eq:27}
\end{equation}
is very important in hydrodynamics, nevertheless, it is also of some relevance for elasticity, and in any case it is generally well-defined in any infinite continuum. One can also introduce its material representation:
\begin{equation}
\widehat{\Omega}^{A}\!_{B}=\frac{\partial\hat{v}^{A}}{\partial a^{B}}.\label{eq:28}
\end{equation}

\textbf{Remark}: as mentioned, we are using only Cartesian rectangular coordinates. Nevertheless, the above formulas remain valid in general coordinates, including curvilinear ones, when the usual partial derivatives are replaced by the covariant ones.

The tensorial ``membership'' of (\ref{eq:27}), (\ref{eq:28}) is, 
\begin{equation}
\Omega\in V\otimes V^{\star}\simeq L\left(V\right),\qquad\widehat{\Omega}\in U\otimes U^{\star}\simeq L\left(U\right).\label{eq:29}
\end{equation}

The $g$-skew-symmetric part of (\ref{eq:27}) describes the local field of angular velocity, 
\begin{equation}
\omega^{i}\!_{j}=\frac{1}{2}\left(\Omega^{i}\!_{j}-\Omega_{j}\!^{i}\right)=\frac{1}{2}\left(\Omega^{i}\!_{j}-g_{jk}\:g^{il}\:\Omega^{k}\!_{l}\right),\label{eq:30}
\end{equation}
and the $g$-symmetric part refers to deformation velocity, 
\begin{equation}
d^{i}\!_{j}=\frac{1}{2}\left(\Omega^{i}\!_{j}+\Omega_{j}\!^{i}\right)=\frac{1}{2}\left(\Omega^{i}\!_{j}+g_{jk}\:g^{il}\:\Omega^{k}\!_{l}\right),\label{eq:31}
\end{equation}
the quantity very often used in mechanics of viscous fluids. Its material representation, 
\begin{equation}
D_{AB}=d_{ij}\:\frac{\partial y^{i}}{\partial a^{A}}\frac{\partial y^{j}}{\partial a^{B}},\label{eq:32}
\end{equation}
i.e., strain rate, may be expressed as: 
\begin{equation}
D_{AB}=\frac{1}{2}\frac{d}{dt}\:G_{AB}.\label{eq:33}
\end{equation}

If $G_{AB}$ is expressed as a function of Euler variables $y^{i}$, then the usual time derivative is to be replaced by the substantial one.

An important thing is the study of transformations and symmetries. Let $A:\: M\rightarrow M$, $B:\: N\rightarrow N$ be, respectively, diffeomorphisms of the physical space onto itself and of the material space onto itself. Roughly speaking, they describe the spatial and material transformations. In particular, spatial and material symmetries are described in this way. Transformations $A$, $B$ act on configurations $\Phi$ respectively through the left and right superposition: 
\begin{equation}
\phi\rightarrow A\circ\Phi\circ B=\left(L_{A}\circ R_{B}\right)\left(\phi\right)=\left(R_{B}\circ L_{A}\right)\left(\phi\right);\label{eq:34}
\end{equation}
in the last two symbols $L_{A}$ and $R_{B}$ are to be meant as left and right regular translations by $A$ and $B$. In particular, the homogeneity and isotropy (translational and rotational invariance) of space and material are described in terms of such mappings.

\section{Affine bodies. Basic geometry, symmetries and canonical formalism}

Now let us go definitely to our main subject, i.e., to affine constraints. This means that the set of all a priori admissible configurations $\Phi$ is confined to $AfI\left(N,M\right)$, the $n\left(n+1\right)$-dimensional manifold of affine isomorphisms of $N$ onto $M$. To be more precise, $AfI\left(N,M\right)$ is a non-connected open submanifold of $Af\left(N,M\right)$. Obviously, there are two open connected components in $AfI\left(N,M\right)$. They are mutually disjoint, nevertheless they infinitesimally approach each other along $Af_{sing}\left(N,M\right)\subset Af\left(N,M\right)$, i.e., along the closed subset of $AfI\left(N,M\right)$ consisting of dimension-degenerating affine mappings of $N$ into $M$, i.e., such ones which ``glue'' different material points. Such transformations are non-admissible in continuum mechanics, nevertheless, at least some of them may be acceptable in mechanics of discrete affine bodies. Nothing bad happens if, e.g., all atoms of a four-atomic molecule in three-dimensional space happen at some moment co-planarly or collinearly placed. However below we shall not discuss such details, although they are interesting in themselves. If orientations of affine spaces $N$, $M$, i.e., those of linear spaces $U$, $V$ are fixed, then the mentioned two connected components of $AfI(N,M)$ consist respectively of orientation-preserving and orientation-reversing affine isomorphisms. Only if we assume that $N=M$, then the two connected components are defined intrinsically, without fixing orientation.

If we use rectilinear, in particular rectangular coordinates $a^{K}$, $y^{i}$ respectively in $N$ and $M$, then affine transformations, i.e., configurations of affine body, are analytically described in terms of linear-nonhomogeneous transformations, 
\begin{equation}
y^{i}=x^{i}+\varphi^{i}\hspace{0in}_{K}\:a^{K}.\label{eq:35}
\end{equation}
The quantities $x^{i}$ and $\varphi^{i}\hspace{0in}_{K}$ are generalized coordinates; motion is described by their time-dependence.

For the sake of completeness, let us remind that $\Phi:N\rightarrow M$ is, by definition, an affine transformations if it preserves all affine concepts, like straight-lines, parallelism etc. This means that there exist a linear mapping $L\left[\Phi\right]:U\rightarrow V$, denoted also for obvious reasons by $DL$, such that 
\begin{equation}
\overrightarrow{\Phi(a)\Phi(b)}=L[\Phi]\:\overrightarrow{ab}\label{eq:36}
\end{equation}
 for any $a,b\in N$. Obviously, the arrow symbol denotes the vector $\overrightarrow{ab}\in U$ originating at $a\in N$ and terminating at $b\in N$. For brevity, the same symbol is used in $(M,N)$. It is clear that for any chain of affine spaces we have 
\begin{equation}
L[\Phi_{1}\circ\Phi_{2}]=L[\Phi_{1}]L[\Phi_{2}],\label{eq:37}
\end{equation}
 assuming of course that superpositions are well-defined. More generally,
\begin{equation}
L[\Phi_{1}\circ\Phi_{2}\circ\ldots\circ\Phi_{K}]=L[\Phi_{1}]L[\Phi_{2}]\cdots L[\Phi_{K}]\label{eq:38}
\end{equation}
under the same assumption.

We shall use the standard abbreviations $GAf(M)$, $Af(M)$, $GL(V)$, $L(V)$, respectively for $AfI(M,M)$, $Af(M,M)$, $LI(V,V)$, $L(V,V)$. Then, obviously, the operation $L$ acts as a homomorphism of $GAf(M)$ onto $GL(V)$ (epimorphism of $GAf(M)$ onto $GL(V)$). Its kernel is identical with the group of translations $T(M)\subset GAf(M)$. Let us remind, this group is isomorphic with the linear space $V$ as an additive Abelian group and consists of transformations $t_{v}:M\rightarrow M$, $v\in V$ such that 
\begin{equation}
t_{v}(a)=b\qquad\textrm{if}\qquad\overrightarrow{ab}=v.\label{eq:39}
\end{equation}

Obviously, $T(M)$ is a normal divisor of $GAf(M)$. The connected components of unity, i.e., subgroups of orientation-preserving isomorphisms are denoted by $GAf^{+}(M)$, $GL^{+}(V)$; their cosets orientation-inverting isomorphisms are denoted by $GAf^{-}(M)$, $GL^{-}(V)$ (obviously, they are not subgroups).

Analytical representation (\ref{eq:35}) preassumes some choice of affine reference frames $(\mathfrak{O};E_{1},\ldots,E_{2}$), $(\mathfrak{o},e_{1}\ldots e_{n})$ respectively in $N$ and $M$. The points $\mathfrak{O}\in N$, $\mathfrak{o}\in M$ are origins and $E_{A}\in U$, $e_{i}\in V,$ $A=1...,n$, $i=1...n$ are basic vectors. The elements of dual bases will be denoted as usual by $E^{A}\in U^{\star}$, $e^{i}\in V^{\star}$. The parameters $x^{i}$ refer to translational motion. They are just coordinates of the current position $\Phi\left(\mathfrak{O}\right)\in M$ of a fixed material point $\mathfrak{O}\in N$. The variables $\varphi^{i}\hspace{0in}_{K}$ are generalized coordinates of relative/internal motion. When $\mathfrak{O}\in N$ is fixed once for all, then our configuration space $AfI(N,M)$ may be identified with the Cartesian product 
\begin{equation}
Q=M\times LI(U,V)=Q_{tr}\times Q_{int}\label{eq:40}
\end{equation}
where, obviously $LI(U,V)$ denotes the set of all linear isomorphisms of $U$ onto $V$; it is an open submanifold in $L(U,V)$, the linear space of linear mapping of $U$ into $V$. Concerning the term ``all'' used above, again the previously mentioned provisos are maintained, depending on whether the continuous or discrete system is meant.

Usually $\mathfrak{O}$ is chosen as the center of co-moving, i.e., Lagrangian (therefore constant) mass distribution. If $\mu$ is a positive measure on $N$ describing this distribution, then $\mathfrak{O}\in N$ is uniquely defined by 
\begin{equation}
\int\overrightarrow{\mathfrak{O}A}\:d\mu(A)=0.\label{eq:41}
\end{equation}

Then $\mathfrak{o_{\phi}}:=\Phi(\mathfrak{O})\in M$ is the current position of the center of mass in $M$, thus it satisfies: 
\begin{equation}
\int\overrightarrow{\mathfrak{o}_{\phi}y}\:d\mu_{\phi}(y)=0,\label{eq:42}
\end{equation}
where $\mu_{\phi}$ is the $\phi$-transport of the measure $\mu$
to $M$.

\textbf{Remark:} It is easy to commit a dangerous mistake. Equations (\ref{eq:41}), (\ref{eq:42}) are equivalent only for affine configurations $\Phi$. If $\Phi$ is non-affine, then in general the current (Euler) center of mass $\mathfrak{o}_{\Phi}\in M$ as defined by (\ref{eq:42}) does not coincide with $\Phi(\mathfrak{O})$. This is just one of definitions of affine mappings: center of mass is their invariant.

We use the (more or less) standard symbols $UAf(M)$, $SAf(M)$, $UL(V)$, $SL(V)$ respectively for the unimodular affine group of $M$, the special affine group of $M$, the unimodular group of $V$ and the special linear group of $V$. Let us remind that (by definition) the unimodular subgroups preserve the volume, and their special $S$-subgroups preserve the volume and orientation.

The fixed metric tensors $\eta\in U^{\star}\otimes U^{\star}$, $g\in V^{\star}\otimes V^{\star}$ give rise to the isometry subgroups $E(N,\eta)\subset GAf(N)$, $O(U,\eta)\subset GL(U)$, $E(M,g)\subset GAf(M)$, $O(V,g)\subset GL(V)$. The symbol ``$E$'' above refers to the Euclidean group, respectively in the $\eta$- and $g$-sense. The corresponding matrix elements of $a=L[A]$, $b=L[B]$, cf. (\ref{eq:34}) satisfy then 
\begin{equation}
g_{ij}=g_{kl}\:a^{k}\!_{i}\:a^{l}\!_{j},\qquad\eta_{AB}=\eta_{KL}\:b^{K}\!_{A}\:b^{L}\!_{B},\label{eq:43}
\end{equation}
i.e., $a\in GL(V)$,$b\in GL(U)$ are respectively $g$-orthogonal and $\eta$-orthogonal linear automorphisms. As usual, the connected components of group identity, i.e., the subgroups of orientation-preserving mappings, are denoted by 
\begin{eqnarray*}
SE(N,\eta)=E^{+}(N,\eta) & ,\qquad & SE(M,g)=E^{+}(M,g),\\
SO(U,\eta)=O^{+}(U,\eta) & ,\qquad & SO(V,g)=O^{+}(V,g).
\end{eqnarray*}

And their cosets consisting of orientation inverting mappings are so denoted with the label ``$-$'' instead ``$+$''. Let as mention, this time the ``$+$'' and ``$-$''-components are finitely-separated in the general affine and linear groups, not infinitesimally as previously.

The set of affine isometries, i.e, Euclidean mappings of $(N,\eta)$ onto $(M,g)$ will be denoted by $E(N,\eta;\: M,g)$. Analytically, in terms of (\ref{eq:35}), the membership $\Phi\in E(N,\eta;M,g)$ means that 
\begin{equation}
\eta_{AB}=g_{ij}\:\varphi^{i}\!_{A}\:\varphi^{j}\!_{B},\label{eq:44}
\end{equation}
 i.e., using the modern geometric notation, 
\begin{equation}
\eta=\Phi^{\star}g.\label{eq:45}
\end{equation}
Obviously, (\ref{eq:15}) becomes then 
\[
G=\eta,\qquad C=g.
\]

This means that $L[\Phi]$ is an element of $O(U,\eta;\: V,g)$, i.e., it is an orthogonal mapping of $(U,\eta)$ onto $(V,g)$. Just like orthogonal groups, the set $O(N,\eta;\: M,g)$ is non-connected; it is a disjoint union of two connected manifolds. If some orientation standards are fixed in $N$, $M$ (in $U$, $V$), those are manifolds of orientation-preserving and orientation-inverting isometries. The same concerns $E(U,\eta;\: V,g)$. The metrical rigid body configuration space is then identical with the manifold of orientation-preserving isometries.

In mechanics of affine bodies, when $\phi\in AfI(N,M)$, the transformation group (\ref{eq:34}) is restricted to one with $A$, $B$ being elements of $GAf(M)$, $GAf(N)$ respectively. And similarly, when some of the afore mentioned constraints are imposed on $\Phi$, then in (\ref{eq:34}) transformations $A$, $B$ are confined to run over appropriate subgroups of $GAf(M)$, $GAf(N)$.

We are particularly interested in geometry and dynamics of internal degrees of freedom, i.e., in what is going on in the manifold $Q_{int}=LI(U,V)$. There we are dealing with $GL(V)$ and $GL(U)$ acting as internal transformation groups. In analogy to (\ref{eq:34}) we have then the following transformations of $Q_{int}$ generated by $a\in GL(V)$, $b\in GL(U)$: 
\begin{equation}
\varphi\rightarrow a\:\varphi \:b.\label{eq:46}
\end{equation}

This is just exactly a consequence of (\ref{eq:34}) obtained by the application of the projection $L:Af(N,M)\rightarrow L(U,V)$ $a$, $b$ being images of $A$, $B$.

Affine constrains simplify the structure of many expressions used in mechanics of continua. The placement (\ref{eq:5}) becomes constant both in the physical space and in the body; it equals the internal part of the configuration, 
\begin{equation}
\chi^{i}\hspace{0in}_{K}=\varphi^{i}\!_{K}.\label{eq:47}
\end{equation}

Deformation tensors are also spatially and materially constant, for example the expressions (\ref{eq:13}), (\ref{eq:14}), (\ref{eq:17}) for the Green and Cauchy tensors become: 
\begin{equation}
\begin{array}{cc}
G_{KL}=g_{ij}\:\varphi^{i}\!_{K}\:\varphi^{j}\!_{L}, & C_{ij}=\eta_{KL}\:\varphi^{-1\ K}\!_{i}\:\varphi^{-1\ L}\!_{j},\\
\\
G^{-1\ KL}=\varphi^{-1\ K}\!_{i}\:\varphi^{-1\ L}\!_{j}\:g^{ij}, & C^{-1\ ij}=g_{ij}\:\varphi^{i}\!_{K}\:\varphi^{j}\!_{L}\eta^{KL}.
\end{array}\label{eq:48}
\end{equation}

The quantities $\Omega$, $\widehat{\Omega}$ became what in our earlier papers \cite{Gol2004} \cite{JJS1975} \cite{JJS1982} \cite{IJJSco2004} \cite{IIJJSco2005} was called affine velocity, respectively in spatial and co-moving representation.

Expression (\ref{eq:9}) becomes the following affine (``linear-inhomogeneous'') function of the Euler variables $y^{i}$: 
\begin{equation}
v^{i}(t,y)=\frac{dx^{i}}{dt}+\frac{d\varphi^{i}\!_{A}}{dt}\left.\varphi^{-1}\right.^{A}\!_{j}\left(y^{j}-x^{j}\right)=V(tr)^{i}+\Omega^{i}\!_{j}\left(y^{j}-x^{j}\right),\label{eq:49}
\end{equation}
therefore 
\begin{equation}
\frac{\partial v^{i}}{\partial y^{j}}=\Omega^{i}\hspace{0in}_{j}=\frac{d\varphi^{i}\hspace{0in}_{A}}{dt}\left.\varphi^{-1}\right.^{A}\hspace{0in}_{j},\label{eq:50}
\end{equation}
compare this with (\ref{eq:27}). In our earlier papers the quantity (\ref{eq:50}) was referred to as affine velocity; Eringen in his papers about micromorphic continua used the term ``gyration''. Obviously, $\Omega$ as defined in (\ref{eq:50}), i.e., in coordinates-free form, 
\begin{equation}
\Omega=\frac{d\varphi}{dt}\left.\varphi^{-1}\right.\label{eq:51}
\end{equation}
is an element of the tensor space $V\otimes V^{\star}\simeq L(V)$; $\Omega^{i}\hspace{0in}_{j}$ are spatial, or laboratory-referred components. The co-moving (material) representation, mentioned in (\ref{eq:28}) for general (not necessarily affine) bodies is given by 
\begin{equation}
\widehat{\Omega}=\varphi^{-1}\frac{d\varphi}{dt}=\varphi^{-1}\:\Omega\:\varphi\in U\otimes U^{\star}\simeq L(U),\label{eq:52}
\end{equation}
i.e., analytically, 
\begin{equation}
\widehat{\Omega}^{A}\!_{B}=\left.\varphi^{-1}\right.^{A}\!_{i}\:\frac{d\varphi^{i}\!_{B}}{dt}=\left.\varphi^{-1}\right.^{A}\!_{i}\:\Omega^{i}\!_{j}\:\varphi^{j}\!_{B}.\label{eq:53}
\end{equation}

In spite we are dealing here with the special case of (\ref{eq:27}) (\ref{eq:28}), the quantities $\Omega$, $\widehat{\Omega}$ are geometrically very peculiar. They are elements of the Lie algebras of $GL(V)$, $GL(U)$. When affine constraints are replaced by stronger ones, ruled by some subgroups of $GL(V)$, $GL(U)$, then $\Omega$, $\widehat{\Omega}$ become elements of the corresponding Lie subalgebras. For example, if we take $Q_{int}=O(U,\eta;\: V,g)$, or rather, its connected component, i.e., the configuration space of the usual rigid body (gyroscope) without translational degrees of freedom, then $\Omega$, $\widehat{\Omega}$ are respectively $g$-skew-symmetric and $\eta$-skew-symmetric angular velocities, 
\begin{equation}
\Omega^{i}\!_{j}=-\Omega_{j}\!^{i}=-g_{ja}\:g^{ib}\:\Omega^{a}\!_{b},\qquad\widehat{\Omega}^{A}\!_{B}=-\widehat{\Omega}_{B}\!^{A}=-\eta_{BC}\:\eta^{AD}\:\widehat{\Omega}^{C}\!_{D};\label{eq:54}
\end{equation}
they are elements of the commutator Lie algebras of $SO(V,g)'$, $SO(U,\eta)'$ of $SO(V,g)$, $SO(U,\eta)$. Another important example is that of incompressible affine body, when the internal configuration space $Q_{int}$ consists of volume-preserving (and in the continuum case orientation-preserving) linear mappings of $U$ onto $V$. This manifold is invariant under (\ref{eq:46}) with $a$, $b$ restricted to the subgroups $U(V)$, $\mathcal{U}(U)$, or, in the continuum case, to $SU(V)'$, $SL(U)'$. Then $\Omega$ and $\widehat{\Omega}$ as elements of Lie algebras $SL(V)'$, $SL(U)'$ are trace-less, 
\begin{equation}
Tr(\Omega)=\Omega^{i}\!_{i}=0,\qquad Tr(\widehat{\Omega})=\widehat{\Omega}^{A}\!_{A}=0.\label{eq:55}
\end{equation}

As all the mentioned groups are non-Abelian (except the planar orthogonal group when, $n=2$), the objects $\Omega$, $\widehat{\Omega}$ are non-holonomic velocities in the sense of Boltzmann, i.e., they are not time derivatives of any generalized coordinates.

In certain formulas it is convenient to use the co-moving representation of translational velocity, namely the quantity $\hat{v}(tr)\in U$ with components 
\begin{equation}
\hat{v}(tr)^{A}=e^{A}\!_{i}\:v(tr)^{i}=e^{A}\!_{i}\:\frac{dx^{i}}{dt}.\label{eq:56}
\end{equation}

Canonical momenta conjugate to generalized velocities $v(tr)^{i}$ and $\mathcal{V}^{i}\!_{A}=d\varphi^{i}\!_{A}/dt$, are denoted respectively as $p(tr)_{i}$, $P^{A}\!_{i}$; their tensorial membership $p(tr)\in V^{\star}$, $P\in U\otimes V^{\star}\simeq L(U,V)$, and the duality of $\mathcal{V}\in V\otimes U^{\star}$ and $P\in U\otimes V^{\star}$ is meant in the sense: 
\begin{equation}
\left\langle P,\mathcal{V}\right\rangle =Tr\left(P\mathcal{V}\right)=P^{A}\!_{i}\:\mathcal{V}^{i}\!_{A}.\label{eq:57}
\end{equation}

Canonical translational momenta conjugate to $\hat{v}\in U$ will be dented by $\hat{p}(tr)\in U^{\star}$; their components are given by 
\begin{equation}
\hat{p}(tr)_{A}=p(tr)_{i}\:\varphi^{i}\!_{A}.\label{eq:58}
\end{equation}

The non-holonomic momenta conjugate to $\Omega$, $\widehat{\Omega}$ will be denoted respectively by $\Sigma\in L(V)\simeq V\star V^{\star}\simeq L(V)$, $\widehat{\Sigma}\in L(U)\simeq U\otimes U^{\star}$. They are given by 
\begin{equation}
\Sigma^{i}\!_{j}=\varphi^{i}\!_{A}\:P^{A}\!_{j},\quad\widehat{\Sigma}^{A}\!_{B}=P^{A}\!_{i}\:\varphi^{i}\!_{B}=\left.\varphi^{-1}\right.^{A}\!_{i}\:\Sigma^{i}\!_{j}\:\varphi^{j}\!_{B}.\label{eq:59}
\end{equation}

Obviously, the duality between $\Sigma$, $\widehat{\Sigma}$ and $\Omega$, $\widehat{\Omega}$ is meant in the sense of trace: 
\begin{equation}
\left\langle \Sigma,\Omega\right\rangle =\left\langle \widehat{\Sigma},\widehat{\Omega}\right\rangle =Tr\left(\Sigma\:\Omega\right)=Tr\left(\widehat{\Sigma}\:\widehat{\Omega}\right)=\Sigma^{i}\!_{j}\:\Omega^{j}\!_{i}=\widehat{\Sigma}^{A}\!_{B}\:\widehat{\Omega}^{B}\!_{A}.\label{eq:60}
\end{equation}

The transformations (\ref{eq:46}) act on the above quantities according to the rules: 
\begin{equation}
\begin{array}{c}
\Omega\rightarrow a\:\Omega \:a^{-1},\quad\Sigma\rightarrow a\:\Sigma \:a^{-1},\quad\widehat{\Omega}\rightarrow b^{-1}\:\widehat{\Omega}\:b,\quad\widehat{\Sigma}\rightarrow b^{-1}\:\widehat{\Sigma}\:b,\\
\\
p(tr)\rightarrow p(tr)\circ a^{-1},\qquad\hat{p}(tr)\rightarrow\hat{p}(tr)\circ b;
\end{array}\label{eq:61}
\end{equation}
analytically the last formulas have the form 
\begin{equation}
p(tr)'_{i}=p(tr)_{j}\left.a^{-1}\right.^{j}\!_{i},\qquad\hat{p}(tr)'_{A}=\hat{p}(tr)_{B}b^{B}\!_{A}.\label{eq:62}
\end{equation}

Transformation rules for deformation tensor are more complicated and there are some subtle points about them. Namely, for the Cauchy tensor we have 
\begin{equation}
C\left[a\varphi\right]_{ij}=C\left[\varphi\right]_{kl}\left.a^{-1}\right.^{k}\!_{i}\left.a^{-1}\right.^{l}\!_{j},\label{eq:63}
\end{equation}
i.e, symbolically, 
\begin{equation}
C\left[a\varphi\right]=a_{\star}\:\varphi.\label{eq:64}
\end{equation}

On the other hand, for a general $b\in GL(U)$, there is no explicit expression for $C\left[\varphi b\right]$ as an algebraic function of $C\left[\varphi\right]$. But for isometries of $\left(U,\eta\right)$ we have obviously 
\begin{equation}
C\left[\varphi b\right]=C\left[\varphi\right],\quad b\in O\left(U,\eta\right).\label{eq:65}
\end{equation}

Transformation properties of the Green deformation tensor are dual to the above ones. So, for a general $a\in GL(V)$ there is no concise relationship between $G\left[a\varphi\right]$ and $G\left[\varphi\right]$ although for isometries of $\left(V,g\right)$ we have obvious invariance rule 
\begin{equation}
G\left[a\varphi\right]=G\left[\varphi\right],\quad a\in O\left(V,g\right).\label{eq:66}
\end{equation}

On the other hand, for a general $b\in GL\left(U\right)$, the following covariance is satisfied: 
\begin{equation}
G\left[\varphi b\right]_{KL}=G\left[\varphi\right]_{MN}b^{M}\!_{K}\:b^{N}\!_{L},\label{eq:67}
\end{equation}
i.e., symbolically, 
\begin{equation}
G\left[\varphi b\right]=b^{\star}\:G\left[\varphi\right].\label{eq:68}
\end{equation}

By their very definition, the deformation invariants (\ref{eq:23}) are preserved only by isometries, 
\begin{equation}
\mathcal{K}_{p}\left[a\varphi b\right]=\mathcal{K}_{p}\left[\varphi\right],\quad a\in O\left(V,g\right),\quad b\in O\left(U,\eta\right),\label{eq:69}
\end{equation}
but not under the larger subgroups of $GL\left(V\right)$, $GL\left(U\right)$. Concerning the last statement, an exception does exist, namely, the ``isochoric'' invariant $\mathcal{K}_{n}\left[\varphi\right]$, which is preserved by (\ref{eq:46}) with $a\in SL\left(V\right)$, $b\in SL\left(U\right)$, or, when admitting the orientation-inverting mappings, $a$, $b$ being unimodular $a\in UL\left(V\right)$, $b\in UL\left(U\right)$.

The above transformation rules are fundamental for analysis of symmetries of equations of motion, Lagrangians, etc.

Another very important tool is the system of basic Poisson brackets. First of all, let us mention that the quantities $\Sigma^{i}\!_{j}$, $\widehat{\Sigma}^{A}\!_{B}$ are Hamiltonian generators of (\ref{eq:46}), respectively for the $a$- and $b$-transformations. Because of this they are called the canonical components of affine spin, respectively in the spatial and co-moving (material) representation. Their doubled $g$-skew-symmetric and $\eta$-skew-symmetric parts generate respectively the action of $O(V,g)$ and $O(U,\eta)$ through (\ref{eq:46}). They are respectively components of the canonical spin and vorticity \cite{Dys1968},
\begin{eqnarray}
S^{i}\!_{j} & = & \Sigma^{i}\!_{j}-\Sigma_{j}\!^{i}=\Sigma^{i}\!_{j}-g_{jk}\:\Sigma^{k}\!_{l}\:g^{li},\label{eq:70}\\
V^{A}\!_{B} & = & \widehat{\Sigma}^{A}\!_{B}-\widehat{\Sigma}_{B}\!^{A}=\widehat{\Sigma}^{A}\!_{B}-\eta_{BC}\:\widehat{\Sigma}^{C}\!_{D}\:g^{DA}.\label{eq:71}
\end{eqnarray}

\textbf{Remark-warning}: Unlike the relationships between $\Sigma$ and $\widehat{\Sigma}$, $V^{A}\,_{B}$ are NOT co-moving components of spin, 
\begin{equation}
V^{A}\!_{B}\neq\left.\varphi^{-1}\right.^{A}\!_{i}\:S^{i}\!_{j}\:\varphi^{j}\!_{B},\label{eq:72}
\end{equation}
unless motion is metrically-rigid ($\varphi\in O\left(U,\eta;\: V,g\right)$ is an isometry), and in the latter case $V^{A}\!_{B}$ are just the co-moving components of spin.

One can also introduce the translational, i.e., orbital affine momentum with respect to some fixed spatial point $\mathfrak{o}\in M$ and the total affine momentum with respect to that origin. They are respectively given by 
\begin{equation}
\Lambda\left(\mathfrak{o}\right)^{i}\!_{j}=x^{i}p\left(tr\right)_{j},\qquad J\left(\mathfrak{o}\right)^{i}\!_{j}=\Lambda\left(\mathfrak{o}\right)^{i}\!_{j}+\Sigma^{i}\!_{j};\label{eq:73}
\end{equation}
when there is no danger of misunderstanding, the ``label'' $\mathfrak{o}\in M$ is omitted.

Obviously, the quantities $J\left(\mathfrak{o}\right)^{i}\!_{j}$ are Hamiltonian generators of $\mathfrak{o}$-centered affine mappings (ones preserving $\mathfrak{o}$) (\ref{eq:34}) with $B=id_{N}$ analytically given by 
\begin{equation}
'y^{i}=A^{i}\!_{j}\:y^{j},\quad\textrm{i.e.,}\quad'x^{i}=A^{i}\!_{j}\:x^{j},\quad'\varphi^{i}\!_{K}=A^{i}\!_{j}\:\varphi^{j}\!_{K};\label{eq:74}
\end{equation}
obviously, it is assumed here that coordinates $y^{i}$ vanish at the fixed origin $\mathfrak{o}\in M$.

Translations in $M$ are generated by canonical translational momenta $p\left(tr\right)_{i}$ as Hamiltonian generators. Let us now quote the mentioned basic Poisson brackets 
\begin{eqnarray}
\left\{ \Sigma^{i}\!_{j},\Sigma^{k}\!_{l}\right\}  & = & \delta^{i}\!_{l}\:\Sigma^{k}\!_{j}-\delta^{k}\!_{j}\:\Sigma^{i}\!_{l}\label{eq:75}\\
\left\{ \widehat{\Sigma}^{A}\!_{B},\widehat{\Sigma}^{C}\!_{D}\right\}  & = & \delta^{C}\!_{B}\:\widehat{\Sigma}^{A}\!_{D}-\delta^{A}\!_{D}\:\widehat{\Sigma}^{C}\!_{B}\label{eq:76}\\
\left\{ \Sigma^{i}\!_{j},\widehat{\Sigma}^{A}\!_{B}\right\}  & = & 0\label{eq:77}\\
\left\{ \widehat{\Sigma}^{A}\!_{B},\hat{p}_{C}\right\}  & = & \delta^{A}\!_{C}\:\hat{p}_{B}\label{eq:78}\\
\left\{ J^{i}\!_{j},p_{k}\right\}  & = & \left\{ \Lambda^{i}\!_{j},p_{k}\right\} =\delta^{i}\!_{k}\:p_{j}.\label{eq:79}
\end{eqnarray}

If $F$ depends only on the configuration variables $\left(x^{i},\varphi^{i}\!_{K}\right)$, then 
\begin{equation}
\left\{ F,\:\Sigma^{i}\!_{j}\right\} =\varphi^{i}\!_{K}\frac{\partial F}{\partial\varphi^{j}\!_{K}},\quad\left\{ F,\Lambda^{i}\!_{j}\right\} =x^{i}\frac{\partial F}{\partial x^{j}},\quad\left\{ F,\widehat{\Sigma}^{K}\!_{L}\right\} =\varphi^{i}\!_{L}\frac{\partial F}{\partial\varphi^{i}\!_{K}}\label{eq:80}
\end{equation}
and if both $F$, $G$ depend only on configuration, then obviously
\begin{equation}
\left\{ F,G\right\} =0.\label{eq:81}
\end{equation}

Using the basic rules, the standard properties of Poisson bracket as a Lie bracket, and an additional important property: 
\begin{equation}
\left\{ f\left(K\right),G\right\} =f'\left(K\right)\left\{ K,G\right\} ,\label{eq:82}
\end{equation}
one can easily calculate any Poisson bracket and write down equations of motion as a system of equations of the form 
\begin{equation}
\frac{dF}{dt}=\left\{ F,H\right\} ,\label{eq:83}
\end{equation}
where $H$ is a Hamiltonian and $F$ runes over some maximal functionally independent system of phase space functions. This way of deriving equations of motion is much more effective than the direct use of Lagrange equations. Obviously, only non-dissipative, variational models may be studied in terms of the standard Lagrange or Hamiltonian formalism. Nevertheless, once derived in this way, equations of motion may be easily generalized to ones admitting dissipation, by introducing some more or less phenomenological friction terms.

\section{Affine dynamics. Generalized forces, balance laws, d'Alembert principle, affine Euler equations.}

Before going any further, we must introduce some additional concepts, more ``touchable'' from the practical mechanical point of view. The co-moving (Lagrangian) mass distribution is described by the time-independent positive measure $\mu$ on the material space $N$. Any configuration $\Phi:N\rightarrow M$, not necessarily affine one, gives rise to the current (Euler) mass distribution in the physical space $M$. It is described by the positive measure $\mu_{\Phi}$ on $M$, the $\Phi$-transport of $\mu$. Obviously, it is time-dependent because
$\Phi$ is so. The total mass of the body is obviously given by 
\begin{equation}
m=\mu(N)=\underset{N}{\int}d\mu(a)=\mu_{\Phi}(M)=\underset{M}{\int}d\mu_{\Phi}(x),\label{eq:84}
\end{equation}
roughly speaking, the monopole moment of $\mu$ or $\mu_{\Phi}$. Higher-order material (Lagrangian, co-moving) moments of inertia are given by the family of constant tensors in $N$, $J(k)\in\underset{k}{\otimes}U$; analytically they are given by 
\begin{eqnarray}
J(1)^{K} & = & \int a^{K}d\mu(a)\nonumber \\
J(2)^{KL} & = & \int a^{K}a^{L}d\mu(a)\nonumber \\
 & \vdots\label{eq:85}\\
J(k)^{A_{1}\ldots A_{k}} & = & \int a^{A_{1}}\cdots a^{A_{k}}d\mu(a).\nonumber \\
 & \vdots\nonumber 
\end{eqnarray}

All those tensors are symmetric. There are three important special cases: $J(o)=m$, $J(1)\in U$, $J(2)\otimes U\otimes U$. Usually, and so we do in this paper, the origin of material (Lagrange) coordinates is chosen in the reference center of mass, so be definition, cf. also (\ref{eq:41}) 
\begin{equation}
J(1)=0.\label{eq:86}
\end{equation}

All the inertial tensors quoted here contain some important information: roughly speaking, in typical simple situations, the knowledge of the system of all $J(k)$-s is essentially equivalent to the knowledge of $\mu$. Nevertheless, in the special case of affine motion it is only $J(2)\in U\otimes U$ that is relevant for dynamics. Then it is denoted simply as $J$ without the label ``$2$'' and referred to as the ``inertial tensor''. It is not the same what is called ``tensor of inertia'' in mechanics of rigid bodies, however, they are essentially equivalent concepts. More precisely, they are linear functions of each other. I describes the inertia of rotational and homogeneously-deformative modes of motion. In some formulas we need also the covariant inverse of $J$ denoted by $J^{-1}\in U^{*}\otimes U^{*}$,
\begin{equation}
J^{AC}\left.J^{-1}\right._{CB}=\delta^{A}\!_{B}.\label{eq:87}
\end{equation}

\textbf{Remark}; a very important warning, like one concerning (\ref{eq:18}): Do not confuse $J^{-1}$ with $J_{\eta}\in U^{*}\in U^{*}$ obtained form $J$ by the $\eta$-lowering of indices, 
\begin{equation}
\left.J_{\eta}\right._{AB}:=\eta_{AK}\:\eta_{BL}\:J^{KL}.\label{eq:88}
\end{equation}
In general $J_{\eta}\neq J^{-1}$. In certain formulas one uses also the mixed tensor $\widehat{J}_{\eta}\in U\otimes U^{*}\simeq L(U)$ obtained from $J$ by the $\eta$-lowering of the second index, 
\begin{equation}
\left.\widehat{J}_{\eta}\right.^{A}\!_{B}:=J^{AC}\:\eta_{CB}.\label{eq:89}
\end{equation}

One must be careful and avoid confusing $J^{-1}$, $J$, $J_{\eta}$, $\widehat{J}_{\eta}$. An important points is that $J^{-1}$, $J$ are purely affine concepts, whereas $J_{\eta}$, $\widehat{J}_{\eta}$ are partially metrical.

In certain considerations one uses also the Eulerian version of $J$. More precisely, there are two such versions: $J[\phi,\mathfrak{o}]$, $J[\varphi]$, both being symmetric tensors, elements of $V\otimes V$. Obviously, $\mathfrak{o}\in M$ denotes here the fixed origin in $M$ given by coordinates $y^{i}=0$ . $J[\phi,\mathfrak{o}]$ denotes the Eulerian inertial tensor with respect to the fixed spatial origin, and $J[\varphi]$ is the Eulerian inertial tensor related to the current position of the center of mass in the physical space, $\mathfrak{o}_{\phi}\in M$. Let us remind that in affine motion (and only then) $\mathfrak{o}_{\phi}=\phi(\mathfrak{O})$. It is easy to show that the following formulas hold: 
\begin{eqnarray}
J[\varphi]^{ij} & = & \varphi^{i}\!_{A}\:\varphi^{j}\!_{B}\:J^{AB},\label{eq:90}\\
J[\Phi,\mathfrak{o}]^{ij} & = & m\:x^{i}x^{j}+J[\varphi]^{ij}.\label{eq:91}
\end{eqnarray}

Unlike $J$, which is constant and depends only on geometry of the mass distribution in the body, $J[\varphi]$ is configuration-dependent, therefore, also time-dependent. Nevertheless, $J[\varphi]$ is sometimes useful as a subsystem of alternative generalized coordinates.

Eulerian multipoles may be also introduced for higher values of $k$. However, for affine bodies we do not need them, and for non-affine configurations they do not admit the nice orbital-internal splitting (\ref{eq:91}).

In dynamics of multiparticle systems, including continua, it is also convenient to use multipole moments for distributions of other physical quantities like linear momentum and forces. When dealing with affine motion we need only monopole and dipole moments of those distributions in both Euler (spatial) and Lagrange (co-moving) representations.

In a general non-constrained motion the $l$-th order multipole moment of the distribution of linear momentum, calculated with respect to the fixed spatial origin $\mathfrak{o}\in M$ is analytically given by 
\begin{eqnarray}
\mathcal{K}_{\mathfrak{o}}(l)^{i_{1}\ldots i_{l}\: j} & = & \int y^{i_{1}}\ldots y^{i_{l}}v^{j}(y)d\mu_{\phi}(y)=\label{eq:92}\\
 & = & \int y(a)^{i_{1}}\ldots y(a)^{i_{l}}V^{j}(a)d\mu(a).\nonumber 
\end{eqnarray}

As usual, $\mu_{\phi}$ denotes the $\phi$-transport of the measure $\mu$ from $N$ to $M$, and $v$, $\mathcal{V}$ are, respectively, the Euler and Lagrange velocity fields (\ref{eq:9}) (\ref{eq:8}); the coordinates $y^{i}$ are assumed to vanish at the origin $\mathfrak{o}\in M$. The tensor $\mathcal{K}_{\mathfrak{o}}(l)$ is symmetric in the first $l$-tuple of indices. It depends explicitly on the choice of $\mathfrak{o}\in M$. One uses also quantities $\mathcal{K}_{int}(l)$ for which the fixed reference point $\mathfrak{o}$ is replaced by the current position of the center of mass $\mathfrak{o}_{\phi}$, 
\begin{equation}
\mathcal{K}_{int}(l)^{i_{1}\ldots i_{l}\: j}=\int\left(y^{i_{1}}-x^{i_{1}}\right)\cdots\left(y^{i_{l}}-x^{i_{l}}\right)v^{i}(y)d\mu_{\phi}(y).\label{eq:93}
\end{equation}

Being tensor in $V$, $\mathcal{K}_{\mathfrak{o}}(l)$, $\mathcal{K}_{int}(l)$ are Euler-like quantities. In certain formulas it is more convenient to use the multipole moments with respect to Lagrange variables $a^{B}$; we denote them by $\widetilde{\mathcal{K}}(l)$. 
\begin{equation}
\widetilde{\mathcal{K}}(l)^{A_{1}\ldots A_{l}\: j}=\int a^{A_{1}}\ldots a^{A_{l}}\mathcal{V}^{j}(a)d\mu(a).\label{eq:94}
\end{equation}

This is a mixed tensorial quantity, namely, the $l$-th order symmetric contravariant tensor in $U$ and the usual vector in $V$. In spite of its being injected in the ``abstract'' material space, in many formulas $\widetilde{\mathcal{K}}(l)$ is more convenient and effective than the purely spatial $\mathcal{K}(l)$.

Obviously, the monopole moments are identical with the total linear momentum in its kinematical version ``inertia $\times$ velocity; we shall use the symbol: 
\begin{equation}
k^{i}:=\mathcal{K}_{\mathfrak{o}}(0)^{i}=\widetilde{\mathcal{K}}(0)^{i}=m\frac{dx^{i}}{dt}=mv(tr)^{i}.\label{eq:95}
\end{equation}

As is well-known, in analytical mechanics there is a subtle distinction between canonical linear momentum $p(tr)_{i}$ and kinematical linear momentum $k^{i}$. The relationships between them is a dynamical concept, not kinematical one. Below we shall return to this problem.

The dipole moments $\mathcal{K}_{\mathfrak{o}}(1)$, $\mathcal{K}_{int}(1)$, $\widetilde{\mathcal{K}}(1)$ have to do with velocities of rotational and homogeneously deformative motion, combined multiplicatively with appropriate internal objects. The doubled skew-symmetric parts of $\mathcal{K}_{\mathfrak{o}}(1)$, $\mathcal{K}_{int}(1)$ represent respectively the total angular momentum with respect to the fixed origin $\mathfrak{o}\in M$ and the spin angular momentum (one with respect to the center of mass), both in kinematical versions (vector product of the radius vectors and kinematical linear momenta).

In affine motion we obtain for $l=1$: 
\begin{equation}
\mathcal{K}_{\mathfrak{o}}\!^{ij}=m\:x^{i}v(tr)^{j}+\varphi^{i}\!_{A}\frac{d\varphi^{j}\!_{B}}{dt}J^{AB}.\label{eq:96}
\end{equation}
This is just the additive splitting into translational (``orbital'') and internal (``affine spin'') parts, 
\begin{equation}
\left.\mathcal{K}_{\mathfrak{o}\, tr}\right.^{ij}=m\:x^{i}v(tr)^{j},\:\left.\mathcal{K}_{int}\right.^{ij}=\varphi^{i}\!_{A}\frac{d\varphi^{j}\!_{B}}{dt}J^{AB}.\label{eq:97}
\end{equation}

The internal part does not depend on the choice of $\mathfrak{o}$. To avoid the crowd of symbols we denote it simply by $\mathcal{K}^{ij}$ without the label ``$int$''. The corresponding translational and internal angular momenta, both meant in the kinetic sense, are denoted respectively by: 
\begin{equation}
\mathcal{L}_{\mathfrak{o}}\!^{ij}=\left.\mathcal{K}_{\mathfrak{o}\, tr}\right.^{ij}-\left.\mathcal{K}_{\mathfrak{o}\, tr}\right.^{ji},\quad\mathcal{S}^{ij}=\left.\mathcal{K}\right.^{ij}-\left.\mathcal{K}\right.^{ji}.\label{eq:98}
\end{equation}

The total angular momentum with respect to $\mathfrak{o}\in M$ is denoted by 
\begin{equation}
\mathfrak{J}_{\mathfrak{o}}\!^{ij}=\mathcal{L}_{\mathfrak{o}}\!^{ij}+\mathcal{S}^{ij}.\label{eq:99}
\end{equation}

Obviously, in the physical dimension $n=3$, the skew-symmetric tensors (\ref{eq:98}) (\ref{eq:99}) are identified with the corresponding axial vectors with components $\mathcal{L}_{\mathfrak{o}}\!^{i}$, $\mathcal{S}^{i}$, $\mathfrak{J}_{\mathfrak{o}}\!^{i}$.

The nice and intuitive translational-internal splitting like (\ref{eq:96}) holds also for non-affine bodies, but only for the dipole moments, i.e., for $\mathcal{K}_{\mathfrak{o}}(1)$, no longer for $\mathcal{K}_{\mathfrak{o}}(\ell)$ with $\ell>1$. This fact has some deep geometric reasons. Unlike this, all Lagrange moments $\widetilde{\mathcal{K}}(\ell)$ split in this way.

It is convenient to express also the distribution of forces in terms of its multipole moments. These moments play the role of generalized forces responsible for the dynamics of some collective models of motion. As expected, the monopole and dipole moments of forces are sufficient for describing affine motion.

Multipole moments $N_{o}(l)$, $N_{int}(l)$, $\widetilde{N}(l)$ of the distribution of forces are given by formulas obtained form (\ref{eq:92}) (\ref{eq:93}) (\ref{eq:94}) by substituting instead velocity field the density of forces per unit mass. This density at a given particle $a\in N$ and the time instant $t\in\mathbb{R}$ will be denoted by $\mathcal{F}^{i}[t,a;\Phi,\frac{\partial\Phi}{\partial t}]$; analytically $\mathcal{F}^{i}[t,a;y^{i},\frac{\partial y^{i}}{\partial t}]$. The dependence on configuration $\Phi$ (analytically $y^{j}$ as functions of $a^{k}$) and generalized velocity ($\frac{\partial y^{j}}{\partial t}$ as functions of $a^{k}$) may be functional, in particular non-local in space and time (memory); this is the reason of using the brackets symbols $[\:]$. However, in applications we have in mind, mainly simple bodies (usually elastic or viscoelastic), $\mathcal{F}^{i}$ will be of the form $\mathcal{F}^{i}\left(t,a;y^{j}(t,a),\chi^{j}\!_{K}(t,a),\frac{\partial y^{j}}{\partial t}(t,a)\right)$, i.e., local and memory-free. Quite independently on any particular model, the meaning of $\mathfrak{\mathcal{F}}^{i}$ is that the quantity
\begin{equation}
dF^{i}\left[t,a;\Phi,\frac{\partial\Phi}{\partial t}\right]:=\mathcal{F}^{i}\left[t,a;\Phi,\frac{\partial\Phi}{\partial t}\right]d\mu(a)\label{eq:100}
\end{equation}
is the force acting on the material mass element $d\mu(a)$ at the time instant $t$, when configuration is given by $\Phi$. Substituting $\mathcal{F}^{j}$ instead $\mathcal{V}^{j}$ in formulas (\ref{eq:92}) (\ref{eq:93}) (\ref{eq:94}), we obtain, as mentioned, generalized forces $N_{\mathfrak{o}}(l)$, $N_{int}(l)$, $\widetilde{N}(l)$ controlling the dynamics of some collective modes. And again for $l=1$ (and only then) we have the orbital-internal splitting, 
\begin{equation}
N_{\mathfrak{o}}\!^{ij}=N_{\mathfrak{o}\, tr}\!^{ij}+N_{int}\!^{ij}=x^{i}F^{j}+N_{int}\!^{ij},\label{eq:101}
\end{equation}
where the monopole 
\begin{equation}
F^{j}=\int\mathcal{F}^{j}d\mu(a)\label{eq:102}
\end{equation}
is the total force acting on the system (in a sense on its center of mass).

The quantity $N_{\mathfrak{o}}^{ij}$ will be also referred to as an affine moment of forces or ``affine torque'' with respect to the origin $\mathfrak{o}\in M$, and $N_{int}^{ij}$ is an ``internal affine torque''; it is related to the center of mass as the ``origin of lever''. Their doubled skew-symmetric parts are usual torques (moment of forces), 
\begin{equation}
\mathcal{N}_{\mathfrak{o}}\!^{ij}:=N_{\mathfrak{o}}\!^{ij}-N_{\mathfrak{o}}\!^{ji},\qquad\mathcal{N}_{int}\!^{ij}:=N_{int}\!^{ij}-N_{int}\!^{ji}.\label{eq:103}
\end{equation}

In affine motion, when configurations are given by (\ref{eq:35}) and placements by (\ref{eq:36}), all these generalized forces are functions of $x^{i},\frac{dx^{i}}{dt};\varphi^{i}\!_{K},\frac{d\varphi^{i}\!_{K}}{dt}$ and possibly of time $t$ explicitly, 
\begin{equation}
F^{i}\left(t;x^{k},\frac{dx^{k}}{dt};\varphi^{k}\!_{A},\frac{d\varphi^{k}\!_{A}}{dt}\right),\qquad N^{ij}\left(t;x^{k},\frac{dx^{k}}{dt};\varphi^{k}\!_{A},\frac{d\varphi^{k}\!_{A}}{dt}\right).\label{eq:104}
\end{equation}

The power $\mathcal{P}$ of distribution of forces is given by 
\begin{equation}
\mathcal{P}=\int g_{ij}\mathcal{F}^{i}\left[t,a;y(\cdot),\frac{\partial y}{\partial t}(\cdot)\right]\frac{\partial y^{j}}{\partial t}(t,a)d\mu(a).\label{eq:105}
\end{equation}
It is easy to show that in affine motion this becomes 
\begin{equation}
\mathcal{P}=g_{ij}\:F^{i}\:v^{j}+g_{mj}\:N^{im}\:\Omega^{j}\!_{i}=G_{AB}\:\widehat{F}^{A}\:\hat{v}^{B}+G_{CB}\:\widehat{N}^{AC}\:\widehat{\Omega}^{B}\!_{A},\label{eq:105a}
\end{equation}
where for simplicity the arguments of $F$, $N$ are omitted (cf (\ref{eq:104})), $v^{j}$ is an abbreviation for $v(tr)^{j}$, and $\Omega$ denotes ``gyration'', i.e., affine velocity, cf. (\ref{eq:49}), (\ref{eq:50}), (\ref{eq:51}). For simplicity the label ``$tr$'' at $N^{ij}$ is omitted.

Non-constrained equations of motion have the form: 
\begin{equation}
\frac{\partial^{2}y^{i}}{\partial t^{2}}(t,a)=\mathcal{F}^{i}\left[t,a;\: y^{j}(\cdot,\cdot),\frac{\partial y^{j}}{\partial t}(\cdot,\cdot)\right],\label{eq:106}
\end{equation}
and the non-constrained kinetic energy is given by 
\begin{equation}
T=\frac{1}{2}\int g_{ij}\:\frac{\partial y^{i}(t,a)}{\partial t}\frac{\partial y^{j}(t,a)}{\partial t}\:d\mu(a).\label{eq:106a}
\end{equation}

When affine constraints are imposed, then $y^{i}$ must be expressed in terms of generalized coordinates $(x^{i},\varphi^{i}\!_{A})$ like in (\ref{eq:35}), and, according to the d'Alembert principle, the right-hand side of (\ref{eq:106}) must be modified by introducing the distribution $\mathcal{F}_{R}\!^{i}$ of reaction forces maintaining the constraints, 
\begin{equation}
\frac{\partial^{2}y^{i}}{\partial t^{2}}(t,a)=\mathcal{F}^{i}\left[t,a;\: y^{j}(\cdot,\cdot),\frac{\partial y^{j}}{\partial t}(\cdot,\cdot)\right]+\mathcal{F}_{R}\!^{i}\left[t,a;\: y^{j}(\cdot,\cdot),\frac{\partial y^{j}}{\partial t}(\cdot,\cdot)\right].\label{eq:107}
\end{equation}

Substituting to (\ref{eq:106a}) the parametric description of constraints, (\ref{eq:35}), one obtains: 
\begin{equation}
T=T_{tr}+T_{int}=\frac{m}{2}\:g_{ij}\:\frac{dx^{i}}{dt}\frac{dx^{j}}{dt}+\frac{1}{2}\:g_{ij}\:\frac{d\varphi^{i}\!_{A}}{dt}\frac{d\varphi^{j}\!_{B}}{dt}\:J^{AB},\label{eq:107a}
\end{equation}
where, obviously, $T_{tr},T_{int}$ denote respectively the kinetic energies of translational and internal motion. This splitting, i.e., the absence of translational-internal interference terms is based on the assumption (\ref{eq:86}) which tells us that the origin of Lagrange coordinates is chosen at the reference center of mass. Expressions (\ref{eq:35}), as any constraints, may be, or rather should, be automatically substituted to the kinetic energy (\ref{eq:106a}), however, it would be absolutely wrong to substitute them to reactions-free equation of the unconstrained motion (\ref{eq:106}). They are to be substituted to (\ref{eq:107}) together with constitutive laws for reactions.

Constraints are assumed to be ideal (passive), i.e., the power of reactions $\mathcal{P}_{R}$((\ref{eq:105}) with $\mathcal{F}^{i}\!_{R}$ substituted instead $\mathcal{F}^{i}$) does vanish on any virtual velocity compatible with (\ref{eq:35}), 
\begin{equation}
\mathcal{V}^{i}=\frac{dx^{i}}{dt}+\frac{d\varphi^{i}\!_{K}}{dt}\: a^{K},\label{eq:108}
\end{equation}
thus 
\begin{equation}
\mathcal{P}\!_{R}=g_{ij}\left.F\!_{R}\right.^{i}v^{j}+g_{mj}\:N_{R}\!^{im}\:\Omega^{j}\!_{i}=G_{AB}\left.\widehat{F}\!_{R}\right.^{A}\hat{v}^{B}+G_{CB}\:\widehat{N}_{R}\!^{AC}\:\widehat{\Omega}^{B}\!_{A},\label{eq:109}
\end{equation}
for any $(v^{j},\Omega^{i}\!_{j})$. But this means that the monopole and dipole moments of (\ref{eq:107}) are free of reactions; the total force of reactions and their affine torque do vanish,  
\begin{equation}
F_{R}\!^{i}=0,\qquad N_{R}\!^{ij}=0.\label{eq:110}
\end{equation}

And finally we obtain the effective system of ordinary second-order differential equations imposed on the time dependence of generalized coordinates $x^{i},\varphi^{i}\!_{A}$: 
\begin{eqnarray}
m\:\frac{d^{2}x^{i}}{dt^{2}} & = & F^{i}\left(t;\: x^{j},\frac{dx^{j}}{dt};\varphi^{j}\!_{B},\frac{d\varphi^{j}\!_{B}}{dt}\right),\label{eq:111}\\
\varphi^{i}\!_{K}\:\frac{d^{2}\varphi^{j}\!_{L}}{dt^{2}}\:J^{KL} & = & N^{ij}\left(t;\: x^{j},\frac{dx^{j}}{dt};\varphi^{j}\!_{B},\frac{d\varphi^{j}\!_{B}}{dt}\right),\label{eq:112}
\end{eqnarray}
or, in the balance form: 
\begin{equation}
\frac{dk^{i}}{dt}=F^{i},\label{eq:113}
\end{equation}
\begin{equation}
\frac{dK^{ij}}{dt}=N^{ij}+\frac{d\varphi^{i}\!_{A}}{dt}\frac{d\varphi^{j}\!_{B}}{dt}\:J^{AB}=\Omega^{i}\!_{k}\:\Omega^{j}\!_{m}\: J[\varphi]^{km}+N^{ij}.\label{eq:114}
\end{equation}

As previously, $K^{ij}$ written without any label is an abbreviation for $K_{int}(2)^{ij}$, the internal affine momentum (``hyperspin''); similarly, $N^{ij}$ denotes the affine torque (``hyperforce'', ``dipole of forces'') $N_{int}(2)^{ij}$ the ``lever arm'' of which originates at the center of mass.

There is an important difference between (\ref{eq:113}) and (\ref{eq:114}). Namely, if the total force vanishes, $F^{i}=0$, then (\ref{eq:113}) becomes the conservation law of the linear momentum (in kinematical form). Unlike this, (\ref{eq:114}) does not become the conservation law for affine spin even if the affine torque vanishes. An obstacle is given by the second term on the right-hand side of (\ref{eq:114}). If one takes into account (\ref{eq:113}), then for any choice of the spatial origin $\mathcal{\mathfrak{o}}\in M$, (\ref{eq:114}) may be written as: 
\begin{equation}
\frac{dK_{\mathfrak{o}}\!^{ij}}{dt}=\left.N\!_{\mathfrak{o}}\right.^{ij}+m\left.v(tr)\right.^{i}\left.v(tr)\right.^{j}+\frac{d\varphi^{i}\!_{A}}{dt}\frac{d\varphi^{j}\!_{B}}{dt}\:J^{AB}.\label{eq:115}
\end{equation}
Equations (\ref{eq:114}) (\ref{eq:115}) may be concisely written in the following suggestive form, a bit symbolic one: 
\begin{equation}
\frac{dK^{ij}}{dt}=N^{ij}+2\:\frac{\partial T_{int}}{\partial g_{ij}},\label{eq:116}
\end{equation}
\begin{equation}
\frac{dK\!_{\mathfrak{o}}\!^{ij}}{dt}=N\!_{\mathfrak{o}}\!^{ij}+2\:\frac{\partial T}{\partial g_{ij}}.\label{eq:117}
\end{equation}

The parametric dependence of $T_{int}$, $T$ on the metric tensor $g\in V^{*}\otimes V^{*}$ and the resulting terms on the right hand sides, just visualize the fact that the non-conservation of $K^{ij}$, $K\!_{\mathfrak{o}}\!^{ij}$, even in the interaction-free case, follows from the metrical breaking of affine symmetry of degrees of freedom. And it is very intuitive that the internal and total angular momenta $\mathcal{S}^{ij}$, $\mathfrak{J}\!_{\mathfrak{o}}\!^{ij}$; i.e., the doubled skew-symmetric parts of $K^{ij}$, $K\!_{\mathfrak{o}}\!^{ij}$, are conserved quantities when the corresponding affine torques $N^{ij}$, $N\!_{\mathfrak{o}}\!^{ij}$ do vanish; moreover, they are conserved when affine torques are symmetric tensors, i.e., when the usual torques (\ref{eq:103}) do vanish. This is just the conservation of kinematical angular momentum, or its balance, when $\mathcal{N}^{ij}$, $\mathcal{N}\!_{\mathfrak{o}}\!^{ij}$ are non-vanishing. There is an obvious similarity to the relationship between conservation of angular momentum and symmetry of the Cauchy stress tensor in continuum mechanics. This is quite natural because the volume density of contact forces in continuum is given by the divergence of the Cauchy stress field $\sigma^{ij}$: 
\begin{equation}
f^{i}=\sigma^{ij}\!,j.\label{eq:118}
\end{equation}

The corresponding expression in terms of Lagrange variables has the form: 
\begin{equation}
\widetilde{f}^{i}=T^{Ki}\!,_{K}=\frac{\partial}{\partial a^{K}}\:T^{Ki},\label{eq:119}
\end{equation}
where $T$ denotes the first Piola-Kirchhoff stress tensor. Substituting this to the definitions of $N^{ij}$, $\widetilde{N}^{Ki}$ performing integration by parts and taking into account the conditions at infinity, one obtains: 
\begin{equation}
N^{ji}=-\int\sigma^{ij},\quad\widetilde{N}^{Ki}=-\int T^{Ki},\label{eq:120}
\end{equation}
\[
\widehat{N}^{AB}=-\int T^{AB}
\]
where, $T^{AB}$ is the second Piola-Kirchhoff stress tensor. Obviously, integration is performed respectively with respect to the spatial and material rectangular coordinates. Expression in terms of general coordinates is obvious, for example in (\ref{eq:119}) (\ref{eq:120}) one must then replace the usual partial differentiation by the covariant one.

It is interesting to rewrite the balance equations (\ref{eq:113}) (\ref{eq:114}) in terms of the co-moving and mixed representations. Using the co-moving components: 
\begin{eqnarray}
\hat{k}^{A}=\left.\varphi^{-1}\right.^{A}\!_{i}\:k^{i} & , & \widehat{K}^{AB}=\left.\varphi^{-1}\right.^{A}\!_{i}\left.\varphi^{-1}\right.^{B}\!_{j}\:K^{ij},\nonumber \\
\label{eq:121}\\
\widehat{F}^{A}=\left.\varphi^{-1}\right.^{A}\!_{i}\:F^{i} & , & \widehat{N}^{AB}=\left.\varphi^{-1}\right.^{A}\!_{i}\left.\varphi^{-1}\right.^{B}\!_{j}\:N^{ij},\nonumber 
\end{eqnarray}
we can rewrite (\ref{eq:113}) (\ref{eq:114}) as follows: 
\begin{equation}
\frac{d\hat{k}^{A}}{dt}=-\hat{k}^{B}\:J^{-1}\!_{BC}\:\widehat{K}^{CA}+\widehat{F}^{A},\qquad\frac{d\widehat{K}^{AB}}{dt}=-\widehat{K}^{AB}\:J^{-1}\!_{CD}\:\widehat{K}^{DB}+\widehat{N}^{AB}.\label{eq:122}
\end{equation}

Let us also mention the mixed, spatial-material form of the internal balance, i.e., second subsystems of (\ref{eq:114}) (\ref{eq:122}), namely: 
\begin{equation}
\frac{d\widetilde{K}^{Ai}}{dt}=\frac{d^{2}\varphi^{i}\!_{B}}{dt^{2}}\:J^{BA}=\widetilde{N}^{Ai}.\label{eq:123}
\end{equation}

In a sense, (\ref{eq:122}) are ``affine Euler equations''. As said, ``in a sense'' only, because the non-dynamical (forces-independent) terms on the right-hand side do not vanish even in the case of highest inertial symmetry, when 
\begin{equation}
J^{AB}={\rm I}\:\eta^{AB}.\label{eq:124}
\end{equation}

Using the co-moving components of velocities, we can write (\ref{eq:122}) as follows: 
\begin{equation}
m\:\frac{d\hat{v}^{A}}{dt}=-m\:\widehat{\Omega}^{A}\!_{B}\:\hat{v}^{B}+\widehat{F}^{A},\quad\frac{d\widehat{\Omega}^{B}\!_{C}}{dt}\:J^{CA}=-\widehat{\Omega}^{B}\!_{D}\:\widehat{\Omega}^{D}\!_{C}\:J^{CA}+\widehat{N}^{AB}.\label{eq:125}
\end{equation}
This follows from the obvious fact that the second of equations (\ref{eq:97}) may be alternatively written as: 
\begin{equation}
\widehat{K}^{AB}=\widehat{\Omega}^{B}\!_{C}\:J^{CA},\quad K^{ij}=\Omega^{j}\!_{m}\:J[\varphi]^{mi},\label{eq:126}
\end{equation}
$J\left[\varphi\right]$ given by (\ref{eq:90}). Obviously, the first formula in (\ref{eq:126}) is more convenient, because the relationship between $\widehat{\Omega}$ and $\widehat{K}$ is based on constant coefficients.

The balance formulation of equations of motion, (\ref{eq:113}) (\ref{eq:114}) (\ref{eq:115}) (\ref{eq:116}) (\ref{eq:122}) is very convenient when discussing some additional constraints imposed on affine motion. The point is that often one means constraints expressed mathematically in some geometric terms. The most natural of them are: 
\begin{enumerate}
\item Metrically-rigid body, i.e., gyroscopic motion. Then the affine velocity $\Omega$ is $g$-skew-isometric, and, equivalently, $\widehat{\Omega}$ is $\eta$-skew-isometric: 
\begin{eqnarray}
\Omega^{i}\!_{j}+\Omega_{j}\!^{i} & = & \Omega^{i}\!_{j}+g_{jk}\:\Omega^{k}\!_{m}\:g^{mi}=0,\label{eq:127}\\
\widehat{\Omega}^{A}\!_{B}+\widehat{\Omega}_{B}\!^{A} & = & \widehat{\Omega}^{A}\!_{B}+\eta_{BC}\:\eta^{AD}\:\widehat{\Omega}^{C}\!_{D}=0,\label{eq:128}
\end{eqnarray}
i.e., $\Omega$, $\widehat{\Omega}$ are elements of the Lie algebras $SO(V,g)'$, $SO(U,\eta)'$ of the corresponding orthogonal groups. And, obviously, $\varphi$ permanently remains within the manifold of linear isometries $O(U,\eta;V,g)$; more precisely within one of its connected components. There are $\frac{1}{2}n(n-1)$ degrees of freedom of internal (relative) motion . 
\item Shape-preserving motion, i.e., superposition of gyroscopic and dilatational motion. This means that $\varphi$ permanently remains within the manifold $\mathbb{R}^{+}O(U,\eta;V,g)$, of linear-conformal mappings. Therefore, $\Omega$, $\widehat{\Omega}$ are respectively elements of the Lie algebras $\mathbb{R}\oplus SO(V,g)'$, $\mathbb{R}\oplus SO(U,\eta)'$ of the linear-conformal groups $\mathbb{R}^{+}SO(V,g)$, $\mathbb{R}^{+}SO(U,\eta)$. This means that they split uniquely into sums of skew-symmetric and identity transformations, 
\begin{equation}
\Omega=\omega+\alpha{\rm Id}_{V}\quad,\quad\widehat{\Omega}=\widehat{\omega}+\alpha{\rm Id}_{U}.\label{eq:129}
\end{equation}
Here $\omega$, $\widehat{\omega}$ are respectively $g$-skew-symmetric and $\eta$-skew-symmetric, cf. (\ref{eq:127}), (\ref{eq:128}), and ${\rm Id}_{V}$, ${\rm Id}_{U}$ are identity mappings in $V$, $U$. There are $\frac{1}{2}n(n-1)+1$ internal degrees of freedom.  
\item \label{enu:case 3}Incompressible affine body. Then $\varphi$ preserves all volume standards, and affine velocities $\Omega$, $\widehat{\Omega}$ are traceless mappings, i.e., satisfy two equivalent conditions: 
\begin{equation}
\Omega^{i}\!_{i}=0\quad,\quad\widehat{\Omega}^{A}\!_{A}=0.\label{eq:130}
\end{equation}
This means that they are respectively elements of the Lie algebras $SL(V)'$, $SL(U)'$ of special linear groups $SL(V)'$, $SL(U)$. There are $\left(n^{2}-1\right)$ internal degrees of freedom.  
\item \label{enu:case 4}Purely dilatational body. There is only one internal degree of freedom. The configuration space may be represented as a manifold of linear isomorphisms $\varphi=\lambda\varphi_{o}$, where $\lambda$ runs over $\mathbb{R}^{+}$, and $\varphi_{o}$ is some fixed isometry. The particular choice of $\varphi_{o}$ is non-essential. Obviously, affine velocities are then one-dimensional objects proportional to the identity mappings: 
\begin{equation}
\Omega=\alpha{\rm Id}_{V}=\frac{d\lambda}{dt}\:{\rm Id}_{V}\quad,\quad\widehat{\Omega}=\alpha{\rm Id}_{U}=\frac{d\lambda}{dt}\:{\rm Id}_{U}.\label{eq:131}
\end{equation}
\item Rotation-less motion, i.e., purely deformative motion. This must be something completely opposite, complementary, to the gyroscopic motion, i.e., to the item (1). The simplest possibility is to replace (\ref{eq:127}) by the condition that $\Omega$ is $g$-symmetric, 
\begin{equation}
\Omega^{i}\!_{j}-\Omega_{j}\!^{i}=\Omega^{i}\!_{j}-g_{jk}\:\Omega^{k}\!_{m}\:g^{mi}=0.\label{eq:132}
\end{equation}
In any case this is the most natural and geometrically unique possibility of defining the rotation-free motion in $V$. But some important novelty appears now. Namely, constraints (\ref{eq:132}) are non-holonomic, unlike the all formerly quoted. The geometric reason for that is such that $g$-symmetric operators do not form a Lie algebra. Commutators of such operators are just $g$-skew-symmetric. There are no restrictions on configurations, but only ones imposed on virtual velocities. The arena of motion is given by some submanifold in the $2n^{2}$-dimensional space of Newtonian states which are parametrized by $\left(\varphi^{i}\!_{A},\dot{\varphi}^{i}\!_{A}\right)$ or equivalently by $\left(\varphi^{i}\!_{A},\Omega^{i}\!_{j}\right)$. The mentioned submanifold is parametrized by $\varphi^{i}\!_{A}$ and, let us say, independent components of $\left(\Omega^{i}\!_{j}+\Omega_{j}\!^{i}\right)$, $i\leq j$. This gives together $n^{2}+\frac{1}{2}n\left(n+1\right)=\frac{1}{2}n\left(3n+1\right)$ independent state variables. Such a rotation-free motion may occur, e.g., when one deals with suspensions in a very viscous fluids. In analogy to the item (1) one can also try to define rotation-free motion in co-moving terms, i.e., as the $\eta$-symmetry of $\widehat{\Omega}$,
\begin{equation}
\widehat{\Omega}^{A}\!_{B}-\widehat{\Omega}_{B}\!^{A}=\widehat{\Omega}^{A}\!_{B}-\eta_{BC}\:\widehat{\Omega}^{C}\!_{D}\:\eta^{DA}=0.\label{eq:133}
\end{equation}
But now a new surprise appears, namely equations (\ref{eq:132}) (\ref{eq:133}) are non-equivalent, unlike (\ref{eq:127}) (\ref{eq:128}). Indeed, (\ref{eq:132}) implies the $G$-symmetry of $\widehat{\Omega}$,
\begin{equation}
\widehat{\Omega}^{A}\!_{B}-G_{BC}\:\widehat{\Omega}^{C}\!_{D}\left.G^{-1}\right.^{DA}=0\label{eq:134}
\end{equation}
rather then its $\eta$-symmetry (\ref{eq:133}). And conversely, (\ref{eq:133}) implies the $C$-symmetry of $\Omega$, not the $g$-symmetry given by (\ref{eq:132}), 
\begin{equation}
\Omega^{i}\!_{j}-C_{jk}\Omega^{k}\!_{m}\left.C^{-1}\right.^{mi}=0.\label{eq:135}
\end{equation}
And again, (\ref{eq:135}) is something else then (\ref{eq:127}). Both models are interesting from the formal point of view of analytical mechanics. But it seems that it is rather (\ref{eq:132}) that is physically more applicable. 
\end{enumerate}

Equations of affine motion with the above all constraints 1 - 5 may be derived from the d'Alembert principle, which tells us that for ideal constraints the moments of reactions $N_{R}$ do not do any work on virtual velocities, i.e., on $\Omega$-s satisfying (\ref{eq:127})-(\ref{eq:135}). For gyroscopic constraints the virtual affine velocities $\Omega$ are $g$-skew-symmetric, thus the reaction moments $N_{R}$ are symmetric and the effective, reactions-free equations of internal motion are given by the skew-symmetric part of the balance laws (\ref{eq:114}), i.e., by the skew-symmetric part of (\ref{eq:112}), therefore, 
\begin{equation}
\frac{dK^{ij}}{dt}-\frac{dK^{ji}}{dt}=N^{ij}-N^{ji},\quad\text{ i.e.},\quad\frac{d\mathcal{S}^{ij}}{dt}=\mathcal{N}^{ij}.\label{eq:136}
\end{equation}
This is the balance law for spin. Explicitly we have 
\begin{eqnarray}
\varphi^{i}\!_{A}\:\frac{d^{2}\varphi^{j}\!_{B}}{dt^{2}}\:J^{AB}-\varphi^{j}\!_{A}\:\frac{d^{2}\varphi^{i}\!_{B}}{dt^{2}}\:J^{AB} & = & \mathcal{N}^{ij},\label{eq:137}\\
g_{ij}\:\varphi^{i}\!_{A}\:\varphi^{j}\!_{B} & = & \eta_{AB},\label{eq:138}
\end{eqnarray}
where the last formula is the final, integrated form of (\ref{eq:127}). Equations of gyroscopic motion consist of the joint system (\ref{eq:137})\&(\ref{eq:138}). Let us stress, it would be wrong to substitute constraint equations (\ref{eq:138}) simply to (\ref{eq:112}) or (\ref{eq:114}).

The implicit equations (\ref{eq:138}) may be transformed to the parametric form where $\varphi^{i}\!_{A}$ are expressed as functions of some generalized coordinates $q^{\alpha}$ on $O\left(u,\eta;V,g\right)$, e.g., Euler angles, canonical coordinates of the first kind (rotation pseudo-vector if $n=3$, or rotation bivector for the general $n$), etc. The functions $\varphi^{i}\!_{A}\left(q^{\alpha}\right)$ may be simply substituted to (\ref{eq:137}), resulting in a system of $\frac{1}{2}n\left(n-1\right)$ equations for $\frac{1}{2}n\left(n-1\right)$ coordinates $q^{\alpha}\left(t\right)$. Let us stress again that such a substitution to (\ref{eq:112}) (\ref{eq:114}) would be wrong. This pattern is to be followed in discussion of all other mentioned holonomic constraints.

Equivalently, the gyroscopic balance equations (\ref{eq:136}) may be written in co-moving terms, as a skew-symmetric part of (\ref{eq:122}), (\ref{eq:125}), e.g., 
\begin{eqnarray}
\frac{d\widehat{\Omega}^{B}\!_{C}}{dt}J^{CA}-\frac{d\widehat{\Omega}^{A}\!_{C}}{dt}J^{CB} & = & -\widehat{\Omega}^{B}\!_{D}\:\widehat{\Omega}^{D}\!_{C}\:J^{CA}+ \widehat{\Omega}^{A}\!_{D}\:\widehat{\Omega}^{D}\!_{C}\:J^{CB}+\label{eq:139}\\
 & + & \widehat{N}^{AB}-\widehat{N}^{BA}.\nonumber 
\end{eqnarray}

Equations of motion of the body with frozen shape consist of (\ref{eq:136})/(\ref{eq:137}) and the $g$-trace of (\ref{eq:114})/(\ref{eq:112}), 
\begin{equation}
g_{ij}\frac{dK^{ij}}{dt}=g_{ij}\:N^{ij}+g_{ij}\:\frac{d\varphi^{i}\!_{A}}{dt}\frac{d\varphi^{j}\!_{B}}{dt}\:J^{AB},\label{eq:140}
\end{equation}
i.e., 
\begin{equation}
g_{ij}\:\varphi^{i}\!_{K}\frac{d^{2}\varphi^{j}\!_{L}}{dt^{2}}\:J^{KL}=g_{ij}\:N^{ij}.\label{eq:141}
\end{equation}
The system (\ref{eq:136}) (\ref{eq:140}), i.e., (\ref{eq:137}) (\ref{eq:141}) contains $\frac{1}{2}n\left(n-1\right)+1$ independent equations and it must be completed, e.g., by the parametric representations of $\varphi$ satisfying: 
\begin{equation}
\varphi^{i}\!_{A}=\lambda\Psi^{i}\!_{A}\quad,\quad\lambda\in\mathbb{R}^{+}\quad,\quad\Psi\in O\left(U,\eta;V,g\right),\label{eq:142}
\end{equation}
therefore, 
\begin{equation}
G_{AB}=g_{ij}\:\varphi^{i}\!_{A}\:\varphi^{j}\!_{B}=\lambda^{2}\:\eta_{AB}.\label{eq:143}
\end{equation}
Here $\lambda$ is dilatational generalized coordinate, the remaining ones may be chosen, e.g., as Euler angels, rotation bivector (axial vector when $n=1$), etc, parameterizing $\Psi\in O\left(U,\eta;V,g\right)$. This parametrization, $\varphi=\lambda\Psi\left(q^{\alpha}\right)$, $\alpha=\frac{1}{2}n\left(n-1\right)$ may be directly substituted to the system (\ref{eq:136})/(\ref{eq:140}) (\ref{eq:137})/(\ref{eq:141}) quite automatically, the reaction forces do not occur there.

Let us observe some important novelty, in co-moving representation the contraction in the trace of (\ref{eq:122}) in principle is not performed with the help of the metric $\eta_{AB}$ but instead, with the help of Green deformation tensor $G$. However, this does not influence anything, because in view of (\ref{eq:143}) one con simply divide both sides by $\lambda^{2}$ and write: 
\begin{equation}
\eta_{AB}\:\frac{d\widehat{K}^{AB}}{dt}=-\widehat{K}^{AC}\left.J^{-1}\right._{CD}\widehat{K}^{DB}\:\eta_{AB}+\eta_{AB}\:\widehat{N}^{AB}.\label{eq:144}
\end{equation}

The effective, i.e., reactions-free equations of isochoric (incompressible) motion (\ref{enu:case 3}) are given by the $g$-trace-less part of \ref{eq:114}, \ref{eq:112} 
\begin{eqnarray}
\frac{d}{dt}\left(K^{ij}-\frac{1}{n}\:g_{ab}\:K^{ab}g^{ij}\right) & = & \left(N^{ij}-\frac{1}{n}\:g_{ab}\:N^{ab}g^{ij}\right)\label{eq:145}\\
 & + & \left(\frac{d\varphi^{i}\!_{A}}{dt}\frac{d\varphi^{j}\!_{B}}{dt}-\frac{1}{n}\:g_{ab}\:\frac{d\varphi^{a}\!_{A}}{dt}\frac{d\varphi^{b}\!_{B}}{dt}\:g^{ij}\right)J^{AB}.\nonumber 
\end{eqnarray}
There are $\left(n^{2}-1\right)$ independent equations in this system, and it must be completed by substituting constraints equation: 
\begin{equation}
\det\left[\varphi^{i}\!_{A}\right]=\sqrt{\frac{\det\left[\eta_{AB}\right]}{\det\left[g_{ij}\right]}}.\label{eq:146}
\end{equation}
And again, when using the co-moving representation (\ref{eq:122}) we must take the traceless part in the sense of the Green deformation tensor $G$, not in the sense of $\eta$, i.e., 
\begin{eqnarray}
 &  & \left(\frac{d\widehat{K}^{AB}}{dt}-\frac{1}{n}\:G_{CD}\:\frac{d\widehat{K}^{CD}}{dt}\left.G^{-1}\right.^{AB}\right)=\label{eq:147}\\
 & = & \left(\widehat{K}^{AC}\left.J^{-1}\right._{CD}\widehat{K}^{DB}-\frac{1}{n}\:G_{KL}\widehat{K}^{LC}\left.J^{-1}\right._{CD}\widehat{K}^{DK}\left.G^{-1}\right.^{AB}\right)+\nonumber \\
 & + & \widehat{N}^{AB}- \frac{1}{n}\:G_{CD}\widehat{N}^{CD}\left.G^{-1}\right.^{AB}.\nonumber 
\end{eqnarray}

Just this form is compatible with the d'Alembert principle, and now the use of $G$ instead $\eta$ is essential, not cosmetic.

For the purely dilatational motion (\ref{enu:case 4}) d'Alembert principle implies elimination of reactions by taking as equation of motion the $g$-trace of (\ref{eq:114})/(\ref{eq:112}), 
\begin{equation}
g_{ij}\:\frac{dK^{ij}}{dt}=g_{ij}\:N^{ij}+g_{ij}\:\frac{d\varphi^{i}\!_{A}}{dt}\frac{d\varphi^{j}\!_{B}}{dt}\:J^{AB},\label{eq:148}
\end{equation}
i.e., 
\begin{equation}
g_{ij}\:\varphi^{i}\!_{K}\frac{d^{2}\varphi^{j}\!_{L}}{dt^{2}}\:J^{KL}=g_{ij}\:N^{ij}.\label{eq:149}
\end{equation}
Substituting: 
\begin{equation}
\varphi^{i}\!_{A}=\lambda\Psi^{i}\!_{A},\qquad G_{AB}=\rho\:\eta_{AB}=\lambda^{2}\:\eta_{AB},\label{eq:150}
\end{equation}
where $\Psi$ is a fixed reference isometry, we obtain: 
\begin{equation}
\eta_{KL}\:J^{KL}\lambda\:\frac{d^{2}\lambda}{dt^{2}}=g_{ij}\:N^{ij},\label{eq:151}
\end{equation}
where, obviously, after the substitution of constraints $g_{ij}$ $N^{ij}$ becomes the function of $\left(\rho,\frac{d\rho}{dt}\right)$; equivalently, of $\left(\lambda,\frac{d\lambda}{dt}\right)$. As usual, the explicit dependence on time $t$ is also admitted.

As usual, when we use the co-moving representation (\ref{eq:122}), the trace scalar is in principle meant in the sense of $G$, 
\begin{equation}
G_{AB}\frac{d\widehat{K}^{AB}}{dt}=-G_{AB}\widehat{K}^{AC}\left.J^{-1}\right._{CD}\widehat{K}^{DB}+G_{AB}\widehat{N}^{AB},\label{eq:152}
\end{equation}
but (\ref{eq:150}) enables one to contract (\ref{eq:122}) simply with the help of $\eta$, 
\begin{equation}
\eta_{AB}\frac{d\widehat{K}^{AB}}{dt}=-\eta_{AB}\widehat{K}^{AC}\left.J^{-1}\right._{CD}\widehat{K}^{DB}+\eta_{AB}\widehat{N}^{AB}.\label{eq:153}
\end{equation}
Finally, for the rotation-free motion, i.e., for non-holonomic constraints (\ref{eq:132}), d'Alembert principle implies the symmetric part of (\ref{eq:114})/(\ref{eq:112}): 
\begin{eqnarray}
\frac{dK^{ij}}{dt}+\frac{dK^{ji}}{dt} & = & N^{ij}+N^{ji}+2\:\frac{d\varphi^{i}\!_{A}}{dt}\frac{d\varphi^{j}\!_{B}}{dt}J^{AB},\label{eq:154}\\
\varphi^{i}\!_{A}\frac{d^{2}\varphi^{j}\!_{B}}{dt^{2}}J^{AB}+\varphi^{j}\!_{A}\frac{d^{2}\varphi^{i}\!_{B}}{dt^{2}} & = & N^{ij}+N^{ji}.\label{eq:155}
\end{eqnarray}
This is to be completed by equations of non-holonomic constraints (\ref{eq:132}); reactions are automatically eliminated by the symmetrization in (\ref{eq:155})/(\ref{eq:154}).

The purely dynamical term on the right hand sides of (\ref{eq:155}) (\ref{eq:154}), $N^{ij}+N^{ji}$ is symmetric. The symmetric part of the affine moment of reactions vanishes, therefore, reaction forces do not enter those equations, $\left.N\!_{R}\right.^{ij}+\left.N\!_{R}\right.^{ji}=0$. The skew-symmetry of $\left.N\!_{R}\right.^{ij}$ is due to the fact that it must be $g$-dual to all virtual velocities compatible with constraints (\ref{eq:132}) therefore, 
\begin{equation}
\left.N\!_{R}\right.^{ij}\omega_{ji}=0,\label{eq:156}
\end{equation}
for any twice covariant symmetric tensor $\omega_{ij}=\omega_{ji}$.

There is however some delicate point when transforming (\ref{eq:154}) (\ref{eq:155}) to the co-moving representation. Namely, the twice contravariant symmetric tensor $N^{ij}+N^{ji}$ is then transformed into symmetric tensor $\widehat{N}^{AB}+\widehat{N}^{BA}$, and the corresponding co-moving reaction moment $\widehat{N}_{R}$ is skewsymmetric just like the spatial one, $\widehat{N}_{R}\!^{AB}+\widehat{N}_{R}\!^{BA}=0$. And one might suspect some mistake or misunderstanding because one cannot prove that $\widehat{N}_{R}$ is dual to $\eta$-symmetric co-moving affine velocities. But everything is correct, one cannot prove, because it is not true. Namely, if deformation occurs, then equation (\ref{eq:156}), more precisely the system: 
\begin{equation}
N_{R}\!^{ik}\left(g_{kj}\:\Omega^{j}\!_{i}\right)=0,\qquad g_{kj}\:\Omega^{j}\!_{i}-g_{ij}\:\Omega^{j}\!_{k}=0\label{eq:157}
\end{equation}
does not imply that 
\begin{equation}
\widehat{N}_{R}\!^{AC}\left(\eta_{CD}\:\widehat{\Omega}^{D}\!_{B}\right)=0,\qquad\eta_{CD}\:\widehat{\Omega}^{D}\!_{B}-\eta_{BD}\:\widehat{\Omega}^{D}\!_{C}=0.\label{eq:158}
\end{equation}
The point is that the $g$-symmetry of $\Omega$ does not imply the $\eta$-symmetry of $\widehat{\Omega}$; instead, $\widehat{\Omega}$ is $G$-symmetric and in virtue of (\ref{eq:157}) we have 
\begin{equation}
\widehat{N}_{R}\!^{AC}\left(G_{CD}\:\widehat{\Omega}^{D}\!_{B}\right)=0,\qquad G_{CD}\:\widehat{\Omega}^{D}\!_{B}-\eta_{BD}\:\widehat{\Omega}^{D}\!_{C}=0.\label{eq:159}
\end{equation}
instead the wrong formula (\ref{eq:158}).

A similar problem appears when materially-non-rotational constraints (\ref{eq:133}) are discussed. And something rather strange is obtained from the d'Alembert principle. Namely, the affine moment of reactions, $N_{R}\!^{ij}$ must be such that 
\begin{equation}
N_{R}\!^{ik}\:g_{kj}\:\Omega^{j}\!_{i}=0,\label{eq:160}
\end{equation}
for any affine velocity $\Omega$ such that $\widehat{\Omega}$ is $\eta$-symmetric, i.e., satisfies (\ref{eq:134}). But this means that $\Omega$ is $C$-symmetric, i.e., satisfies (\ref{eq:135}). Let us introduce the mixed tensor $\mathfrak{D}\in V^{*}\otimes V$ given analytically by 
\begin{equation}
\mathfrak{D}_{k}\!^{j}:=g_{km}C^{-1\ mj}.\label{eq:161}
\end{equation}
The duality between reaction moments $N_{R}\!^{ij}$ and affine velocities $\Omega$ satisfying (\ref{eq:135}) implies that the tensor $\widetilde{N}$ defined by 
\begin{equation}
\widetilde{N}_{R}\!^{ij}:=N_{R}\!^{ik}\:\mathfrak{D}_{k}\!^{j},\label{eq:162}
\end{equation}
must be skew-symmetric: 
\begin{equation}
\widetilde{N}_{R}\!^{ij}+\widetilde{N}_{R}\!^{ji}=0.\label{eq:163}
\end{equation}

Therefore, the procedure for obtaining reactions-free equations of motion is as follows: Take (\ref{eq:114})/(\ref{eq:112}) and transform it by multiplying on the right by $\mathfrak{D}$: 
\begin{eqnarray}
\frac{dK^{im}}{dt}\:\mathfrak{D}_{m}\!^{j} & = & \widetilde{N}^{ij}+\frac{d\varphi^{i}\!_{A}}{dt}\frac{d\varphi^{m}\!_{B}}{dt}\:J^{AB}\:\mathfrak{D}_{m}\!^{j},\label{eq:164}\\
\varphi^{i}\!_{K}\:\frac{d^{2}\varphi^{m}\!_{L}}{dt^{2}}\:J^{AB}\:\mathfrak{D}_{m}\!^{j} & = & \widetilde{N}^{ij}.\label{eq:165}
\end{eqnarray}
Finally we take the symmetric parts of these tensor equations. In this way one obtains reactions-free equations of motion with non-holonomic constraints (\ref{eq:133}); everything based on the assumption of validity of the d'Alembert principle. Physically rather strange and mathematically rather obscure model. But it should be investigated if the analysis is to be complete. Similarly, when using the co-moving representation (\ref{eq:122}), one should multiply it on the right by the co-moving representation of (\ref{eq:161}), 
\begin{equation}
\widehat{\mathfrak{D}}_{A}\!^{B}:=\varphi^{k}\!_{A}\:\varphi^{-1\ B}\!_{j}\:\mathfrak{D}_{k}\!^{j}=G_{AC}\:\eta^{CB}.\label{eq:166}
\end{equation}
Then one obtains: 
\begin{equation}
\frac{\widehat{K}^{AC}}{dt}\:\widehat{\mathfrak{D}}_{C}\!^{B}=-\widehat{K}^{AM}J^{-1}\!_{MN}\widehat{K}^{NC}\:\widehat{\mathfrak{D}}_{C}\!^{B}+\widehat{N}^{AC}\:\widehat{\mathfrak{D}}_{C}\!^{B}.\label{eq:167}
\end{equation}

Taking the symmetric part of (\ref{eq:167}) we obtain the effective reaction-free equations of materially rotationless motion, because the d'Alembert principle implies that 
\begin{equation}
\widehat{N}_{R}\!^{AC}\:\widehat{\mathfrak{D}}_{C}\!^{B}+\widehat{N}_{R}\!^{BC}\:\widehat{\mathfrak{D}}_{C}\!^{A}=0.\label{eq:168}
\end{equation}

\section{The link between kinematical and canonical concepts. Hamiltonian and dissipative models.}

Unfortunately, kinematical and canonical, i.e., Hamiltonian, concepts are often confused in mechanics and no sufficient attention is paid both to distinctions and links between them. The finite-dimensional model we are dealing with here enables one to understand those details correctly and sheds some light on the more difficult problems one is faced with in non-constrained continuum mechanics. In section 3 after the basic discussion of kinematical quantities like various measures of deformation and velocities we introduce phase-space concepts like canonical translational momentum, affine momentum, affine spin, angular momentum and metrical (usual) spin. The geometric meaning of canonical linear momenta, canonical affine momenta, etc. was that of Hamiltonian generators of natural groups of affine transformations acting in the physical space and in the material space (in the body itself). The basic Poisson brackets were given and it was seen they were expressed by the known structure constants of affine and linear groups. And then equations of motion of conservative systems were given in the canonical form (\ref{eq:83}) based on Poisson brackets, does not matter where the Hamiltonian $H$ was taken from. And, as a rule, equations of internal motion are given by the balance law of $\Sigma$ or $\widehat{\Sigma}$, 
\begin{equation}
\frac{d}{dt}\:\Sigma^{i}\!_{j}=\left\{ \Sigma^{i}\!_{j},H\right\} ,\qquad\frac{d}{dt}\:\widehat{\Sigma}^{A}\!_{B}=\left\{ \widehat{\Sigma}^{A}\!_{B},H\right\} .\label{eq:166a}
\end{equation}

This is the way of thinking of theoretical physicists working in fundamental problems. But at the same time, it is clear that usually in mechanics of continua one proceeds in another way, using kinematical quantities like linear momentum $k^{i}$ or internal angular momentum $S^{i}$, where, according to the school wisdom, for the material point $k^{i}=mv^{i}$, $v^{i}$ denoting the translational velocity, and angular momentum is the vector product of the radius vector and linear momentum. And one begins from the system of Newton equations, and constraints are taken into account on the basis of d'Alembert principle. The relationship between kinematical quantities $\left(k^{i},\hat{k}^{A},K^{ij},\widehat{K}^{AB}\right)$ and canonical ones $\left(p_{i},\hat{p}_{A},\Sigma^{i}\!_{j},\widehat{\Sigma}^{A}\!_{B}\right)$ is based on Legendre transformation. Denoting the background Lagrangian of the non-dissipative mechanics by $L$, we have 
\begin{equation}
p_{i}=\frac{\partial L}{\partial v^{i}},\ \hat{p}_{A}=\frac{\partial L}{\partial\hat{v}^{A}},\ \Sigma^{i}\!_{j}=\frac{\partial L}{\partial\Omega^{j}\!_{i}},\ \widehat{\Sigma}^{A}\!_{B}=\frac{\partial L}{\partial\widehat{\Omega}^{B}\!_{A}},\ P^{A}\!_{i}=\frac{\partial L}{\partial V^{i}\!_{A}},\label{eq:167a}
\end{equation}
depending on if Lagrangian is expressed respectively on velocity arguments in the version $v^{i}$, $\hat{v}^{A}$, $\Omega^{j}\!_{i}$, $\widehat{\Omega}^{B}\!_{A}$; $V^{i}\!_{A}=\dot{\varphi}^{i}\!_{A}$. If Lagrangian has the potential form: 
\begin{equation}
L=T-V=T_{tr}+T_{int}-V\left(x^{i},\varphi^{i}\!_{A}\right),\label{eq:168a}
\end{equation}
$T$, $T_{int}$, $T_{tr}$ given by the usual formula (\ref{eq:107a}), then Legendre transformation expresses canonical momenta $p_{i}$, $P^{A}\!_{i}$ as the following functions of generalized velocities:
\begin{equation}
p_{i}=m\:g_{ij}\frac{dx^{j}}{dt}=g_{ij}\:k^{j},P^{A}\!_{i}=g_{ij}\:\frac{d\varphi^{j}\!_{B}}{dt}\:J^{AB}.\label{eq:169}
\end{equation}

Kinetic energy (\ref{eq:107a}) and Legendre transformation (\ref{eq:167}) (\ref{eq:169}) may be written in some alternative forms, very useful in theoretical analysis, e.g., 
\begin{eqnarray}
T_{tr} & = & \frac{m}{2}\:G_{AB}\:\hat{v}^{A}\:\hat{v}^{B},\label{eq:170}\\
T_{int} & = & \frac{1}{2}\:G_{AB}\:\widehat{\Omega}^{A}\!_{K}\: \widehat{\Omega}^{B}\!_{L}\:J^{KL}= \frac{1}{2}\:g_{ij}\:\Omega^{i}\!_{k}\:\Omega^{j}\!_{l}\:J[\varphi]^{kl},\label{eq:171}\\
\hat{p}_{A} & = & m\:G_{AB}\:\hat{v}^{B}=G_{AB}\:\hat{k}^{B},\label{eq:172}\\
\widehat{\Sigma}^{A}\!_{B} & = & G_{BD}\:\widehat{\Omega}^{D}\!_{C}J^{CA},\quad\Sigma^{i}\!_{j}=g_{jk}\:\Omega^{k}\!_{m}\:J[\varphi]^{mj}.\label{eq:173}
\end{eqnarray}

Let us notice that the second equation of (\ref{eq:97}) may be alternatively written as follows: 
\begin{equation}
K^{ij}=\Omega^{j}\!_{b}J[\varphi]^{bi},\qquad\widehat{K}^{AB}=\widehat{\Omega}^{B}\!_{C}J^{CA}.\label{eq:174}
\end{equation}
therefore, 
\begin{equation}
\Sigma^{i}\!_{j}=K^{ib}g_{bj},\qquad\widehat{\Sigma}^{A}\!_{B}=\widehat{K}^{AC}G_{CB}.\label{eq:175}
\end{equation}

Now we have the complete ``dictionary'' between two versions of concepts like linear momentum, affine momentum, angular momentum, affine spin and just spin. Those are kinematical and canonical (Hamiltonian) versions. The relationship between them depends on the particular choice of dynamical variational model, i.e., on the choice of Lagrangian. The above formulas are valid for the classical potential class of models. If we admitted in $L$ some velocity-dependent terms in addition to the kinetic energy, the relationship would be different. For example, if magnetic forces are present, Lagrangian differs from (\ref{eq:168}) by terms linear in generalized velocities. Then Legendre transformation expresses kinematical quantities as affine, i.e., linear-nonhomogeneous, functions of kinematical ones. Inverting Legendre transformation, i.e., expressing kinematical quantities as functions of canonical ones, and substituting them to the energy function, 
\begin{equation}
E=T+V=T_{tr}+T_{int}+V(x,\varphi),\label{eq:176}
\end{equation}
one obtains Hamiltonian 
\begin{equation}
H=\mathfrak{T}+V=\mathfrak{T}_{tr}+\mathfrak{T}_{int}+V(x,\varphi).\label{eq:177}
\end{equation}

Obviously, the kinetic term of the Hamiltonian is given by: 
\begin{equation}
\mathfrak{T}=\mathfrak{T}_{tr}+\mathfrak{T}_{int}=\frac{1}{2m}g^{ij}p_{i}p_{j}+\frac{1}{2}J_{AB}^{-1}P^{A}\!_{i}P^{B}\!_{j}g^{ij}.\label{eq:178}
\end{equation}
In analogy to (\ref{eq:170}), (\ref{eq:171}) we can also use the following suggestive expressions: 
\begin{eqnarray}
\mathfrak{T}_{tr} & = & =\frac{1}{2m}G^{-1AB}\hat{p}_{A}\hat{p}_{B},\label{eq:179}\\
\mathfrak{T}_{int} & = & \frac{1}{2}J_{AB}^{-1}\widehat{\Sigma}^{A}\!_{K}\widehat{\Sigma}^{B}\!_{L}G^{-1KL}=\frac{1}{2}J[\varphi]^{-1}\!_{ij}\Sigma^{i}\!_{k}\Sigma^{j}\!_{l}g^{kl}.\label{eq:180}
\end{eqnarray}

Let us observe that the quadratic forms (\ref{eq:107a}) (\ref{eq:178}) have constant coefficients. Unlike this, (\ref{eq:170}) (\ref{eq:171}) (\ref{eq:179}) (\ref{eq:180}) have configuration-dependent coefficients, however, the geometric objects $\widehat{\Omega}$, $\Omega$, $\widehat{\Sigma}$, $\Sigma$ are more suggestive than $\frac{d\varphi^{i}\!_{A}}{dt}$, $P^{A}\!_{i}$. The reason is that they are Lie-algebraic objects relevant for the structure of our configuration space; for example $\Sigma^{i}\!_{j}$, $\widehat{\Sigma}^{A}\!_{B}$ are Hamiltonian generators of $GL(V)$, $GL(U)$ acting on the internal configuration space.

Roughly speaking, equation (\ref{eq:175}) tells us that the kinematical and canonical affine spin, $K\in V\otimes V$, $\Sigma\in V\otimes V^{*}\simeq L(V)$ are related to each other by the $g$-shift of the second index. The metric tensor $g$ is fixed, constant and in appropriate coordinates its components are given by the Kronecker symbol, therefore, analytically this is a rather cosmetic difference. The co-moving objects $\widehat{\Sigma}\in U\otimes U^{*}\simeq L(U)$, $\widehat{K}\in U\otimes U$ are also related to each other by some shifting of the second index, however the shifting tensor depends on generalized coordinates, it is simply the Green deformation tensor.

The most important geometric and physical content of equations of motion is summarized in balance equations for Hamiltonian generators $\left(p_{i},\Sigma^{i}\!_{j}\right)$, or $\left(\hat{p}_{A},\widehat{\Sigma}^{A}\!_{B}\right)$,
\begin{eqnarray}
\frac{dp_{i}}{dt}=\left\{ p_{i},H\right\} =-\frac{\partial H}{\partial x^{i}} & , & \qquad\frac{d\Sigma^{i}\!_{j}}{dt}=\left\{ \Sigma^{i}\!_{j},H\right\} ,\label{eq:181}\\
\frac{d\hat{p}_{A}}{dt}=\left\{ \hat{p}_{A},H\right\}  & , & \qquad\frac{d\widehat{\Sigma}^{A}\!_{B}}{dt}=\left\{ \widehat{\Sigma}^{A}\!_{B},H\right\} .\label{eq:182}
\end{eqnarray}

The right-hand sides may be calculated with the help of the basic Poisson brackets (\ref{eq:75})-(\ref{eq:81}). The fundamental properties of this operation must be used, e.g., Lie-algebraic rules, and first of all, the separation rule (\ref{eq:82}). Substituting to (\ref{eq:181}) (\ref{eq:182}) the Legendre transformation, e.g, (\ref{eq:169}) for the potential systems, or better its equivalent forms (\ref{eq:174}) (\ref{eq:175}), one obtains second-order differential equations for the time-dependence of generalized coordinates $\left(x^{i},\varphi^{i}\!_{A}\right)$. The balance laws (\ref{eq:181}) for the potential models (\ref{eq:168}) have the form: 
\begin{equation}
\frac{dp_{i}}{dt}=-\frac{\partial V}{\partial x^{i}}=-g_{ij}F^{j},\qquad\frac{d\Sigma^{i}\!_{j}}{dt}=\left\{ \Sigma^{i}\!_{j},\mathcal{T}\right\} +N^{i}\!_{j}[V]\label{eq:183}
\end{equation}
where 
\begin{equation}
N^{i}\!_{j}[V]=\left\{ \Sigma^{i}\!_{j},V\right\} =-\varphi^{i}\!_{A}\frac{\partial V}{\partial\varphi^{j}\!_{A}},\qquad F^{i}=-g^{ik}\frac{\partial V}{\partial x^{k}}.\label{eq:184}
\end{equation}

The differential operator which acts on $V$ on the right-hand side of (\ref{eq:184}) equals the minus generator of left $GL(V)$-mappings acting on the argument of $V$. Calculating the Poisson brackets $\left\{ \Sigma^{i}\!_{j},\mathcal{T}\right\} $, $g$-raising the covariant indices, and substituting (\ref{eq:169}), (\ref{eq:174}), (\ref{eq:175}) to (\ref{eq:183}), one obtains just (\ref{eq:113}) (\ref{eq:114}) or equivalently (\ref{eq:122}) with
\begin{equation}
N^{ij}=N^{i}\!_{k}[V]\:g^{kj},\qquad\widehat{N}^{AB}=\varphi^{-1A}\!_{i}\:\varphi^{-1B}\!_{j}N^{ij}.\label{eq:185}
\end{equation}

The distinction between tensors $N^{i}\!_{j}[V]$, and $N^{ij}$ is, obviously, just as one between $\Sigma^{i}\!_{j}$ and $K^{ij}$, of a rather ``cosmetic'' character. Indeed, the shift of indices is performed with the use of a fixed tensor $g$, the matrix of which in appropriate coordinates coincides with the Kronecker symbol. (According to the commonly accepted convention we could use in principle the same kernel symbol and write simply $\Sigma^{ij}$ instead $K^{ij}$; we do not do it, because certain misunderstandings would be possible). And again some delicate problems appear when we compare the spatial and material descriptions. Namely, in the co-moving representation, (\ref{eq:183}) becomes 
\begin{equation}
\frac{d\hat{p}_{A}}{dt}=-\frac{\partial V}{\partial x^{i}}\:\varphi^{i}\!_{A},\qquad\frac{d\widehat{\Sigma}^{A}\!_{B}}{dt}=\left\{ \widehat{\Sigma}^{A}\!_{B},\mathcal{T}\right\} +\widehat{N}^{A}\!_{B}[V]\label{eq:186}
\end{equation}
where 
\begin{equation}
\widehat{N}^{A}\!_{B}[V]=\left\{ \widehat{\Sigma}^{A}\!_{B},V\right\} =-\varphi^{i}\!_{B}\frac{\partial V}{\partial\varphi^{i}\!_{A}}.\label{eq:187}
\end{equation}

Substituting here the Legendre transformation and (\ref{eq:174}) (\ref{eq:175}), we obtain just the representation (\ref{eq:122}). It must be stressed however that $\widehat{N}^{AB}$ in (\ref{eq:122}) is not the $\eta$-raised version of $\widehat{N}^{A}\!_{B}[V]$ in (\ref{eq:187}), but instead it is its $G$-raised version: 
\begin{equation}
\widehat{N}^{AB}=\widehat{N}^{A}\!_{C}[V]\:G^{-1CB}\neq\widehat{N}^{A}\!_{C}[V]\:\eta^{CB}.\label{eq:188}
\end{equation}

And here really the use of two different kernel symbols would be more adequate, however, we are afraid of the crowd of symbols and of changing them in the course of writing. To obtain (\ref{eq:122}) one must perform some calculations, for example show that 
\begin{eqnarray}
\frac{dG_{AB}}{dt} & = & J^{-1}\!_{AD}\widehat{K}^{DC}G_{CB}+J^{-1}\!_{BD}\widehat{K}^{DC}G_{CA}=\label{eq:189}\\
 & = & G_{BC}\:\widehat{\Omega}^{C}\!_{A}+G_{AC}\:\widehat{\Omega}^{C}\!_{B}=\left(\Omega_{ij}-\Omega_{ji}\right)\varphi^{i}\!_{A}\:\varphi^{j}\!_{B}.\nonumber 
\end{eqnarray}
where, obviously,
\[
\Omega_{ij}:=g_{ik}\Omega^{k}\!_{j}.
\]
In analogy to this we have 
\begin{equation}
\frac{dC_{ij}}{dt}=-\left(\widehat{\Omega}_{AB}+\widehat{\Omega}_{BA}\right)\varphi^{-1A}\!_{i}\:\varphi^{-1B}\!_{j},\qquad\widehat{\Omega}_{AB}:=\eta_{AC}\:\widehat{\Omega}^{C}\!_{B}.\label{eq:190}
\end{equation}
The proof is easy and we do not quote it here. The formulas (\ref{eq:189}) (\ref{eq:190}) are very suggestive and reveal some more about the geometric content of tensors $G$, $C$, $\widehat{\Omega}$, $\Omega$ and their mutual relationships. What concerns kinematical content of formulas (\ref{eq:189}) (\ref{eq:190}), compare them with the statement (\ref{eq:33}) valid for the general, non-constrained continua. Affine framework sheds some light on geometry hidden behind analytical formulas of continuum mechanics.

When the system is potential, e.g., when one deals with the hyperelastic affine body, then equations of internal motion are given by (\ref{eq:112}) or its alternative forms like (\ref{eq:114}), (\ref{eq:122}) with $N^{ij}$, $\widehat{N}^{AB}$ given by (\ref{eq:184}) (\ref{eq:185}) (\ref{eq:186}) (\ref{eq:187}) (\ref{eq:188}). Obviously, one can admit also more general forms of the dependence of $N$, $\widehat{N}$ on the configuration $\varphi$, e.g., ones describing the elastic but not necessarily hyperelastic affine dynamics.

One can reasonably expect that the most useful and realistic models are those combining some potential term with some purely dissipative, viscous one, 
\begin{equation}
N^{ij}=N[\mathcal{V}]^{ij}+N_{diss}^{ij}=-\varphi^{i}\!_{A}\:\frac{\partial\mathcal{V}}{\partial\varphi^{k}\!_{A}}\:g^{kj}+N_{diss}^{ij}.\label{eq:191}
\end{equation}

For the isotropic internal viscous friction we have 
\begin{eqnarray}
N_{int.diss}^{ij} & = & -\text{Vol}_{o}\sqrt{\frac{\det[g_{ij}]}{\det[\eta_{AB}]}}\det\left[\varphi^{i}\!_{A}\right]\left(\eta\left(\Omega^{ij}+\Omega^{ji}\right)\right)+\label{eq:192}\\
 & - & \text{Vol}_{o}\sqrt{\frac{\det[g_{ij}]}{\det[\eta_{AB}]}}\det\left[\varphi^{i}\!_{A}\right]\left(\left(\zeta-\frac{2\eta}{n}\right)\Omega^{k}\!_{k}g^{ij}\right).\nonumber 
\end{eqnarray}
The last formula is written in a somewhat pretentious, but geometrically correct way. Obviously, $n$ is the spatial dimension, physically $n=3$, in some problems $n=2$, $\Omega^{ij}=\Omega^{i}\!_{k}g^{kj}$ and $Vol_{o}$ denotes the reference volume of the body.

Traditional symbols $\eta$, $\zeta$ are used for coefficients of linear and isotropic internal friction. The square-root-term reduces to unitary when orthonormal bases are used in $V$, $U$. Geometrically the square root of $\det[g_{ij}]$ is the scalar density of weight one in $V$, the square root of $\det[\eta_{AB}]$ is the scalar density of weight one in $U$, and $\det\left[\varphi^{i}\!_{A}\right]$ has a double structure: it is scalar density of weight minus one in $V$ and scalar density of weight one in $U$. The total product of determinant expressions 
\begin{equation}
\mathcal{D}\varphi=\sqrt{\frac{\det[g_{ij}]}{\det[\eta_{AB}]}}\det\left[\varphi^{i}\!_{A}\right]\label{eq:193}
\end{equation}
is a scalar quantity, as it should be. The formula (\ref{eq:192}) is obtained from (\ref{eq:120}) when the textbook formula for the linear and isotropic stress tensor of viscous friction is used, 
\begin{equation}
\sigma_{vis}^{ij}=2\eta \:d^{ij}+\left(\zeta-\frac{2\eta}{n}\right)g_{ab}\:d^{ab}g^{ij},\quad d^{ij}=\frac{1}{2}\left(\Omega^{ij}+\Omega^{ji}\right).\label{eq:194}
\end{equation}
Generalizations to anisotropic and non-linear models are obvious. $N_{int.diss}^{ij}$ must be then some anisotropic or/and nonlinear tensor function of $d^{kl}$.

Another model of dissipative affine torque should be used in problems of external friction, e.g., when one discussed an affine motion of suspensions. The simplest models are ones linear in $\Omega$, 
\begin{equation}
N^{ij}=-F^{ijkl}\:\Omega_{kl},\qquad\Omega_{kl}=g_{km}\:\Omega^{m}\!_{l},\label{eq:195}
\end{equation}
where $F$ is a constant fourth-order tensor. In isotropic models we have 
\begin{equation}
N^{ij}=-k\:\Omega^{ij}-l\:\Omega^{ji}-p\:\Omega^{a}\!_{a}g^{ij},\label{eq:196}
\end{equation}
i.e., 
\begin{equation}
F^{ijmn}=kg^{im}g^{jn}+lg^{jm}g^{in}+pg^{ij}g^{mn}.\label{eq:197}
\end{equation}
Nonlinear modifications are structurally obvious. Let us notice that the metric tensor may be partially eliminated by putting: 
\begin{equation}
N^{ij}=-F^{ij}\!_{m}\!^{n}\:\Omega^{m}\!_{n},\quad\text{or}\quad N^{i}\!_{j}=-F^{i}\!_{jm}\!^{n}\:\Omega^{m}\!_{n},\label{eq:198}
\end{equation}
i.e., in the isotropic case, 
\begin{eqnarray}
N^{i}\!_{j} & = & -k\:\Omega^{i}\!_{j}-l\:\Omega_{j}\!^{i}-p\:\Omega^{a}\!_{a}\:\delta^{i}\!_{j},\label{eq:199}\\
F^{i}\!_{jm}\!^{n} & = & k\:\delta^{i}\!_{m}\:\delta_{j}\!^{n}+l\:g_{jm}\:g^{in}+p\:\delta^{i}\!_{j}\:\delta_{m}\!^{n}.\label{eq:200}
\end{eqnarray}

Obviously, in the mentioned applications it is rather natural to expect that it is mainly the $g$-skew-symmetric part of $\Omega^{i}\!_{j}$, i.e., (\ref{eq:30}), that contributes to the external friction. Then we have 
\begin{equation}
F^{i}\!_{j}\!^{mn}=-F^{i}\!_{j}\!^{nm},\quad\text{i.e.,}\quad F^{i}\!_{jm}\!^{n}=-g^{na}g_{mb}\:F^{i}\!_{ja}\!^{b}.\label{eq:201}
\end{equation}
In the isotropic case this means that in (\ref{eq:199}) (\ref{eq:200}) we have: $l=-k$, $p=0$, therefore, 
\begin{equation}
N^{i}\!_{j}=-k\left(\Omega^{i}\!_{j}-\Omega_{j}\!^{i}\right)=-k\left(\Omega^{i}\!_{j}-g^{ia}g_{jb}\:\Omega^{b}\!_{a}\right)=-k\:\omega^{i}\!_{j},\label{eq:202}
\end{equation}
where $\omega$ is the angular velocity (\ref{eq:30}).

\section{Symmetries and conservation laws}

The complete, systematic description of symmetries, conservation laws and their mutual relationships is based on variational principles and Hamiltonian formalism. Nevertheless, many partial results may be obtained within the more general Newton-d'Alembert framework, including the dynamics of non-conservative systems, in particular dissipative ones. Obviously, even if not used explicitly, the Lagrangian and symplectic concepts are always somehow hidden behind the treatment.

Let us begin with translational invariance. In mechanics of affine bodies the problem of material translational invariance (material homogeneity) becomes diffused, and as a matter of fact, it disappears. There are two reasons for that. The first one is that on the level of dynamics one deals with global quantities obtained as mean values, integral averages performed over the material space. The second reason is that all formulas we use, in particular the one for kinetic energy, are expressed in Lagrangian coordinates vanishing at the center of mass. This fixed point and finite size of the body break translational symmetry in the material space.

Obviously, translational symmetry in the physical space is still well-defined. Equations of motion (\ref{eq:111}) (\ref{eq:112}) are invariant under spatial translations when the total force $F^{i}$ and affine torque $N^{ij}$ do not depend on the spatial position $x\in M$. Let us observe that this does not imply the conservation of kinematical linear momentum even if $F^{i}$ do not depend on internal generalized coordinates $\varphi^{i}\!_{k}$, on their generalized velocities and on the time variable $t$. Indeed, if they depend only on translational velocities $v^{i}=\frac{dx^{i}}{dt}$, equations of translational motion are translationally-invariant in $M$, but as seen from (\ref{eq:113}), $k^{i}$ is not a constant of motion. It becomes a conserved quantity only if the total force vanishes, $F^{i}=0$. If $F^{i}$ is constant but non-vanishing (homogeneous field of forces), then equations of translational motion are translationally-invariant, however $K^{i}$ is not a constant of motion, either. This is obvious, because the center of mass motion is then uniformly accelerated. Nevertheless, there is some explicitly time-dependent constant of motion somehow corresponding to translational symmetry. This is 
\begin{equation}
\varkappa^{i}:=k^{i}-F^{i}t.\label{eq:203}
\end{equation}
Obviously, $\frac{1}{m}\varkappa^{i}$ is the initial velocity at the time instant $t=0$. Galilean boosts do not preserve equations of motion, but they preserve their general solution, i.e., the set of uniformly accelerated motions (and separately preserve the subset of uniform motions). There is some prescription which associates with this symmetry some time-dependent constants of motion, namely: 
\begin{equation}
\xi^{i}=x^{i}-\frac{k^{i}}{m}\:t+\frac{F^{i}}{2m}\:t^{2}.\label{eq:204}
\end{equation}
It is seen that those are initial coordinates at the initial time instant $t=0$. The simple formulas have a very interesting interpretation in terms of symplectic geometry and Hamiltonian mechanics, however, there is no place here for a more detailed discussion.

Much more interesting are problems concerning dynamical symmetries of internal degrees of freedom. To simplify discussion we neglect translational degrees of freedom and considerate internal dynamics as autonomous one.

Let us now begin with the spatial internal transformations, i.e., with (\ref{eq:46}) where we put: $b=Id_{U}$. Equations of internal motion, i.e., (\ref{eq:112}) and their byproducts, are invariant under such transformations, i.e., their general solution is transformed onto itself if and only if the affine torque satisfies the following transformation rule: 
\begin{equation}
N^{ij}\left(a\varphi,a\frac{d\varphi}{dt},t\right)=a^{i}\!_{k}a^{j}\!_{l}N^{kl}\left(\varphi,\frac{d\varphi}{dt},t\right),\label{eq:205}
\end{equation}
i.e., the prescription for $N$ as a function of state variables is ``transparent'' under the left action of any $a\in GL(V)$. In terms of the co-moving description this simply means that 
\begin{equation}
\widehat{N}^{AB}\left(a\varphi,a\frac{d\varphi}{dt},t\right)=\widehat{N}^{AB}\left(\varphi,\frac{d\varphi}{dt},t\right).\label{eq:206}
\end{equation}

If such a rule is to be satisfied for all $a\in GL(V)$, then, obviously, $\widehat{N}$ must be built exclusively of quantities with the capital (material) indices. It is clear that without additional geometric objects, such a prescription does not exist. The only purely material objects we have then at disposal is $\widehat{\Omega}$ and its tensorial byproducts $\widehat{\Omega}^{m}$, $m$ being non-negative integer; there are also invariantly defined scalars $Tr\left(\widehat{\Omega}^{p}\right)$. But the only second-order tensors built of $\widehat{\Omega}$ are mixed ones in $U$, whereas $\widehat{N}$ must be twice contravariant. The second index of $\widehat{N}^{A}\!_{B}$ may be obviously raised with the help of $\eta$, the material metric, 
\begin{equation}
\widehat{N}^{AB}=\widehat{N}^{A}\!_{C}\eta^{CB};\label{eq:207}
\end{equation}
but one must be aware that the occurrence of $\eta$ restricts the material $GL(U)$-symmetry to isometries $O(U,\eta)$. But well, we are fighting now for $GL(V)$-symmetries. Therefore, any material tensors might be formally fixed and used for producing $\widehat{N}^{AB}$ from $\widehat{\Omega}^{K}\!_{L}$, e.g., in the simplest case we might use the scheme 
\begin{equation}
\widehat{N}^{AB}=T^{AB}\!_{K}\!^{L}\left(Tr\widehat{\Omega}^{p}\right)\widehat{\Omega}^{K}\!_{L},\label{eq:208}
\end{equation}
etc. Obviously, such strange models of $\widehat{N}$ are completely useless for describing the elastic-like behavior. For such purposes we need the Green deformation tensor $G_{KL}$. Incidentally, let us remind that $G$ may be also used for shifting the material indices, like $\eta$ in (\ref{eq:207}) although this is not always physically motivated.

Summarizing: Realistic dynamical models compatible with the idea of invariance under spatial transformations have the form: 
\begin{equation}
\widehat{N}^{AB}=\widehat{N}^{AB}\left(G_{KL},\widehat{\Omega}^{C}\!_{D}\right).\label{eq:209}
\end{equation}

Incidentally, let us remind the isotropic constitutive laws for unconstrained continua, when the second Piola-Kirchhoff stress tensor is expressed as a function of the Green deformation tensor and the material representation of the velocity gradient. Their analogy to (\ref{eq:209}) is obvious and certainly non-accidental, namely, in virtue of (\ref{eq:120}), (\ref{eq:209}) is obtained as the material average of the unconstrained constitutive law.

The spatial invariance is then automatically reduced to $g$-isometries $E(M,g)$ because $G$ is algebraically built of the spatial metric tensor $g$, cf. (\ref{eq:66}) and the preceding comments. In the hyperelastic case, when the formula (\ref{eq:187}) holds, the potential energy of $V$ of the internal $O(V,g)$-invariant dynamics is given by some function of $G$, $V=\mathcal{W}(G)$. Then, roughly speaking, $\widehat{N}$ is the derivative of $\mathcal{W}$ with respect to $G$, 
\begin{equation}
\widehat{N}(\varphi)=-2D_{G}\mathcal{W},\qquad\widehat{N}^{AB}=-\frac{\partial\mathcal{W}}{\partial G_{AB}},\label{eq:210}
\end{equation}
or, more precisely ($G_{AB}$ are not independent variables because $G$ is symmetric), 
\begin{equation}
\left.\frac{d}{dx}\mathcal{W}(G+x\varepsilon)\right|_{t=0}=\frac{1}{2}\widehat{N}^{AB}\varepsilon_{AB},\label{eq:211}
\end{equation}
for any symmetric $\varepsilon$. By the very construction, $\widehat{N}^{AB}$ obtained in this way is symmetric, and, obviously, so is $N^{ij}$. Therefore, (\ref{eq:114}) implies that spin (internal angular momentum) is a conserved quantity, 
\begin{equation}
\frac{dS}{dt}=\frac{d}{dt}\left(K^{ij}-K^{ji}\right)=0.\label{eq:212}
\end{equation}

If $\widehat{N}$ is derived from the potential $V$, like in (\ref{eq:208}), this conservation law is a consequence of the Noether theorem. Indeed, equations of motion are then derived from the Lagrangian 
\begin{equation}
L_{int}=T_{int}-V_{int}=\frac{1}{2}\:g_{ij}\:\frac{d\varphi^{i}\!_{A}}{dt}\frac{d\varphi^{j}\!_{B}}{dt}-\mathcal{W}(G).\label{eq:213}
\end{equation}

And this Lagrangian is invariant under all transformations $\varphi\mapsto a\varphi$, $a\in O(V,g)$; the resulting Euler-Lagrange equations are so as well, however the invariance of Lagrangian itself is something more. And it is just the invariance of Lagrangian that implies the conservation of angular momentum. The invariance of equations of motion alone, i.e., condition (\ref{eq:209}) does not imply spin conservation. Nevertheless, if $\widehat{N}^{AB}$ is symmetric, this conservation law is satisfied, even if $\widehat{N}$ is structurally non-variational, even if dissipative forces occur. It is seen that the relationship between symmetries and constants of motion is a rather delicate matter when beyond the variational framework.

It is seen that Noether theorem excludes higher spatial symmetries than the isometry group. This is because the metric tensor $g_{ij}$ is explicitly present in Lagrangian, both in the kinetic energy and in the potential term, where it enters via the Green tensor $G_{AB}$.

Obviously, to construct any explicit prescription for the dependence of $\widehat{N}$ on $G$ and $\widehat{\Omega}$ in (\ref{eq:209}), one must use some constitutive tensors in the material space $U$. As a rule, this restricts the a priori material group $GL(U)$ to some proper subgroup.

A typical example is the anisotropic nonlinear Hooke law, 
\begin{equation}
N^{AB}=C^{ABKL}E_{KL}=\frac{1}{2}C^{ABKL}\left(G_{KL}-\eta_{KL}\right).\label{eq:209a}
\end{equation}

It is linear in $E$, if $C$ is configuration-independent, but it is explicitly non-linear in generalized coordinates $\varphi^{i}\!_{K}$. There is an obvious analogy with the constitutive laws for anisotropic continua subject to large elastic deformations, e.g., polymer media. However, usually in such realistic nonlinear models one prefers rather the situation when nonlinearity a appears already on the level of the very relationship between $E$ and $N$.

The relationship (\ref{eq:209a}) between $E$ and $N$ is invariant under such material mappings which preserve the tensors $C$, $\eta$. They form a subgroup of the material orthogonal group $O(U,\eta)$.

The simplest possibility is, however to rely only on the material metric $\eta$. Then (\ref{eq:209a}) becomes the isotropic ``nonlinear Hooke law'', and $C$ is given by: 
\begin{equation}
C^{ABKL}=\lambda\eta^{AK}\eta^{BL}+\mu\eta^{AB}\eta^{KL}.\label{eq:209b}
\end{equation}

Such a model is invariant under $O(U,\eta)$ and it is the simplest model with the symmetry group $O(V,g)\times O(U,\eta)$ acting through (\ref{eq:46}).

It was mentioned in this paper that in certain formulas the material indices are shifted with the help of the Green tensor $G[\varphi]$. One can modify (\ref{eq:209a}) along such lines, and namely replacing the constant (configuration-independent) constitutive tensor $C$ by the following configuration-dependent tensor $\widetilde{C}$:
\begin{equation}
\widetilde{C}^{ABKL}=\lambda G^{-1AK}G^{-1BL}+\mu G^{-1AB}G^{-1KL}.\label{eq:209c}
\end{equation}
We do not quote the corresponding formula for $N$. In spite of the formal similarity of (\ref{eq:209b}), (\ref{eq:209c}), the second model is completely different from the previous one, in particular, its nonlinearity is much stronger.

It is convenient to use the mixed tensor $\widehat{G}$ (\ref{eq:20}). Being linear mapping in $U$, the tensors $\widehat{G}$, $\widehat{\Omega}$ may be multiplied by each other and give rise to monomials like $\widehat{G}^{a}\:\widehat{\Omega}^{b}\:\widehat{G}^{c}\:\widehat{\Omega}^{d}$, etc., where $a,b,c,d$, etc. are integers, non-negative ones when used with $\widehat{\Omega}$. The traces of those monomials are scalars in $U$, some $O(U,\eta)$-invariants. Combining the monomials with coefficients depending on the mentioned scalars, one obtains some mixed tensors, elements of $U\otimes U^{*}\cong L(U)$, correctly defined as functions of $\widehat{G}$, $\widehat{\Omega}$. Then raising their second indices with the help of $\eta^{AB}$ or $G^{AB}$, one obtains twice contravariant material tensors, elements of $U\otimes U$, just the $\widehat{N}^{AB}(G,\widehat{\Omega})$ in (\ref{eq:209}). Seemingly one might think about infinite series defining (\ref{eq:209}), however, the number of essential monomials will be finite in virtue of the Cayley-Hamilton theorem.

The class of models described above and based on using merely the material metric $\eta$ for producing $\widehat{N}$ from $G$ and $\widehat{\Omega}$ in (\ref{eq:209}), is distinguished among other ones by the very geometry of the material space. Nevertheless, other constitutive material tensors are also admissible. The peculiarity of the situation when the prescription (\ref{eq:209}) uses only the material metric $\eta$ as a tool for ``gluing'' $\widehat{G}$, $\widehat{\Omega}$ into $\widehat{N}$ is that one deals then simultaneously with the spatial $g$-isotropy and material $\eta$-isotropy.

Let us now just consider the problem of material affine symmetry and corresponding conservation laws. As mentioned, the only thing to be discussed is the action of center-affine material group isomorphic with $GL(U)$. Again, if we start from the invariance of equations of motion, without any reference to Lagrangian (existing or not), we conclude that the general solution is transformed onto itself if and only if the following two conditions hold: 
\begin{eqnarray}
b^{K}\!_{M}\:b^{L}\!_{N}J^{MN} & = & J^{KL},\label{eq:214}\\
N^{ij}\left(\varphi b,\frac{d\varphi}{dt}\:b,t\right) & = & N^{ij}\left(\varphi,\frac{d\varphi}{dt},t\right),\label{eq:215}
\end{eqnarray}
for any $b\in GL(U)$. This is the invariance under (\ref{eq:46}) with $a=Id_{V}$. Therefore, $b$ must be $J^{-1}$-orthogonal, $b\in O(U,J^{-1})$, and $N$ must be a function of $J[\varphi]$, $\Omega$, where $J[\varphi]$ is given by (\ref{eq:90}) and represents the internal inertia with respect to the space-fixed reference frame, 
\begin{equation}
N^{ij}=N^{ij}\left(J[\varphi]^{ab},\Omega^{k}\!_{l}\right).\label{eq:216}
\end{equation}

In the hyperelastic case this means that the internal potential energy depends on $\varphi$ through $J[\varphi]$; $\mathcal{V}=\mathcal{V}\left(J[\varphi]\right)$. Obviously, the only possibility to construct scalars from the twice contravariant tensor $J[\varphi]^{ab}$ is to use some other tensor in $V$, e.g., twice covariant one (the simplest possibility). Similarly, in the elastic case one must use some additional tensor objects in $V$ to construct $N^{ij}$ from $J[\varphi]^{ab}$. Obviously, the most natural possibility is just $g$ itself, the metric tensor of the physical space.

Usually we are interested in problems of spatial and material isotropy, i.e., invariance of equations of motion under (\ref{eq:46}) with $a$, $b$ running over the group $O(V,g)$, $O(U,\eta)$ respectively. The material isotropy of equations of motion is possible only when the inertial tensor is isotropic, i.e., 
\begin{equation}
J^{AB}=I\eta^{AB}.\label{eq:217}
\end{equation}

Then, obviously, in (\ref{eq:216}) $J[\varphi]$ is to be replaced by the inverse Cauchy tensor $C[\varphi]^{-1}$, thus, we can write:
\begin{equation}
N^{ij}=N^{ij}\left(C[\varphi]_{ab},\Omega^{a}\!_{b}\right).\label{eq:218}
\end{equation}
In the case of hyperelastic body we have, in analogy to (\ref{eq:210}):
\begin{equation}
N(\varphi)=-2D_{C}\mathcal{V},\qquad N^{ij}=-\frac{\partial\mathcal{W}}{\partial C_{ij}};\label{eq:219}
\end{equation}
the potential energy $\mathcal{V}$ depends on $\varphi$ through $C$, $\mathcal{V}(\varphi)=\mathcal{W}(C)$.

Let us summarize the above invariance analysis of affine dynamics: 
\begin{enumerate}
\item Equations of motion (\ref{eq:112}) are invariant under internal spatial rotations when the affine torque $\widehat{N}$ is an algebraic function of $G$, $\widehat{\Omega}$, 
\begin{equation}
\widehat{N}=\widehat{N}\left(G,\widehat{\Omega}\right).\label{eq:220}
\end{equation}
More precisely, one should explicitly insert into this expression some constitutive material tensors ${\bf C}_{U}$ in the space $U$,
\begin{equation}
\widehat{N}=\widehat{N}\left(\widehat{G},\widehat{\Omega};C_{U}\right).\label{eq:221}
\end{equation}
Those tensors are algebraically necessary for producing the quantity $\widehat{N}$ from $\left(\widehat{G},\widehat{\Omega}\right)$. And physically they give an account of the structure of internal interactions. 
\item Equations of internal motion (\ref{eq:112}) are invariant under material rotations in $U$ when the inertial tensor $J$ is isotropic, i.e., (\ref{eq:217}) holds, $J^{AB}=I\eta^{AB}$, and the Eulerian torque $N$ is algebraically built of $C$, $\Omega$, 
\begin{equation}
N=N(C,\Omega).\label{eq:222}
\end{equation}
And again some spatial constitutive tensors ${\bf C}_{V}$ in $V$ are algebraically necessary for prescribing $N$ as a function of $C$, $\Omega$, so more precisely, we should write: 
\begin{equation}
N=N(C,\Omega;{\bf C}_{V}).\label{eq:223}
\end{equation}
\item Equations of motion are simultaneously spatially and materially isotropic when both (\ref{eq:222}) (\ref{eq:223}) hold and are equivalent. But this means that in (\ref{eq:221}) ${\bf C}_{V}$ is built algebraically of $\eta$ and in (\ref{eq:223}) ${\bf C}_{V}$ is built algebraically of $g$, thus we have two equivalent representations: 
\begin{equation}
\widehat{N}=\widehat{N}(G,\widehat{\Omega};\eta),\qquad N=N(C,\Omega;g),\label{eq:224}
\end{equation}
or, alternatively,
\begin{equation}
\widehat{N}=\widehat{N}(\widehat{G},\widehat{\Omega};\eta),\qquad N=N(\widehat{C},\Omega;g),\label{eq:225}
\end{equation}
where, as usual, $\widehat{G}$, $\widehat{C}$ denote the mixed tensors obtained from $G$, $C$ respectively by the $\eta$-shift and $g$-shift of indices. Unfortunately, there is some disorder in notation, because characters are missing. For example, the ``roof'' symbol is used to denote the co-moving representation, but at the same time, it is also used for the metric-based shift of indices. It would be perhaps better and certainly non-ambiguous to use the symbols like $^{\eta}G$, $^{g}C$ for the $\eta$-raising and $g$-raising of the first index, 
\begin{equation}
\left(^{\eta}G\right)^{A}\!_{B}:=\eta^{AC}G_{CB},\qquad\left(^{g}C\right)^{i}\!_{j}:=g^{ik}C_{kj},\label{eq:226}
\end{equation}
and similarly $\widehat{N}_{\eta}$, $N_{g}$ in the same sense, but also, e.g., $\widehat{\Sigma}^{\eta}$, $\Sigma^{g}$, $\widehat{\Omega}^{\eta}$, $\Omega^{g}$, etc., for the metrical raising of the second index, and similarly for the metrical lowering, 
\begin{eqnarray}
\widehat{\Sigma}^{\eta\ AB} & := & \widehat{\Sigma}^{A}\!_{C}\:\eta^{CB},\quad\widehat{\Omega}^{\eta\! AB}:=\widehat{\Omega}^{A}\!_{C}\:\eta^{CB},\nonumber \\
\Sigma^{g\ ij} & := & \Sigma^{i}\!_{k}\:g^{kj},\quad\Omega^{g\! ij}:=\Omega^{i}\!_{k}\:g^{kj},\label{eq:227}\\
\widehat{N}_{\eta}\!^{A}\!_{B} & := & \widehat{N}^{AC}\eta_{CB},\quad N_{g}\!^{i}\!_{j}:=N^{ik}g_{kj}.\nonumber 
\end{eqnarray}
Obviously, the more indices, the worse nuisance and sometimes it is more convenient to use fewer symbols and to comment them in words or additional formulas, cf. back to the comments to formulas (\ref{eq:185}) (\ref{eq:188}).

But sometimes it is more convenient, or perhaps just necessary to be more pedantic. It is so even in spite of the often used argument that one can always simply use orthonormal bases in which $\eta_{AB}=_{*}\delta_{AB}$, $g_{ij}=_{*}\delta_{ij}$, and the last formulas become trivial. 
\end{enumerate}

Though we witnessed above the situation where, in a sense, indices are moved in the sense of Green or Cauchy deformation tensors. For various reasons not always orthonormal bases are most convenient, moreover we often must work in curvilinear coordinates, and also in curved manifold. But even if we use orthonormal bases, we can easily commit mistakes when forgetting about the geometric status of second-order tensors represented analytically by matrices. Any non-singular quadratic matrix $a$ with real entries induces three transformation rules in the linear spaces of real matrices of the same order: 
\begin{equation}
x\mapsto axa^{-1},\quad x\mapsto axa^{T},\quad x\mapsto a^{-1T}xa^{-1},\label{eq:228}
\end{equation}
corresponding respectively to mixed tensors, twice contravariant tensors, twice covariant tensors. In complex algebra there are also other rules, first of all: 
\begin{equation}
x\mapsto axa^{+},\qquad x\mapsto a^{-1+}xa^{-1}.\label{eq:229}
\end{equation}

So, let us repeat more carefully what we said above about the structure of affine torque for the doubly invariant models, using conventions (\ref{eq:226}) (\ref{eq:227}) and also the following ones concerning the metrical transposition of mixed tensors $X\in U\otimes U^{*}\simeq L(U)$, $Y\in V\otimes V^{*}\simeq L(V)$: 
\begin{eqnarray}
\left(X^{T}\right)^{A}\!_{B} & = & \eta_{BC}\:X^{C}\!_{D}\:\eta^{DA}=X_{B}\!^{A},\\
\left(Y^{T}\right)^{i}\!_{j} & = & g_{jk}\:Y^{k}\!_{m}\:g^{mi}=Y_{j}\!^{i}.\nonumber 
\end{eqnarray}
Obviously, those are metric-dependent operations, to be quite pedantic, we should have written $X^{T(\eta)}$, $Y^{T(g)}$, however, we avoid the crowd of symbols and always keep metric tensors $\eta$, $g$ implicitly assumed.

For velocity-independent internal forces, e.g., elastic ones, the doubly-isotropic affine torques $\widehat{N}$ ($g$-isotropic in physical space, $\eta$-isotropic in the body) have the following algebraic structure: 
\begin{equation}
\widehat{N}_{\eta}=\sum_{a=0}^{n-1}l_{a}\left(\mathcal{K}_{1},\ldots,\mathcal{K}_{n}\right)\left(^{\eta}G\right)^{a},\label{eq:231}
\end{equation}
where, obviously, the matrix exponents are meant, and $\mathcal{K}_{i}$ are basic deformation invariants, e.g., in the form (\ref{eq:23}), $\mathcal{K}_{i}=Tr\left(^{\eta}G^{i}\right)$; $l_{a}$ are scalar functions of $\mathcal{K}_{i}$. Analytically: 
\begin{equation}
\widehat{N}_{\eta}\!^{A}\!_{B}=\sum_{a=0}^{n-1}l_{a}(\mathcal{K})\ \underbrace{^{\eta}G^{A}\!_{C}\ \!^{\eta}G^{C}\!_{D}\ldots\!^{\eta}G^{L}\!_{B}}_{a\,\text{factors}},\label{eq:231a}
\end{equation}
i.e., 
\begin{equation}
\widehat{N}^{AB}=\sum_{a=0}^{n-1}l_{a}(\mathcal{K})\underbrace{\eta^{AC}G_{CD}\eta^{DE}\ldots\eta^{MN}G_{NL}\eta^{LB}}_{a\,\text{factors}\ G}.\label{eq:231b}
\end{equation}

This resembles some known constitutive rules used in continuum mechanics, as expected in view of (\ref{eq:120}). Nevertheless, the above formulas may be derived without averaging continuum mechanics expressions. They apply also to discrete affine bodies like molecules (e.g. fullerens) in an appropriate approximation. Obviously, from the naive point of view one might have expected the infinite series in (\ref{eq:231}), but of course, this series compresses to the finite sum in virtue of the Cayley-Hamilton theorem. Because of the same reason the summation may be extended over any range of integers from some $m\in\mathbb{Z}$ to $(m+n-1)$. Obviously, this changes the functions $l_{a}$. As usual in matrix calculus, for any square matrix $X$, the zero-th exponent is taken to be identity matrix $X^{o}=I$. When written in terms of $M$-spatial geometric objects, the formulas (\ref{eq:231}) (\ref{eq:232}) (\ref{eq:233}) become respectively: 
\begin{equation}
N_{g}=\sum_{a=0}^{n-1}l_{a}\left(\mathcal{K}_{1},\ldots,\mathcal{K}_{n}\right)\left(C_{g}^{-1}\right)^{a},\quad\left(C_{g}^{-1}\right)^{i}\!_{j}=C^{-1im}g_{mj},\label{eq:232}
\end{equation}
or, equivalently, 
\begin{equation}
N_{g}\!^{i}\!_{j}=\sum_{a=0}^{n-1}l_{a}(\mathcal{K})\underbrace{C_{g}^{-1}\!^{i}\!_{k}C_{g}^{-1}\!^{k}\!_{m}\ldots C_{g}^{-1}\!^{r}\!_{j}}_{a\,\text{factors}},\label{eq:232a}
\end{equation}
i.e., 
\begin{equation}
N^{ij}=\sum_{a=0}^{n-1}l_{a}(\mathcal{K})\underbrace{C^{-1\, ik}g_{km}C^{-1\, mn}\ldots C^{-1\, rs}g_{sz}C^{-1\, zj}}_{a\,\text{factors}\, C}.\label{eq:232b}
\end{equation}

If the internal forces are derivable from some potential depending only on deformation invariants, 
\begin{equation}
V=U\left(\mathcal{K}_{1},\ldots,\mathcal{K}_{n}\right),\label{eq:233}
\end{equation}
then the above formulas are characterized by the special form of controlling functions $l_{a}$, namely, one can shaw that 
\begin{equation}
l_{a}(\mathcal{K})=-2a\frac{\partial U}{\partial\mathcal{K}_{a}}.\label{eq:234}
\end{equation}

From the formal point of view, (\ref{eq:232}) is much more general than (\ref{eq:234}), because nothing like the vanishing of ``curl'' 
\begin{equation}
b\:\frac{\partial l_{a}}{\partial\mathcal{K}_{b}}-a\:\frac{\partial l_{b}}{\partial\mathcal{K}_{a}},\label{eq:235}
\end{equation}
is assumed. But, one must mention, quite often some doubts are raised if non-potential velocity-independent forces are physically realistic (although, of course, mathematically well-defined).

Let us go to velocity-dependent affine torques. We mean the dependence on internal velocities. Everything expressed in the formulas (\ref{eq:220}) (\ref{eq:221}) (\ref{eq:222}) (\ref{eq:223}) (\ref{eq:224}) (\ref{eq:225}). The analogues of (\ref{eq:231}) (\ref{eq:232}) are more complicated. Namely, (\ref{eq:231}) is replaced by a sum of monomials like 
\begin{equation}
l_{a}\left(Inv\left(^{\eta}G,\widehat{\Omega},\widehat{\Omega}^{T(\eta)}\right)\right)\left(^{\eta}G\right)^{\alpha}\widehat{\Omega}^{\varkappa}\left(\widehat{\Omega}^{T(\eta)}\right)^{\rho}\ldots\left(^{\eta}G\right)^{\gamma}\widehat{\Omega}^{\lambda}\left(\widehat{\Omega}^{T(\eta)}\right)^{\sigma}.\label{eq:236}
\end{equation}

In words: We take some products of linear operators $\left(^{\eta}G\right)^{\alpha} \widehat{\Omega}^{\varkappa} \left(\widehat{\Omega}^{T(\eta)}\right)^{\rho}$, multiply those monomials by coefficients depending on scalar invariants built of $^{\eta}G$, $\widehat{\Omega}$, $\widehat{\Omega}^{T(\eta)}$ and take the sum of resulting expressions. Obviously, the mentioned scalars are traces of operator monomials. The exponents at $^{\eta}G$ are integers, and those at $\widehat{\Omega}$, $\widehat{\Omega}^{T(\eta)}$ are non-negative integers. As usual, it is sufficient to take exponents from the range $\left(0,\ldots,(n-1)\right)$; this is a consequence of the Cayley-Hamilton theorem.

Similarly, (\ref{eq:232}) is replaced by a sum of the corresponding operator monomials in $L(V)$, 
\begin{equation}
l_{a}\left(Inv\left(^{g}C,\Omega,\Omega^{T(g)}\right)\right)\left(^{g}C\right)^{\alpha}\Omega^{\varkappa}\left(\Omega^{T(g)}\right)^{\rho}\ldots\left(^{g}C\right)^{\gamma}\Omega^{\lambda}\left(\Omega^{T(g)}\right)^{\sigma}.\label{eq:237}
\end{equation}

Literally meant expressions like (\ref{eq:236}) (\ref{eq:237}) are, so-to-speak, ``neurotically'' too general. They represent what algebra does answer to the inquiry concerning the most general doubly isotropic prescriptions for the affine torque. But this is pure algebra, in physics only some special simple models are realistic.

In applications the affine torque $N$, just as its co-moving representation $\widehat{N}$, very often is given as the sum of two terms: one depending only on the configuration $\varphi$, and the other one describing generalized forces which depend in an essential way on velocities. The first term describes in particular the elastic and hyperelastic
behavior. When one deals only with purely internal interactions, it is $g$-isotropic, when the constitution of the object does not distinguish any ``material'' direction, it is $\eta$-isotropic. These two situations are described respectively by (\ref{eq:220}) (\ref{eq:221}), but without the $\widehat{\Omega}$-dependence and by (\ref{eq:222}) (\ref{eq:223}),  so respectively 
\begin{equation}
\widehat{N}=\widehat{N}\left(G;\mathbf{C}_{U}\right),\qquad N=N\left(C;\mathbf{C}_{V}\right).\label{eq:238}
\end{equation}
For the doubly-invariant models this reduces to: 
\begin{equation}
\widehat{N}=\widehat{N}\left(G;\eta\right),\qquad N=N\left(C;g\right).\label{eq:239}
\end{equation}
and this may be alternatively written as (\ref{eq:231}) (\ref{eq:232}). If the system is potential, e.g., hyperelastic, then (\ref{eq:234}) holds.

What concerns velocity-dependent forces, the simplest model in the special case of discretized continua was (\ref{eq:192}), on the basis of linear isotropic viscoelasticity (\ref{eq:194}) (\ref{eq:120}). This does not seem adequate when one deals with discrete systems, and then it is rather more natural to postulate the internal friction term in affine torques in the form: 
\begin{equation}
N_{int.diss}=-\alpha\left(\Omega^{g}+\Omega^{gT}\right)-\beta \:Tr(\Omega)\: g^{-1},\label{eq:240}
\end{equation}
or using the notation (\ref{eq:30}) (\ref{eq:227}), 
\begin{equation}
N_{int.diss\ g}=-\alpha\left(\Omega+\Omega^{T(g)}\right)-\beta \: Tr(\Omega){\rm I}=-2\alpha d-\beta\ Tr(d){\rm I},\label{eq:241}
\end{equation}
obviously, ${\rm I}$ denotes the identity operator in $V$, and $\alpha>0$, $\beta>0$.

Let us notice that in the physical case of weakly compressible (almost isochoric) objects, the formula (\ref{eq:192}) contains higher-order terms, nonlinear in state variables. It is seen that (\ref{eq:240})/(\ref{eq:241}) is but the very special case of (\ref{eq:237}). Using the technique of invariant tensor expressions we can write the general expression for the isotropic fluid-type internal dynamics in the form: 
\begin{equation}
N_{int.diss\ g}=\sum_{a=0}^{n-1}f_{a}\left(\mathcal{L}_{1}\ldots\mathcal{L}_{n}\right)d^{a},\label{eq:242}
\end{equation}
where $\mathcal{L}_{a}$ are scalar invariants built of $d$ according to the standard trace prescription, 
\begin{equation}
\mathcal{L}_{b}=Tr(d^{b}).\label{eq:243}
\end{equation}

The formula (\ref{eq:242}) may also contain the non-dissipative pressure term 
\begin{equation}
N_{pr\ g}=-p\:{\rm I}_{n},\quad\text{i.e.,}\ N_{pr}=-p\:g^{-1}.\label{eq:244}
\end{equation}

Obviously, if the internal friction is anisotropic, the prescription for $N$ as a tensorial function of $d$, must contain some constitutive tensors, e.g., in the linear case: 
\begin{equation}
N_{int.diss}^{ij}=-{\rm V}^{ijab}\:d_{ab},\label{eq:245}
\end{equation}
the shift of indices meant in the metrical $g$-sense.

It is easy to reformulate the above expressions into language of co-moving geometric objects (tensors in $U$).

Finally, let us mention about other, very important dissipative problems, namely, the external surface friction. Applications are obvious: imagine a homogeneously deformable small suspension or inclusion moving in fluid. It is not only translational motion but also the internal motion in $\varphi$-degrees of freedom that is faced with frictional obstacles, we mean the friction between the surface of ``suspension/inclusion'' and the surrounding medium. As usual, the simplest and most natural assumption is that of generalized friction forces linear in generalized velocities, e.g., in the isotropic case, 
\begin{equation}
N_{ext.diss\ g}=-\nu\:\Omega.\label{eq:246}
\end{equation}

This simple expression, however, looks rather not very adequate, because it is physically natural to expect that the internal motion is obstacled different way in the special cases of rotational, shear-like and dilatational motion. So, it is reasonable to suppose something like 
\begin{eqnarray}
N_{ext.diss}^{ij} & = & -\alpha\:\omega^{ij}-\beta\left(d^{ij}-\frac{1}{n}\:g_{ab}\:d^{ab}\:g^{ij}\right)-\gamma \:g_{ab}\:d^{ab}g^{ij}=\nonumber \\
 & = & -\alpha\omega^{ij}-\beta d^{ij}-\left(\gamma-\frac{\beta}{n}\right)g_{ab}\:d^{ab}g^{ij},\label{eq:247}
\end{eqnarray}
where $\alpha$, $\beta$, $\gamma$ are positive constants, and the meaning of symbols $\omega$, $d$ is like in (\ref{eq:30}) (\ref{eq:31}). Obviously, one can also discuss anisotropic models when the above constants $\alpha$, $\beta$, $\gamma$ are replaced by some fourth-order tensors. Another a priori possible modification is the external friction nonlinear in velocities. In the isotropic case this will be again obtained from the combination of operator monomials $(d_{g})^{a}$, $a=0,1,\ldots,n-1$ with coefficients depending on the scalar invariants of $d$, i.e., quantities $Tr(d^{b})$, $a=0,1,\ldots,n-1$.

\section{Towards affine dynamical symmetry}

Two very important and at the same time very delicate points were stressed many times in this paper. Let us repeat and discuss them, to be able to finish this step of investigation with some some conclusions opening the perspective on further developments. 
\begin{enumerate}
\item \label{enu:There-are-two}There are two ways of deriving equations of motion of complex and constrained systems: 
\begin{enumerate}
\item \label{enu:the-procedure-based}the procedure based on Newton equations and d'Alembert principle 
\item \label{enu:the-procedure-based H}the procedure based on the variational Hamiltonian principle, Hamiltonian formalism and Poisson brackets. 
\end{enumerate}
\item \label{enu:Kinematic-of-our}Kinematic of our system is based on affine geometry, however, its dynamic in not invariant under the action of affine group. The highest dynamical invariance we were dealing with above, was that under isometry groups, both in the spatial and material sense. Therefore, there is only partial analogy between our equations of motion and gyroscopic Euler equations. This is at least aesthetically non-satisfactory and disappointing, and brings about some questions concerning the status of dynamical affine symmetry in mechanics and fundamental physics. 
\end{enumerate}
The procedure \ref{enu:the-procedure-based} in the item \ref{enu:There-are-two} is rather more popular among specialists in continuum mechanics. It is directly applicable both to conservative and dissipative systems. The balance form of equations of motion appears there in a rather natural way , nevertheless, there is no direct and systematic relationship between symmetries and conservation laws, only some intuitive hints do exist. Unlike this, the procedure \ref{enu:the-procedure-based H} offers a systematic theory of that relationship; it is based on Noether theorems. On the other hand, dissipative terms of equations of motion are then introduced ``by hand'', and they are always more or less external, exotic corrections, a ``foreign body'' in Hamiltonian framework. But the balance form of equations of motion appears there simply in a canonical way.

The item \ref{enu:Kinematic-of-our} is strongly related to that problem. Namely, the search of affinely-invariant dynamical models is much more easy within the Hamiltonian approach with its direct relationship between symmetries and conservation or balance laws. Namely, it is seen that the first and main obstacle against dynamical affine symmetry is due to the position of the spatial and material (reference) metric tensors $g$, $\eta$ in the ``usual'' expression for the kinetic energy. In other words, it is due to the very particular Euclidean structure of the configuration space, namely, one implemented by Euclidean structures in $M$, $N$, the physical and material spaces, according to (\ref{eq:107a}), this metric $\Gamma$ is given by 
\begin{equation}
\Gamma=mg\oplus\left(g\otimes J\right)\quad,\quad\Gamma^{-1}=\frac{1}{m}\:g^{-1}\oplus\left(g^{-1}\otimes J^{-1}\right).\label{eq:248}
\end{equation}

However, from the purely geometric point of view other metrics on the configuration space are much more natural, ones partially or completely independent on metrics in $M$, $N$.This is geometry, but some physical motivation was also outlined in the Introduction; in any case, physical models may be formulated only when mathematical background is prepared.

The apparently strange formulas (\ref{eq:170}) (\ref{eq:171}), i.e., an alternative expression of (\ref{eq:107a}) is a good starting point. In (\ref{eq:107a}) we were dealing with a quadratic form of generalized velocities with constant coefficients built of $g$ and $J$. Unlike this, in (\ref{eq:170}) (\ref{eq:171}) kinetic energy is expressed as a quadratic form of geometrically nicely-interpretable non-holonomic velocities, however with configuration \textendash\! dependent coefficients built of $\left(G\left[\varphi\right],J\right)$ or $\left(g,J\left[\varphi\right]\right)$. So, the bad alternative: either constant coefficients but representation of velocities non-adapted to geometry of the problem, or conversely-geometric affine velocities but variable coefficients. Why not to take the ``good'' features of both schemes and just to unify, join together their advantages? There are two possibilities of expressions quadratic in $\left(v^{i},\Omega^{i}\!_{j}\right)$ $\left(\hat{v}^{A},\widehat{\Omega}^{A}\!_{B}\right)$ with constant coefficients.

The first one consists in replacing $G_{AB}$ in (\ref{eq:170}) (\ref{eq:171}) by $\eta$: 
\begin{eqnarray}
T_{tr} & = & \frac{m}{2}\:\eta_{AB}\:\hat{v}^{A}\:\hat{v}^{B},\label{eq:249}\\
T_{int} & = & \frac{1}{2}\:\eta_{AB}\:\widehat{\Omega}^{A}\!_{K}\:\widehat{\Omega}^{B}\!_{L}J^{KL}.\label{eq:250}
\end{eqnarray}
The second possibility fixes the spatial metric $g$ and some additional spatial tensor $h\in V^{*}\otimes V^{*}$; the latter one is substituted instead the configuration-dependent $J\left[\varphi\right]$, 
\begin{eqnarray}
T_{tr} & = & \frac{m}{2}\:g_{ij}v^{i}v^{j},\label{eq:251}\\
T_{int} & = & \frac{1}{2}\:g_{ij}\:\Omega^{i}\!_{k}\:\Omega^{j}\!_{l}\:h^{kl}.\label{eq:252}
\end{eqnarray}

Let us observe that (\ref{eq:249}) (\ref{eq:250}) may be alternatively written as follows: 
\begin{eqnarray}
T_{tr} & = & \frac{m}{2}C\left[\varphi\right]_{ij}\frac{dx^{i}}{dt}\frac{dx^{j}}{dt}\label{eq:253}\\
T_{int} & = & \frac{1}{2}C\left[\varphi\right]_{ij}\frac{d\varphi^{i}\!_{A}}{dt}\frac{d\varphi^{j}\!_{B}}{dt}J^{AB}.\label{eq:254}
\end{eqnarray}

This is like (\ref{eq:107a}); the difference is that the usual metric tensor $g$ is replaced by the Cauchy deformation tensor $C[\varphi]$. Because of this there is something like the mentioned similarity to the concept of effective mass in solid state physics.

Similarly, (\ref{eq:251}) (\ref{eq:252}) may be written down as follows: 
\begin{eqnarray}
T_{tr} & = & \frac{m}{2}\:g_{ij}\frac{dx^{i}}{dt}\frac{dx^{j}}{dt},\label{eq:255}\\
T_{int} & = & \frac{1}{2}\:g_{ij}\frac{d\varphi^{i}\!_{A}}{dt}\frac{d\varphi^{j}\!_{B}}{dt}\:h\left[\varphi\right]^{AB}.\label{eq:256}
\end{eqnarray}
where: 
\begin{equation}
h\left[\varphi\right]^{AB}=\left.\varphi^{-1}\right.^{A}\!_{i}\left.\varphi^{-1}\right.^{B}\!_{j}\ h^{ij}\label{eq:257}
\end{equation}
is the co-moving, thus, configuration-dependent representation of the spatial tensor $h$.

Particularly interesting is the special case of maximal symmetry under isometries in $U$, $V$ respectively, when 
\begin{equation}
J^{KL}={\rm I}\eta^{KL},\qquad h^{ij}={\rm I}g^{ij}.\label{eq:258}
\end{equation}

Then we respectively obtain for (\ref{eq:250}) (\ref{eq:252} ) 
\begin{eqnarray}
T_{int} & = & \frac{{\rm I}}{2}\:\eta_{AB}\:\widehat{\Omega}^{A}\!_{K}\:\widehat{\Omega}^{B}\!_{L}\:\eta^{KL},\label{eq:259}\\
T_{int} & = & \frac{{\rm I}}{2}\:g_{ij}\:\Omega^{i}\!_{k}\:\Omega^{j}\!_{l}\:g^{kl}.\label{eq:260}
\end{eqnarray}

An important property of the model (\ref{eq:249}) (\ref{eq:250}) is that it is invariant under the total affine group $GAf\left(M\right)$ in the physical space. What concerns material invariance, (\ref{eq:249}) is invariant under orthogonal group $O(U,\eta)$, and (\ref{eq:250}) is invariant under $O(U,\eta)\cap O(U,J^{-1})$. The latter group becomes simply $O(U,\eta)$, when then internal tensor $J$ is isotropic, i.e., the first of equations (\ref{eq:258}) holds. As seen from (\ref{eq:253}) (\ref{eq:254}), the corresponding metric tensor $\Gamma$ of the configuration space is given by 
\begin{equation}
\Gamma=m\:C[\varphi]\oplus\left(C\left[\varphi\right]\otimes J\right)\quad,\quad\Gamma^{-1}=\frac{1}{m}\:C[\varphi]^{-1}\oplus\left(C\left[\varphi\right]^{-1}\otimes J^{-1}\right).\label{eq:261}
\end{equation}
Unlike (\ref{eq:248}), it is curved; the corresponding geometry in the configuration space $Q$ is essentially Riemannian. It has a large isometry which contains $GAf(M)$ acting through \ref{eq:34} with $B={\rm I}d_{N}$, and $O(U,\eta)\cap O(U,J^{-1})$ (in particular just $O(U,\eta)$ when $J={\rm I}\eta^{-1}$) acting through (\ref{eq:46}) on $Q_{int}$ and trivially on $M$.

Quite symmetrically, the model (\ref{eq:251}) (\ref{eq:252}) is invariant under the total $GL(U)$ acting through (\ref{eq:46}) on our configuration space and it is also invariant under $E(M,g)\cap E(M,h)$ (in particular, under $E(M,g)$ when $h={\rm I}g^{-1}$) acting through (\ref{eq:34}) with $B={\rm Id}_{n}$. The corresponding metric tensor $\Gamma$ on $Q$ is given by 
\begin{equation}
\Gamma=m\:g\oplus\left(g\otimes h\left[\varphi\right]\right)\quad,\quad\Gamma^{-1}=\frac{1}{m}\:g^{-1}\oplus\left(g^{-1}\otimes h\left[\varphi\right]^{-1}\right).\label{eq:262}
\end{equation}

The internal parts of metric tensors (\ref{eq:261}) (\ref{eq:262}), i.e., the corresponding metric tensors on $Q_{int}$ do factorize into tensor products of $V$- and $U$-terms, just like (\ref{eq:248}).

But if we once dare to give up the d'Alembert form (\ref{eq:107a}) there is no reason any longer stick to factorization. The most general kinetic energy (Riemannian metric) on $Q$ splitting into translational and internal parts, and affinely invariant in $M$ (invariant under $GAffM$ acting through (\ref{eq:34}) with $B={\rm Id}_{N}$) has the form: 
\begin{equation}
T=T_{tr}+T_{int}=\frac{m}{2}\:\eta_{AB}\:\hat{v}^{A}\:\hat{v}^{B}+\frac{1}{2}\:\mathcal{L}^{B}\!_{A}\!^{D}\!_{C}\:\widehat{\Omega}^{A}\!_{B}\:\widehat{\Omega}^{C}\!_{D},\label{eq:263}
\end{equation}
where the coefficients $\mathcal{L}$ are constant; obviously they are components of some fourth-order tensor in $U$. Being coefficients of a quadratic from of $\widehat{\Omega}$ they are symmetric in biindices, i.e., 
\begin{equation}
\mathcal{L}^{B}\!_{A}\!^{D}\!_{C}=\mathcal{L}^{D}\!_{C}\!^{B}\!_{A}.\label{eq:264}
\end{equation}

It is clear that the metric tensor $\Gamma$ underlying \ref{eq:263} has the form: 
\begin{equation}
\Gamma=mC\left[\varphi\right]_{ij}dx^{i}\otimes dx^{j}+\mathcal{L}^{B}\!_{A}\!^{D}\!_{C}\left.\varphi^{-1}\right.^{A}\!_{i}\left.\varphi^{-1}\right.^{C}\!_{j}\:d\varphi^{i}\!_{B}\otimes d\varphi^{j}\!_{D}.\label{eq:265}
\end{equation}

It is curved, in none generalized coordinates its components may become constant. Let us observe that the $d\varphi\otimes d\varphi$-part is autonomous, unlike this, the $dx\otimes dx$-part is $\varphi$-dependent.

Similarly, for kinetic energies affinely-invariant in $N$ we have:
\begin{equation}
T=T_{tr}+T_{int}=\frac{m}{2}\:g_{ij}v^{i}v^{j}+\frac{1}{2}\:\mathcal{R}^{j}\!_{i}\!^{l}\!_{k}\:\Omega^{i}\!_{j}\:\Omega^{k}\!_{l},\label{eq:266}
\end{equation}
thus, the corresponding metric tensor $\Gamma$ on $Q$ has the form,
\begin{equation}
\Gamma=m\:g_{ij}\:dx^{i}\otimes dx^{j}+\left.\varphi^{-1}\right.^{A}\!_{j}\left.\varphi^{-1}\right.^{B}\!_{l}\:\mathcal{R}^{j}\!_{i}\!^{l}\!_{k}\:d\varphi^{i}\!_{A}\otimes d\varphi^{k}\!_{B}.\label{eq:267}
\end{equation}
This time the internal and translational part are mutually independent. Obviously, $\mathcal{R}$ is a constant fourth-order tensor in $V$, symmetric in biindices, just like $\mathcal{L}$ (\ref{eq:264}).

There is no model of kinetic energy, i.e., no Riemannian structure on $Q$ which would be affinely-invariant simultaneously in $N$ and $M$. The reason is that the affine group is not semisimple and its translations subgroup is a normal divisor. For example (\ref{eq:263}) (\ref{eq:265}) is affinely-invariant in $M$ but its maximal group of $N$-symmetries consists of isometries in $U$. And conversely, (\ref{eq:266}) (\ref{eq:267}) is affinely-invariant in $N$, but in $M$ it is invariant at most under isometries. The mentioned situations of maximal two-side symmetry occur when the tensors $\mathcal{L}$, $\mathcal{R}$ are algebraically built respectively of $\left(\eta,Id_{U}\right)$ or $\left(g,Id_{V}\right)$.

For kinetic energies (metric tensors on $Q$) affinely-invariant in $M$ and only isometrically invariant in $N$, we have 
\begin{equation}
\mathcal{L}^{B}\!_{A}\!^{D}\!_{C}=\frac{I}{2}\:\eta_{AC}\:\eta^{BD}+\frac{A}{2}\:\delta^{B}\!_{C}\delta^{D}\!_{A}+\frac{B}{2}\:\delta^{B}\!_{A}\delta^{D}\!_{C}.\label{eq:268}
\end{equation}
$I,A,B$ denoting constants-generalized scalar moments of inertia in affine motion.

Similarity, for models affinely-invariant in $N$ and only isometrically invariant in $M$ we have 
\begin{equation}
\mathcal{R}^{j}{}_{i}{}^{l}{}_{k}=\frac{I}{2}\:g_{ik}g^{jl}+\frac{A}{2}\:\delta^{j}\!_{k}\delta^{l}\!_{i}+\frac{B}{2}\:\delta^{j}\!_{i}\delta^{l}\!_{k}\label{eq:269}
\end{equation}
with the same meaning of constants.

Therefore, explicitly we have 
\begin{eqnarray}
T_{int} & = & \frac{I}{2}\:Tr\left(\widehat{\Omega}^{T}\widehat{\Omega}\right)+\frac{A}{2}\:Tr\left(\widehat{\Omega}^{2}\right)+\frac{B}{2}\:Tr\left(\widehat{\Omega}\right)^{2},\label{eq:270}\\
T_{int} & = & \frac{I}{2}\:Tr\left(\Omega^{T}\Omega\right)+\frac{A}{2}\:Tr\left(\Omega^{2}\right)+\frac{B}{2}\:Tr\left(\Omega\right)^{2},\label{eq:271}
\end{eqnarray}
respectively for (\ref{eq:268}) and (\ref{eq:269}). Let us stress that the transposition of $\widehat{\Omega}\in L(U)$, $\Omega\in L(V)$ is always meant in the metrical sense, respectively of the metric tensor $\eta$, $g$: 
\begin{eqnarray}
\left(\widehat{\Omega}^{T}\right)^{A}\!_{B} & = & \eta^{AC}\eta_{BD}\:\widehat{\Omega}^{D}\!_{C}=\widehat{\Omega}_{B}\!^{A}.\label{eq:272}\\
\left(\Omega^{T}\right)^{i}\!_{j} & = & g^{ik}g_{jl}\:\Omega^{l}\!_{k}=\Omega_{j}\!^{i}.\label{eq:273}
\end{eqnarray}

The two last terms in (\ref{eq:270}) (\ref{eq:271}) are pairwise identical, but nevertheless it is convenient to distinguish consequently between the symbols $\widehat{\Omega}$ and $\Omega$ under the corresponding trace expressions. This is not only more ``aesthetic'', but also prevents from some mistakes; though the first terms in (\ref{eq:270}) (\ref{eq:271}) are different, and there it is just forbidden to confuse $\widehat{\Omega}$ with $\Omega$.

It is interesting that there exist models of $T_{int}$ which are simultaneously invariant under $GL(V)$ and $GL(U)$. They correspond to the vanishing value of ${\rm I}$ , i.e., 
\begin{equation}
T_{int}=\frac{A}{2}Tr\left(\widehat{\Omega}^{2}\right)+\frac{B}{2}Tr\left(\widehat{\Omega}\right)^{2}=\frac{A}{2}Tr\left(\Omega^{2}\right)+\frac{B}{2}Tr\left(\Omega\right)^{2}.\label{eq:274}
\end{equation}

Except the singular case $n=1$ such a ``kinetic energy'' (metric tensor on $Q_{int}$) is never positively definite. Nevertheless it may be physically useful and the negative configurations may be interpreted as an alternative description of elastic forces, without any use of potential energy term, just within the framework of purely geodetic models.

The first term in (\ref{eq:274}) has the signature $\left(\frac{1}{2}n(n+1),\frac{1}{2}n(n-1)\right)$. Obviously, this is the main term, and the second one is a merely correction, for $A=0$ the corresponding ``metric'' would be strongly degenerate. For the special case $A=2n$, $B=-2$ more ``generally'' for the ratio $A:B=n:(-1)$, one obtains the Killing ``metric'' on the linear group (or, more ``generally'', something proportional to it). Obviously, this ``metric'' is degenerate and has a one-dimensional singularity corresponding to the dilatational normal divisor of the linear group. For the generic choice of $\left({\rm I},A,B\right)$ the metrics/kinetic energies (\ref{eq:270}) (\ref{eq:271}) are non-degenerate. There exists an open subset of triples $\left({\rm I},A,B\right)\in\mathbb{R}$ for which these metrics are positively definite (Riemannian). With exception of the singular dimension $n=1$, for all such triples ${\rm I}$ must be non-vanishing.

It is instructive to notice that the total kinetic energy corresponding to (\ref{eq:249}) (\ref{eq:270}), i.e., for the $M$-affine and $N$-metrical models may be written down in the alternative forms:
\begin{eqnarray}
T & = & \frac{m}{2}\:\eta_{AB}\:\hat{v}^{A}\:\hat{v}^{B}+\frac{{\rm I}}{2}\:\eta_{KL}\:\eta^{MN}\:\widehat{\Omega}^{K}\!_{M}\:\widehat{\Omega}^{L}\!_{N}+\label{eq:275}\\
 & + & \frac{A}{2}\:\widehat{\Omega}^{I}\!_{J}\:\widehat{\Omega}^{J}\!_{I}+\frac{B}{2}\:\widehat{\Omega}^{I}\!_{I}\:\widehat{\Omega}^{J}\!_{J}=\nonumber \\
 & = & \frac{m}{2}\:C_{ij}\:v^{i}v^{j}+\frac{{\rm I}}{2}\:C_{kl}\:C^{mn}\:\Omega^{k}\!_{m}\:\Omega^{l}\!_{n}+\frac{A}{2}\:\Omega^{i}\!_{j}\:\Omega^{j}\!_{i}+\frac{B}{2}\:\Omega^{i}\!_{i}\:\Omega^{j}\!_{j}.\nonumber 
\end{eqnarray}

Similarity, for the kinetic energy corresponding to (\ref{eq:251}), (\ref{eq:271}), i.e., one affinely invariant in $N$ and metrical in $M$ we have: 
\begin{eqnarray}
T & = & \frac{m}{2}\:G_{AB}\:\hat{v}^{A}\:\hat{v}^{B}+\frac{{\rm I}}{2}\:G_{KL}\:G^{MN}\:\widehat{\Omega}^{K}\!_{M}\:\widehat{\Omega}^{L}\!_{N}+\nonumber \\
 & + & \frac{A}{2}\:\widehat{\Omega}^{I}\!_{J}\:\widehat{\Omega}^{J}\!_{I}+\frac{B}{2}\:\widehat{\Omega}^{I}\!_{I}\:\widehat{\Omega}^{J}\!_{J}=\label{eq:276}\\
 & = & \frac{m}{2}\:g_{ij}\:v^{i}v^{j}+\frac{{\rm I}}{2}\:g_{kl}\:g^{mn}\:\Omega^{k}\!_{m}\:\Omega^{l}\!_{n}+\frac{A}{2}\:\Omega^{i}\!_{j}\:\Omega^{j}\!_{i}+\frac{B}{2}\:\Omega^{i}\!_{i}\:\Omega^{j}\!_{j}.\nonumber 
\end{eqnarray}

A complete description of the scheme of breaking the affine symmetry and reducing it to the metrical one is achieved when some additional metric-dependent terms are admitted. The corresponding expression for the kinetic energy, i.e., for the metric tensor on $Q$, has the form: 
\begin{eqnarray}
T & = & \frac{1}{2}\left(m_{1}G_{AB}+m_{2}\:\eta_{AB}\right)\hat{v}^{A}\:\hat{v}^{B}+ \label{eq:277} \\
&+& \frac{1}{2}\:\left({\rm I}_{1}G_{K L}G^{M N}+{\rm I}_{2}\:\eta_{K\! L}\:\eta^{MN}+ {\rm I}_{3}\:G_{KL}\:\eta^{MN}+\right. \nonumber \\ 
&+&\left. {\rm I}_{4}\:\eta_{KL}\:G^{MN}\right)\widehat{\Omega}^{K}\!_{M}\:\widehat{\Omega}^{L}\!_{N}
 +  \frac{A}{2}\:\widehat{\Omega}^{I}\!_{J}\:\widehat{\Omega}^{J}\!_{I} +\frac{B}{2}\:\widehat{\Omega}^{I}\!_{I}\:\widehat{\Omega}^{J}\!_{J}\nonumber
\end{eqnarray}
or, alternatively, 
\begin{eqnarray}
T & = & \frac{1}{2}\left(m_{1}\:g_{ij}+m_{2}\:C_{ij}\right)v^{i}v^{j}+\nonumber \\
 & + & \frac{1}{2}\left({\rm I}_{1}\:g_{kl}\:g^{mn}+{\rm I}_{2}\:C_{kl}\:C^{mn}\right)\Omega^{k}\!_{m}\:\Omega^{l}\!_{n}+\label{eq:278}\\
 & + & \frac{1}{2}\left({\rm I}_{3}\:g_{kl}\:C^{mn}+{\rm I}_{4}\:C_{kl}\:g^{mn}\right)\Omega^{k}\!_{m}\:\Omega^{l}\!_{n}+\nonumber \\
 & + & \frac{A}{2}\:\Omega^{i}\!_{j}\:\Omega^{j}\!_{i}+\frac{B}{2}\:\Omega^{i}\!_{i}\:\Omega^{j}\!_{j}.\nonumber 
\end{eqnarray}

If translational degrees of freedom are active, then the two-side affine invariance is not possible. It is possible only when we formally put $m_{1}=0$, $m_{2}=0$ (translational degrees of freedom neglected), and in addition ${\rm I}_{1}=0$, ${\rm I}_{2}=0$, ${\rm I}_{3}=0$, ${\rm I}_{4}=0$. Then the metric tensor on $Q_{int}$ is affinely-invariant both on the left (in space) and on the right (in the body). The total affine invariance in space is obtained when $m_{1}=0$, ${\rm I}_{1}=0$, ${\rm I}_{3}=0$, ${\rm I}_{4}=0$. The total affine invariance in the material sense corresponds to the choice: $m_{2}=0$, ${\rm I}_{2}=0$, ${\rm I}_{3}=0$, ${\rm I}_{4}=0$. For any choice of constants in (\ref{eq:277}) (\ref{eq:278}) the corresponding kinetic energy (metric tensor on $Q$) is invariant under spatial and material isometries. All those metrics are curved (essentially Riemannian), except the special case $m_{2}=0$, ${\rm I}_{1}=0$, $I_{2}=0$, ${\rm I}_{4}=0$, $A=0$, $B=0$. In this special case the metric (\ref{eq:277}) (\ref{eq:278}) becomes flat (Euclidean) and reduces to (\ref{eq:107a}), i.e., (\ref{eq:170}), (\ref{eq:171}), or, more precisely, to its particular case $J^{AB}={\rm I}\eta^{AB}$, i.e., the spherically symmetric top subject to homogeneous deformations. Therefore, $m_{1}=m$, ${\rm I}_{3}={\rm I}$, $m$ denoting the usual mass of the body and ${\rm I}$ the scalar inertial moment of the isotropic top.

It is both easy and instructive to write down explicitly the Riemannian metrics $\Gamma$ on the configuration space $Q$, underlying the above kinetic energies. Namely, for (\ref{eq:277}, \ref{eq:278}) they are given by 
\begin{eqnarray}
\Gamma & = & \frac{1}{2}\left(m_{1}\:g_{ij}+m_{2}\:C_{ij}\right)dx^{i}\otimes dx^{j}+\nonumber \\
 & + & \left({\rm I}_{1}\:g_{ij}\left.G^{-1}\right.^{AB}+{\rm I}_{2}\:C_{ij}\:\eta^{AB}+{\rm I}_{3}\:g_{ij}\:\eta^{AB}+{\rm I}_{4}\:C_{ij}\left.G^{-1}\right.^{AB}\right.\label{eq:279}\\
 & + & \left.A\left.\varphi^{-1}\right.^{A}\!_{j}\left.\varphi^{-1}\right.^{B}\!_{i}+B\left.\varphi^{-1}\right.^{A}\!_{i}\left.\varphi^{-1}\right.^{B}\!_{j}\right)d\varphi^{i}\!_{A}\otimes d\varphi^{j}\!_{B}.\nonumber 
\end{eqnarray}
This is the family of metric tensors on $Q$, ordered in a hierarchic way on the basis of their isometry groups. As mentioned, certain choices of constants correspond to isometry groups containing $GL(V)$ acting on the left, $GL(U)$ acting on the right, and sometimes both of them if translational degrees of freedom are neglected.

More general choices of controlling parameters correspond to situations when the isometry groups in $Q$ are smaller and based only on isometries in $M,N$. If they contain transformations induced by affine isomorphisms of $M$ or /and $N$, then certainly the metrics $\Gamma$ are curved (essentially Riemannian). Their interesting feature is that such affine models may describe bonded elastic vibrations without any use of potential energy. The elastic dynamics may be encoded then in the very form of appropriately chosen kinetic energy, as a purely geodetic motion in $Q$. Without affine invariance this would be impossible. In particular, if $m_{2}=0$, ${\rm I}_{1}=0$, ${\rm I}_{2}=0$, ${\rm I}_{4}=0$, $A=0$, $B=0$, the corresponding metrics on $Q$ are flat (Euclidean) and the general solution consists of straight-line in $Q$, evidently non-bounded, non-physical behavior.

When dealing with models admitting hypothetic affine symmetry, at least partial one, we cannot rely upon the d'Alembert principle in its traditional formulation. The only natural procedure one has at disposal then, is based on variational principle and Hamiltonian formalism. Dissipative forces are then postulated as some correction terms motivated by some phenomenological and intuitive guiding hints. This procedure, based on Poisson brackets and Legendre transformation, was described in section 5, however it was specialized there to non-affine models. Let us now write some explicit formulas for metrics affinely invariant in $M$ or in $N$. In the case of affine symmetry in space, (\ref{eq:263}) (\ref{eq:265}), Legendre transformation has the form: 
\begin{equation}
\hat{p}_{A}=m\:\eta_{AB}\:\hat{v}^{B}\quad,\quad\widehat{\Sigma}^{A}\!_{B}=\mathcal{L}^{A}\!_{B}\!^{C}\!_{D}\:\widehat{\Omega}^{D}\!_{C}.\label{eq:280}
\end{equation}
Inverting it we obtain the following formula for the kinetic Hamiltonian, i.e., expression of energy through canonical variables: 
\begin{equation}
\mathcal{T}=\mathcal{T}_{tr}+\mathcal{T}_{int}=\frac{1}{2m}\:\eta^{AB}\:\hat{p}_{A}\hat{p}_{B}+\frac{1}{2}\:\widetilde{\mathcal{L}}^{A}\!_{B}\!^{C}\!_{D}\:\widehat{\Sigma}^{B}\!_{A}\:\widehat{\Sigma}^{D}\!_{C},\label{eq:281}
\end{equation}
where 
\begin{equation}
\widetilde{\mathcal{L}}^{A}\!_{B}\!^{K}\!_{L}\:\mathcal{L}^{L}\!_{K}\!^{C}\!_{D}=\delta^{A}\!_{D}\:\delta^{C}\!_{B}.\label{eq:282}
\end{equation}
The corresponding contravariant inverse of the metric tensor $\Gamma$ (\ref{eq:265}) is given by 
\begin{equation}
\Gamma^{-1}=\frac{1}{m}\left.C[\varphi]^{-1}\right.^{ij}\frac{\partial}{\partial x^{i}}\otimes\frac{\partial}{\partial x^{j}}+ \widetilde{\mathcal{L}}^{A}\!_{B}\!^{C}\!_{D}\:\varphi^{i}\!_{A}\:\varphi^{j}\!_{C}\:\frac{\partial}{\partial\varphi^{i}\!_{B}}\otimes\frac{\partial}{\partial\varphi^{j}\!_{D}}.\label{eq:283}
\end{equation}
Similarity, for the metrically affine models (\ref{eq:266}) (\ref{eq:267}) Legendre transformation is given by 
\begin{equation}
p_{i}=m\:g_{ij}\:v^{j}\quad,\quad\Sigma^{i}\!_{j}=\mathcal{R}^{i}{}_{j}{}^{k}{}_{l}\:\Omega^{l}\!_{k},\label{eq:284}
\end{equation}
and the phase-space expression for kinetic energy becomes 
\begin{equation}
\mathcal{T}=\mathcal{T}_{tr}+\mathcal{T}_{int}=\frac{1}{2m}\:g^{ij}\:p_{i}p_{j}+\frac{1}{2}\:\widetilde{\mathcal{R}}^{a}{}_{b}{}^{c}{}_{d}\:\Sigma^{b}\!_{a}\Sigma^{d}\!_{c},\label{eq:285}
\end{equation}
where 
\begin{equation}
\widetilde{\mathcal{R}}^{a}{}_{b}{}^{k}{}_{l}\:\mathcal{R}^{l}{}_{k}{}^{c}{}_{d}=\delta^{a}\!_{d}\:\delta^{c}\!_{b}.\label{eq:286}
\end{equation}

The corresponding inverse metric has the form: 
\begin{equation}
\Gamma^{-1}=\frac{1}{m}\:g^{ij}\frac{\partial}{\partial x^{i}}\otimes\frac{\partial}{\partial x^{j}}+\widetilde{\mathcal{R}}^{i}{}_{a}{}^{j}{}_{b}\:\varphi^{a}\!_{B}\:\varphi^{b}\!_{D}\:\frac{\partial}{\partial\varphi^{i}\!_{B}}\otimes\frac{\partial}{\partial\varphi^{j}\!_{D}}.\label{eq:287}
\end{equation}

For the general models (\ref{eq:263}) (\ref{eq:266}) it is rather difficult to find the explicit expressions for (\ref{eq:281}) (\ref{eq:285}),i.e., for the inverse coefficients $\widetilde{\mathcal{L}}$, $\widetilde{\mathcal{R}}$. It is also difficult for more specified models (\ref{eq:277}) (\ref{eq:278}) controlled by eight scalar coefficients. However, it may be easily done explicitly for the very special models (\ref{eq:268})/(\ref{eq:270}) and for (\ref{eq:269})/(\ref{eq:271}), and the more so for the simplest models based on (\ref{eq:274}), i.e., corresponding to vanishing $I$. And it is just these particular models which seem to be most interesting in dynamical applications and in theoretical analysis. In the sector of internal variables (relative motion) Legendre transformations for the $\mathcal{L}$-models (\ref{eq:268})/(\ref{eq:270}) and $\mathcal{R}$-models (\ref{eq:269})/(\ref{eq:271}) have respectively the following forms: 
\begin{eqnarray}
\widehat{\Sigma}^{A}\!_{B} & = & \mathcal{L}^{A}\!_{B}\!^{C}\!_{D}\:\widehat{\Omega}^{D}\!_{C},\label{eq:288}\\
\widehat{\Sigma}^{K}\!_{L} & = & {\rm I}\:\eta^{KM}\eta_{LN}\:\Omega^{N}\!_{M}+A\:\widehat{\Omega}^{K}\!_{L}+B\:\delta^{K}\!_{L}\:\widehat{\Omega}^{M}\!_{M},\label{eq:289}\\
\Sigma^{i}\!_{j} & = & \mathcal{R}^{i}{}_{j}{}^{l}{}_{k}\:\Omega^{k}\!_{l},\label{eq:290}\\
\Sigma^{i}\!_{j} & = & {\rm I}\:g^{im}g_{jn}\:\Omega^{n}\!_{m}+A\:\Omega^{i}\!_{j}+B\:\delta^{i}\!_{j}\:\Omega^{m}\!_{m}.\label{eq:291}
\end{eqnarray}

In the sector of translational variables, we have respectively 
\begin{eqnarray}
\hat{p}_{A} & = & m\eta_{AB}\hat{v}^{B},\label{eq:292}\\
p_{i} & = & mg_{ij}v^{j}.\label{eq:293}
\end{eqnarray}

Obviously, (\ref{eq:288}) (\ref{eq:289}) (\ref{eq:292}) may be as well expressed through the spatial $V$-representation, and (\ref{eq:290}) (\ref{eq:291}) (\ref{eq:293}) - through the material $U$-representation. This is, however, non-natural (although sometimes useful in a sense). The constant tensors on the right-hand sides are then replaced by $\varphi$ dependent ones. For example, in (\ref{eq:289}) (\ref{eq:292}) and (\ref{eq:291}) (\ref{eq:293}) the tensors $\eta,g$ are replaced respectively by the Green and Cauchy deformation tensors $G\left[\varphi\right]$, $C\left[\varphi\right]$.

In general it is rather difficult to inverse effectively the formulas (\ref{eq:288}) (\ref{eq:290}), however this may be easily done for the special cases (\ref{eq:289}) (\ref{eq:291}), and it is just these special cases what is particularly interesting both from the point of view of geometry and applications. One can easily obtain then the explicit form of Hamiltonian formalism.

The inverses of (\ref{eq:289}) (\ref{eq:291}) may be respectively expressed as follows: 
\begin{eqnarray}
\widehat{\Omega}^{K}\!_{L} & = & \frac{1}{\widetilde{I}}\:\eta^{KM}\eta_{LN}\widehat{\Sigma}^{N}\!_{M}+\frac{1}{\widetilde{A}}\:\widehat{\Sigma}^{K}\!\!_{L}+\frac{1}{\widetilde{B}}\:\delta^{K}\!_{L}\widehat{\Sigma}^{M}\!_{M},\label{eq:294}\\
\Omega^{i}\!_{j} & = & \frac{1}{\widetilde{I}}\:g^{im}g_{jn}\Sigma^{n}\!_{m}+\frac{1}{\widetilde{A}}\:\Sigma^{i}\!_{j}-\frac{1}{\widetilde{B}}\:\delta^{i}\!_{j}\Sigma^{m}\!_{m},\label{eq:295}
\end{eqnarray}
where the inverse inertial constants $\widetilde{I},\widetilde{A},\widetilde{B}$, are given by: 
\begin{equation}
\widetilde{I}=\frac{1}{I}\left(I^{2}-A^{2}\right)\,,\,\widetilde{A}=\frac{1}{A}\left(A^{2}-I^{2}\right)\,,\,\widetilde{B}=-\frac{1}{B}\left(I+A\right)\left(I+A+nB\right).\label{eq:296}
\end{equation}

The simplest situation is when the internal kinetic energy is affinely-invariant simultaneously on the left and on the right, i.e., when $I=2$. Then, obviously, 
\begin{equation}
\frac{1}{\widetilde{I}}=0\quad\left(\textrm{infinite }\widetilde{I}\right),\quad\widetilde{A}=A,\quad\widetilde{B}=-\frac{1}{B}A\left(A+nB\right).\label{eq:297}
\end{equation}
If there is no $B$-correction term ,$B=0$, then, similarly, 
\begin{equation}
\frac{1}{\widetilde{B}}=0\quad\left(\textrm{infinite }\widetilde{B}\right).\label{eq:298}
\end{equation}

In virtue of (\ref{eq:294}) (\ref{eq:295}), the corresponding kinetic Hamiltonians (geodetic Hamiltonians) are given by: 
\begin{eqnarray}
\mathcal{T}_{int} & = & \frac{1}{2\widetilde{I}}\:\eta_{KL}\widehat{\Sigma}^{K}\!_{M}\widehat{\Sigma}^{L}\!_{N}\:\eta^{NM}+\frac{1}{2\widetilde{A}}\:\widehat{\Sigma}^{K}\!_{L}\widehat{\Sigma}^{L}\!_{K}+\frac{1}{2\widetilde{B}}\:\widehat{\Sigma}^{K}\!_{K}\widehat{\Sigma}^{L}\!_{L}\label{eq:299}\\
\mathcal{T}_{int} & = & \frac{1}{2\widetilde{I}}\:g_{ik}\:\Sigma^{i}\!_{j}\Sigma^{k}\!_{l}g^{jl}+\frac{1}{2\widetilde{A}}\:\Sigma^{i}\!_{j}\Sigma^{j}\!_{i}+\frac{1}{2\widetilde{B}}\:\Sigma^{i}\!_{i}\Sigma^{j}\!_{j}.\label{eq:300}
\end{eqnarray}

When $I=0$ ($\frac{1}{\widetilde{I}}=0$) then the general solution of the geodetic problem is given by exponentials: 
\begin{equation}
\varphi(t)=\exp\left(Et\right)\varphi_{0}=\varphi_{0}\exp\left(\varphi_{0}\!^{-1}E\varphi_{0}t\right)=\varphi_{0}\exp\left(\widehat{E}t\right).\label{eq:301}
\end{equation}
Here $E$ is an arbitrary element of $L(V)$ or equivalently, $\widehat{E}$ is an arbitrary element of $L(U)$, and $\varphi_{0}$ is an arbitrary element of $LI(U,V)$. Roughly speaking, they are constants of motion, or initial conditions : 
\begin{equation}
\varphi(0)=\varphi_{0}\,,\quad\left(\frac{d\varphi}{dt}\right)\left(0\right)=E\varphi_{0}=\varphi_{0}\widehat{E}.\label{eq:302}
\end{equation}

If $I\neq0$, (\ref{eq:301}) is not any longer a general solution of the geodetic problem. Nevertheless, even then there exist special solutions of the type (\ref{eq:301}), so-called stationary solutions. They are special in that initial conditions are subject to certain restrictions; namely, if we use the representation: 
\begin{equation}
\varphi(t)=\varphi_{0}\exp\left(\widehat{E}t\right),\label{eq:303}
\end{equation}
for stationary solutions of (\ref{eq:299}); then $\widehat{E}$ is $\eta$-normal in the sense that: 
\begin{equation}
\left[\widehat{E},\widehat{E}^{\eta T}\right]=0,\label{eq:304}
\end{equation}
where $\widehat{E}^{\eta T}$ is the $\eta$-transpose of $\widehat{E}$,
\begin{equation}
\left(\widehat{E}^{\eta T}\right)^{A}\!_{B}:=\eta^{AC}\eta_{BD}\:E^{D}\!_{C}.\label{eq:305}
\end{equation}
Roughly speaking, $\widehat{E}^{\eta T}$ does commute with $\widehat{E}^{\eta}$. This holds in particular when $\widehat{E}$ is $\eta$-symmetric or $\eta$-antisymmetric: 
\begin{equation}
\widehat{E}^{\eta T}=\pm\widehat{E}^{\eta}.\label{eq:306}
\end{equation}
Similarly, for stationary solutions of (\ref{eq:300}) we have: 
\begin{equation}
\varphi(t)=\exp\left(Et\right)\varphi_{0},\label{eq:307}
\end{equation}
where $\varphi_{0}\in LI(U,V)$ is arbitrary, just like in (\ref{eq:303}) but $E\in L(V)$ is $g$-normal, 
\begin{equation}
\left[E,E^{gT}\right]=0,\label{eq:308}
\end{equation}
where 
\begin{equation}
\left(E^{gT}\right)^{i}\!_{j}:=g^{ik}g_{lj}E^{l}\!_{k}.\label{eq:309}
\end{equation}
This type of ``stationary solutions'' is interesting in itself, just some curious counterpart of stationary rotations in mechanics of anisotropic rigid body.

But this was some kind of digression. What is maximally interesting, these are doubly affinely-invariant geodetic models (\ref{eq:299}) (\ref{eq:300}) with $I=0$ (incidentally, they are identical in both versions of the formula). Then the general solution is given by the matrix exponents (\ref{eq:302}). Of course, one can admit in addition to $\mathcal{T}$ some potentials $V$ and consider other models, then no longer ones admitting exponential solutions. But it is a very curious circumstance that even within purely geodetic framework one can describe strongly nonlinear elastic vibrations. Dynamics is not then encoded in anything like $V$, but just in the kinetic energy, i.e., in the metric tensor of the configuration space. This resembles some properties of the Maupertuis principle. More precisely, this is true for the isochoric (incompressible) part of motion, when we consider only degrees of freedom ruled by the special linear groups $SL(V)$, $SL(U)$. General solution contains then an open subset of bounded motions and an open subset of non-bounded, escaping and collapsing solutions. Roughly speaking we are dealing with some dissociation threshold and bounded, non-linearly vibrating processes. And all this without potential, and analytically based on the properties of exponents! The purely dilatational part of geodetic motion is non-bounded or collapsing, except, of course the constant solution. But this purely dilatational part may be stabilized by introducing some auxiliary dilatational potential in one dimension, some oscillator, potential well, etc.

All this is a very important argument for investigating affinely-invariant dynamical models, although from some point of view they might seem ``exotic'' But they are not more exotic than the concept of effective mass in solid state physics. Analytical tools of the analysis are based on the properties of the matrix exponential map and the polar and two-polar decomposition of the matrix $\varphi$ representing the internal configuration.

First of all, let us notice that the expressions (\ref{eq:299}) (\ref{eq:300}) may be written respectively in the following forms:
\begin{eqnarray}
\mathcal{T}_{int} & = & \frac{1}{2\alpha}\: Tr\left(\widehat{\Sigma}^{2}\right) +\frac{1}{2\beta}\left(Tr \widehat{\Sigma}\right)^{2}-\frac{1}{4\mu} Tr \: \left(V^{2}\right), \label{eq:310}\\
\mathcal{T}_{int} & = & \frac{1}{2\alpha} \:Tr\left(\Sigma^{2}\right) +\frac{1}{2\beta}\left(Tr \Sigma \right)^{2} -\frac{1}{4\mu} \:Tr \left(S^{2}\right) \label{eq:311}
\end{eqnarray}
where, let us remind, the tensors $S$, $V$ denote the canonical spin and vorticity, given by (\ref{eq:70}) (\ref{eq:71}). The constants $\alpha$, $\beta$, $\mu$ are expressed by $A$, $B$, $C$ as follows:
\begin{equation}
\alpha = {\rm I}+A, \quad \beta = - ({\rm I}+A)({\rm I}+A+nB)/B, \quad \mu = ({\rm I}^{2}-A^{2})/{\rm I}.\label{eq:312}
\end{equation}

The formulas (\ref{eq:310}) (\ref{eq:311}) may be also expressed as follows:
\begin{eqnarray}
\mathcal{T}_{int} & = & \frac{1}{2\alpha}\: C(2) +\frac{1}{2\beta}\:C(1)^{2}+\frac{1}{2\mu} \left\|V\right\|^{2}, \label{eq:313}\\
\mathcal{T}_{int} & = & \frac{1}{2\alpha}\: C(2) +\frac{1}{2\beta}\:C(1)^{2}+\frac{1}{2\mu} \left\|S\right\|^{2} \label{eq:314}
\end{eqnarray}
where $C(k)$ denotes the $k$-th degree Casimir quantity, and $\left\|V\right\|$, $\left\|S\right\|$ are the magnitudes of vorticity and spin,
\begin{eqnarray}
\left\|V\right\|^{2}&=& - \frac{1}{2}\: Tr \left(V^{2}\right), \qquad \left\|S\right\|^{2}= - \frac{1}{2}\: Tr \left(S^{2}\right) \label{eq:315}\\
C(k)&=& Tr\left(\Sigma^{k}\right)=Tr\left(\widehat{\Sigma}^{k}\right). \label{eq:316}
\end{eqnarray}

It is seen that the difference in the invariance properties between (\ref{eq:299}) and (\ref{eq:300}), i.e., between (\ref{eq:310}) and (\ref{eq:311}) is reflected only by the last, i.e., third, terms in (\ref{eq:310}) (\ref{eq:311}). For more general models, like (\ref{eq:277}) (\ref{eq:278}), situation is more complicated and the relationship between velocity-based and canonical models is more complicated, although it still does exist. 

Let us stress that (\ref{eq:313}) (\ref{eq:314}) and the doubly (left- and right-) affinely invariant geodetic models (\ref{eq:274}) of internal dynamics differ by constants of motion proportional to $\left\|V\right\|^{2}$ and $\left\|S\right\|^{2}$. They are, so-to-speak, half-affine, i.e., affine in the space and metrical in the body (\ref{eq:313}) or conversely, metrical in the space and affine in the body (\ref{eq:314}). They are special cases of (\ref{eq:277}) (\ref{eq:288}). It may be interesting to consider their combination given in Hamiltonian terms by
\begin{equation}
\mathcal{T}_{int} = \frac{1}{2\alpha}\: C(2) +\frac{1}{2\beta}\:C(1)^{2}+\frac{1}{2\mu} \left\|V\right\|^{2}+ \frac{1}{2\nu} \left\|S\right\|^{2}, \label{eq:317}
\end{equation}
again with $\alpha$, $\beta$, $\mu$, $\nu$ being some inertial constants.

However, for us the most important expression would be one containing only the Casimir $\alpha$-, $\beta$-terms,
\begin{equation}
\mathcal{T}_{int} = \frac{1}{2\alpha}\: C(2) +\frac{1}{2\beta}\:C(1)^{2}, \label{eq:318}
\end{equation}
or merely, the main second-order Casimir term
\begin{equation}
\mathcal{T}_{int} = \frac{1}{2\alpha}\: C(2) = \frac{1}{2\alpha}\: \Sigma^{i}\!_{j}\Sigma^{j}\!_{i} = \frac{1}{2\alpha}\: \widehat{\Sigma}^{A}\!_{B}\widehat{\Sigma}^{B}\!_{A}. \label{eq:319}
\end{equation}

It is just this simplest model where one can observe the mentioned encoding of nonlinear elastic vibrations in negatively-determined part of the kinetic energy expression. To see this explicite, one should use so-called two-polar splitting for the mapping $\varphi \in L(U,V)$describing the internal configuration.

Any linear isomorphism $\varphi$ of $U$ onto $V$ induces in linear spaces $U$, $V$ two metric-like tensors, just the Green and Cauchy deformation tensors $G[\varphi] \in U^{*} \otimes U^{*}$, $C[\varphi] \in V^{*} \otimes V^{*}$. But the material and physical translation spaces $U$, $V$ are endowed also with some fixed, thus also constant in time, metric tensors $\eta \in U^{*} \otimes U^{*}$, $g \in V^{*} \otimes V^{*}$. And, as we saw, these metrics enable one to construct the mixed tensors $\widehat{G}[\varphi] \in U \otimes U^{*}$, $\widehat{C}[\varphi] \in V \otimes V^{*}$. They have eigenvalues $\lambda_{a}$, $\lambda_{a}\!^{-1}$, $a= 1, \ldots, n$. In certain formulas it is also convenient to use the quantities denoted by $Q^{a}$, $q^{a}$, $a=1, \ldots, n$, where 
\begin{equation}
Q^{a}=\exp(q^{a})=\sqrt{\lambda_{a}}. \label{eq:320}
\end{equation}

Obviously, $\lambda_{a}$-s are positive, because $\eta$, $G[\varphi]$ and $g$, $C[\varphi]$ are positively-definite tensors. Using the quantities $Q^{a}$ as diagonal entries, we can construct the diagonal matrix $D= diag (Q^{1}, \ldots, Q^{n})$. It may be identified with some linear isomorphism of ${\mathbb R}^{n}$ onto ${\mathbb R}^{n}$, $D: {\mathbb R}^{n} \rightarrow {\mathbb R}^{n}$.

The quantities $Q^{a}=D^{\underline{a}\underline{a}}$ (no summation convention!) are the basic stretchings of the system. They contain the information about how much the object is elongated/shortened in principal directions. But they do not tell us anything about the orientation of those stretchings in $U$ and $V$. This information is encoded in orthonormal systems of eigenvectors of $\widehat{G}$ and $\widehat{C}$. The eigenvectors satisfy equations
\begin{equation}
\widehat{G}R_{a}= \exp(2q^{a})R_{a}, \qquad \widehat{C}L_{a}= \exp(-2q^{a})L_{a}. \label{eq:321}
\end{equation}

The bases $R_{a}$, $L_{a}$ are assumed to be orthonormal,
\begin{equation}
\eta\left(R_{a}, R_{b}\right)= \eta_{KL}R^{K}\!_{a}R^{L}\!_{b}=\delta_{ab}=g\left(L_{a}, L_{b}\right)=g_{ij}L^{i}\!_{a}L^{j}\!_{b}. \label{eq:322}
\end{equation}

To be more precise, they are assumed to be unit vectors; their orthogonality is a generic situation for the case of simple spectra of operators $\widehat{G}$, $\widehat{C}$.

The dual bases in $U^{*}$and $V^{*}$ will be denoted by $R^{a}$, $L^{a}$, thus
\begin{equation}
\left\langle R^{a}, R_{b}\right\rangle= \delta^{a}\!_{b}, \qquad \left\langle L^{a}, L_{b}\right\rangle= \delta^{a}\!_{b}. \label{eq:323}
\end{equation}

Therefore, the deformation tensors may be expressed as follows:
\begin{eqnarray}
G[\varphi]&=& \sum_{a} \exp\left(2q^{a}[\varphi]\right) R^{a}[\varphi] \otimes R^{a}[\varphi],\nonumber \\
\label{eq:324}\\
C[\varphi]&=& \sum_{a} \exp\left(-2q^{a}[\varphi]\right) L^{a}[\varphi] \otimes L^{a}[\varphi]. \nonumber
\end{eqnarray}

Every mapping of internal configuration $\varphi$ is represented by three kinds of objects, namely, by a pair of metrically rigid bodies with configurations $\left(\ldots, R_{a}[\varphi], \ldots \right)$, $\left(\ldots, L_{a}[\varphi], \ldots \right)$ in $U$ and $V$, and the $n$-tuple of fictitious material points with positions $\left(\ldots, q^{a}[\varphi], \ldots \right)$ in the real axis ${\mathbb R}$. Those subsystems have respectively $n(n-1)/2$, $n(n-1)/2$, $n$ degrees of freedom. It is important to note that this representation is not unique. It is so even in the special case of $\varphi$ with non-degenerate spectra of $\widehat{G}[\varphi]$, $\widehat{C}[\varphi]$, because the permutation of indices $a$ in (\ref{eq:324}) does not affect the sum. When the spectrum of $\widehat{G}[\varphi]$, $\widehat{C}[\varphi]$ is degenerate, the non-uniqueness of (\ref{eq:324}) becomes continuous. This singularity of coordinate system resembles that of spherical or polar coordinates at $r=0$ ($\varphi$, $\vartheta$ completely non-determined), although in concrete details it is much more complicated.

Obviously, the linear frames $L=(\ldots, L_{a}, \ldots)$, $R=(\ldots, R_{a}, \ldots)$ may be canonically identified with some isomorphisms $L: {\mathbb R}^{n} \rightarrow V$ and $R: {\mathbb R}^{n} \rightarrow U$. And, obviously, their dual co-frames $\widetilde{L}=(\ldots, L^{a}, \ldots)$, $\widetilde{R}=(\ldots, R^{a}, \ldots)$ correspond canonically to some linear isomorphisms $L^{-1}:V \rightarrow {\mathbb R}^{n}$ and $R^{-1}: U \rightarrow {\mathbb R}^{n}$. If the diagonal matrix with deformation invariants $Q^{a}$ on diagonal is identified with a linear isomorphism $D$ of ${\mathbb R}^{n}$ onto ${\mathbb R}^{n}$, then, making use of all the above identifications, we can simply write
\begin{equation}
\varphi= LDR^{-1}; \label{eq:325}
\end{equation}
in matrix terms this means: orthogonal times diagonal times orthogonal.

It is clear that the spatial and material isometries preserving orientation in $V$ and $U$, $A \in SO(V,g)$, $B \in SO(U,\eta)$ act on the left on the $L$- and $R$-gyroscopic part of $\varphi$, because the mappings 
\begin{equation}
L \mapsto AL, \qquad R \mapsto BR \label{eq:326}
\end{equation}
imply that $\varphi$ transforms under them as follows:
\begin{equation}
\varphi \rightarrow A\varphi B^{-1}. \label{eq:327}
\end{equation}

Their Hamiltonian generators are respectively the spin (for $A$) and the negative vorticity (for $B$).

Let us now consider in a more detail the structure of $L$- and $R$-rigid bodies. Their ``physical spaces'' are respectively $V$ and $U$. And their material spaces are simply identified with ${\mathbb R}^{n}$. The group $SO(n,{\mathbb R})$ acts on them on the right,
\begin{equation}
L \mapsto LC, \qquad R \mapsto RD \label{eq:328}
\end{equation}
$C,D \in SO(n,{\mathbb R})$. These mapping are not expressible like (\ref{eq:327}), by an action on $\varphi$ as a whole. Nevertheless, locally they are well-defined and so are their Hamiltonian generators. One can define their ``spatial'' (in the $V$, $U$-sense) and co-moving (in the ${\mathbb R}^{n}$-sense) angular velocities and canonical spins.

The ``co-moving'' and ``spatial'' components for the angular velocity of the $L$-top are defined as:
\begin{eqnarray}
\hat{\chi}^{a}\!_{b}&:=&\left\langle L^{a}, \frac{d}{dt} \: L_{b}\right\rangle = L^{a}\!_{i}\frac{d}{dt}\:L^{i}\!_{b},
\label{eq:329}\\
\chi^{i}\!_{j}&:=&\left( \frac{d}{dt} \:L^{i}\!_{a}\right)L^{a}\!_{j}, \quad \textrm{or} \quad \chi = \hat{\chi}^{a}\!_{b}\:L_{a} \otimes L^{b}. \label{eq:330}
\end{eqnarray}
Obviously, $\chi \in SO(V,g)'$, $\hat{\chi} \in SO(n,{\mathbb R})'$. The first latin characters $a$, $b$ are simply labels, i.e., indices in ${\mathbb R}^{n}$, whereas the middle ones $i$, $j$ are tensor indices in $V$. Similarly, the ``co-moving'' and ``$U$-spatial'' components of the angular velocity of the $R$-top are given by analogous formulas:
\begin{eqnarray}
\hat{\vartheta}^{a}\!_{b}&:=&\left\langle R^{a}, \frac{d}{dt} \: R_{b}\right\rangle = R^{a}\!_{M}\frac{d}{dt}\:R^{M}\!_{b},\nonumber \\
\vartheta^{L}\!_{M}&:=&\left( \frac{d}{dt} \:R^{L}\!_{a}\right)R^{a}\!_{M}, \quad \textrm{or} \quad \vartheta = \hat{\vartheta}^{a}\!_{b}\:R_{a} \otimes R^{b}. \label{eq:331}
\end{eqnarray}
Again for this rigid body $U$ plays the role of the ``physical'' space in spite of its being originally ``material''. The ``material'' space for the $R$-top is again ${\mathbb R}^{n}$. The choice of non-holonomic velocities as $\left({\dot{q}}^{a}, {\hat{\chi}}^{a}\!_{b}, {\hat{\vartheta}}^{a}\!_{b}\right)$ or $\left({\dot{q}}^{a}, {\chi}^{i}\!_{j}, {\vartheta}^{L}\!_{M}\right)$ depends on particular purposes. Just as is was the case with the usual, physical rigid bodies, it is convenient to introduce non-holonomic, just Poisson-non-commuting canonical momenta $\left(p_{a}, {\hat{\rho}}^{a}\!_{b}, {\hat{\tau}}^{a}\!_{b}\right)$ or $\left(p_{a}, {\rho}^{i}\!_{j}, {\tau}^{A}\!_{B}\right)$. Here $p_{a}$ are canonical momenta conjugate to deformation invariants $q^{a}$. The quantities ${\hat{\rho}}^{a}\!_{b}$, ${\hat{\tau}}^{a}\!_{b}$ and ${\rho}^{i}\!_{j}$, ${\tau}^{A}\!_{B}$ are canonically conjugate to ${\hat{\chi}}^{a}\!_{b}$, ${\hat{\vartheta}}^{a}\!_{b}$, ${\chi}^{i}\!_{j}$, ${\vartheta}^{L}\!_{M}$. Obviously, $\hat{\rho}, \hat{\tau} \in SO(n, {\mathbb R})'$, $\rho \in SO(V,g)'$, $\tau \in SO(U, \eta)'$. The meaning of the mentioned ``conjugacy'' is encoded in the following duality formulas:
\begin{eqnarray}
& &\left\langle (\rho, \tau, p), (\chi, \vartheta, \dot{q})\right\rangle = \left\langle (\hat{\rho}, \hat{\tau}, p), (\hat{\chi}, \hat{\vartheta}, \dot{q})\right\rangle = \label{eq:332} \\
&=& p_{a}\: \dot{q}^{a} + \frac{1}{2}\:Tr(\rho \chi)+\frac{1}{2}\:Tr(\tau \vartheta) = p_{a}\: \dot{q}^{a} + \frac{1}{2}\:Tr(\hat{\rho} \hat{\chi})+\frac{1}{2}\:Tr(\hat{\tau} \hat{\vartheta}). \nonumber
\end{eqnarray}

The Hamiltonian interpretation of $\rho$ and $-\tau$ as generators of the left- and right-orthogonal translations of $\varphi$ ($V$-spatial and $U$-material ones) implies that they are simply identical with spin and vorticity. They are respectively $g$-skew-symmetric and $\eta$-skew-symmetric parts of $\Sigma$ and $\widehat{\Sigma}$. Unlike $\rho$ and $\tau$, their ${\mathbb R}^{n}$-material representatives $\hat{\rho}$, $\hat{\tau}$, generating (\ref{eq:328}) fail to be constants of motion for geodetic systems and for Lagrangian models with potentials depending only on deformation invariants $q^{a}$. It is also clear that the constants of motion $\rho$, $\tau$ are related to $\hat{\rho}$, $\hat{\tau}$ as follows:
\begin{equation}
\rho= \hat{\rho}^{a}\!_{b}\: L_{a} \otimes L^{b}, \qquad \tau= \hat{\tau}^{a}\!_{b}\: R_{a} \otimes R^{b}. \label{eq:333}
\end{equation}

Let us stress that when dealing with the traditional, orthogonally invariant kinetic energy (\ref{eq:107a}) (\ref{eq:248}), the traditional form of deformation invariants $Q^{a}$ (\ref{eq:320}) is more convenient than $q^{a}$ which are particularly suited to models with affine symmetry.

It was mentioned that the two-polar expansion (\ref{eq:325}) is not unique. At the same time, it is well-known that the usual polar decomposition is unique. It is convenient to mention here about the polar splitting in this description. Namely, $L_{a}[\varphi]$, $R_{a}[\varphi]$ are two orthogonal bases, respectively in the $(V,g)$-sense and $(U, \eta)$-sense. Therefore, there exists the unique isometry $U[\varphi]$ which establishes a one-to-one relationship between them:
\begin{equation}
L_{a}[\varphi] = U[\varphi]R_{a}[\varphi]; \label{eq:334}
\end{equation}
$U[\varphi] \in O(U,\eta;V,g)$. One can show that
\begin{equation}
\varphi = U[\varphi]A[\varphi] = B[\varphi]U[\varphi], \label{eq:335}
\end{equation}
where the linear mapping $A[\varphi] \in GL(U)$, $B[\varphi] \in GL(V)$ are respectively $\eta$-symmetric, $g$-symmetric and positive. Therefore,
\begin{eqnarray}
\eta(x, A[\varphi] y) &=& \eta(A[\varphi]x,  y), \qquad g(w, B[\varphi] z) = g(B[\varphi] w,  z), \nonumber \\
\eta(x, A[\varphi] x) &>& 0, \qquad \qquad \qquad g(w, B[\varphi] w) >0, \label{eq:336}
\end{eqnarray}
for any vectors $x,y,w,z$. In the last two inequalities the vectors $x,w$ are non-vanishing.

The symmetric elements of the left and right polar decomposition are related to each other as follows:
\begin{equation}
B[\varphi] = U[\varphi]A[\varphi]U[\varphi]^{-1}, \label{eq:337}
\end{equation}
i.e, by the $U[\varphi]$-similarity transformation. In the usual matrix representation (\ref{eq:335}) has the form 
\begin{equation}
\varphi = UA=BU. \label{eq:338}
\end{equation}

The symmetric terms, e.g. $A$, may be diagonalized by the orthogonal similarity:
\begin{equation}
A=RDR^{-1}, \quad R-\textrm{orthogonal}. \label{eq:339}
\end{equation}
Then (\ref{eq:338}) becomes
\begin{equation}
\varphi=URDR^{-1}=LDR^{-1}, \qquad U=LR^{-1}. \label{eq:340}
\end{equation}
Although $L,R$ separately taken are not unique, $U$ is so, and therefore, $A,B$ are unique.

Let us define now some objects instead $\hat{\rho}$, $\hat{\tau}$, which enable one to perform a partial diagonalization of the affinely-invariant kinetic energy of internal motion, 
\begin{equation}
M:=-\hat{\rho}- \hat{\tau}, \qquad N=\hat{\rho}- \hat{\tau}. \label{eq:350}
\end{equation}
After this substitution the second-order Casimir invariant,
\begin{equation}
C(2)=Tr \left({\widehat{\Sigma}}^{2}\right), \label{eq:351}
\end{equation}
becomes:
\begin{equation}
C(2)=\sum_{a}p_{a}\!^{2}+ \frac{1}{16}\sum_{a,b} \frac{\left(M^{a}\!_{b}\right)^{2}}{\sinh^{2}\frac{q^{a}-q^{b}}{2}} - \frac{1}{16}\sum_{a,b} \frac{\left(N^{a}\!_{b}\right)^{2}}{\cosh^{2}\frac{q^{a}-q^{b}}{2}}. \label{eq:352}
\end{equation}
The first term in (\ref{eq:352}) admits a nice representation as a sum of the relative and over-all contributions, 
\begin{equation}
\frac{p^{2}}{n}+ \frac{1}{2n}\sum_{a,b}\left(p_{a}-p_{b}\right)^{2}, \label{eq:353}
\end{equation}
where $p$ is the first-degree Casimir invariant,
\begin{equation}
p=C(1)=Tr \left(\Sigma\right)=Tr \left(\widehat{\Sigma}\right). \label{eq:354}
\end{equation}

It is a quantity canonically conjugate to the center of mass of logarithmic deformation invariants,
\begin{equation}
q=\frac{1}{n}\sum_{a}q^{a}, \qquad \left\{q,p\right\}=1. \label{eq:357}
\end{equation}

The corresponding kinetic energy based on the second-order Casimir invariant $C(2)$ as the main term,
\begin{equation}
\mathcal{T}= \frac{1}{2 \alpha}\:C(2) \label{eq:358}
\end{equation}
has, as seen form (\ref{eq:352}) a very peculiar structure. It consists of the first term which formally has a structure of the kinetic energy of the $n$-particle $q^{a}$-objects on $\mathbb R$, and of the next two terms. The first of these terms, i.e., $\sinh^{-2}\frac{q^{a}-q^{b}}{2}$-term describes centrifugal repulsion between invariants, but the second one, i.e., $\cosh^{-2}\frac{q^{a}-q^{b}}{2}$, due to its minus-sign, represents a curious ``centrifugal attraction''. The strengths coefficients $\left(M^{a}\!_{b}\right)^{2}$, $\left(N^{a}\!_{b}\right)^{2}$ are positive, so really one has to do with repulsion and attraction of deformation invariants. Repulsion is singular at the coincidence $q^{a}=q^{b}$, whereas attraction is then finite. But at large $\left|q^{a}-q^{b}\right|$-distances, attraction prevails if $\left|N_{ab}\right| > \left|M_{ab}\right|$. This is the typical shape of ``intermolecular'' forces between deformation invariants interpreted as indistinguishable material points. Admitting the $C(1)$, $\left\|S\right\|^{2}$, $\left\|V\right\|^{2}$-terms, we do not change this interpretation. Before commenting it more, we mention only about some related models.

It is seen that (\ref{eq:352}) (\ref{eq:358})-models resemble the hyperbolic version of the Sutherland lattice. There are both similarities and differences. The main difference is just the occurrence of attractive terms, the negative contributions to (\ref{eq:352}). Their interpretation is slightly obscured because the coupling amplitudes $M^{a}\!_{b}$, $N^{a}\!_{b}$ are not constant, they satisfy together with $q^{a}$ s closed system of equations of motion. These equations may be easily expressed in terms of Poisson brackets between basic state variables $\left(q^{a}, p_{b}, M^{a}\!_{b}, N^{a}\!_{b} \right)$; obviously $\left\{q^{a}, p_{b}\right\}=\delta^{a}\!_{b}$, $\left\{q^{a}, M^{c}\!_{d}\right\}= \left\{q^{a}, N^{c}\!_{d}\right\}=0$, and $\left\{p_{a}, M^{c}\!_{d}\right\}= \left\{p_{a}, N^{c}\!_{d}\right\}=0$. The brackets for $M^{a}\!_{b}, N^{c}\!_{d}$ may be easily obtained from their definition (\ref{eq:350}) and from the standard Poisson rules for ${\hat{\rho}}^{a}\!_{b}$, ${\hat{\tau}}^{c}\!_{d}$ as Hamiltonian generators of the action of $SO(n, {\mathbb R}) \times SO(n, {\mathbb R})$ through (\ref{eq:328}). The family of functions $\left(q^{a}, p_{b}, M^{a}\!_{b}, N^{a}\!_{b} \right)$ forms a Poisson algebra. In the case of non-geodetic models with doubly isotropic potentials, depending only on deformation invariants $q^{a}$, equations of motion may be all obtained in terms of those Poisson brackets as
\begin{equation}
\frac{dF}{dt}= \left\{F, H\right\}; \label{eq:359}
\end{equation}
for $F$ one has to substitute $q^{a}, p_{b}, M^{a}\!_{b}, N^{a}\!_{b}$. 

Obviously, this gives us only a partial description of motion, without the time dependence of gyroscopic variables $L_{a}, R_{a}$. But in practical applications like, e.g., vibrations of molecules, this time dependence is less important. In any case, having solved (\ref{eq:359}), we know the time dependence of $\hat{\rho}, \hat{\tau}$. Then, inverting the Legendre transformation, one can express the time dependence of gyroscopic angular velocities $\hat{\rho}, \hat{\tau}$. And then, finally, for the time dependence of $L_{a}, R_{a}$ one obtains first-order differential equations for the $n$-legs $L_{a}, R_{a}$:
\begin{equation}
\frac{d}{dt} L^{i}\!_{a}= L^{i}\!_{b} {\hat{\chi}}^{b}\!_{a}(t), \qquad \frac{d}{dt} R^{i}\!_{a}= R^{i}\!_{b} {\hat{\vartheta}}^{b}\!_{a}(t). \label{eq:360}
\end{equation}

So, after solving the system (\ref{eq:359}) for $\left(q^{a}, p_{b}, M^{a}\!_{b}, N^{a}\!_{b} \right)$, i.e., after obtaining the main deformation description, we can in principle solve the next, less important step and determine the time dependence of the main axes of inertia. It is so for any doubly-isotropic model.

The very idea is that in (\ref{eq:352}) (\ref{eq:358}). The ``plus'' and ``minus'' terms act in opposite directions, representing, respectively, the repulsion and attraction of deformation invariants even in the completely geodetic, potential-free models. The question is, however, if the idea is correct, because the repulsion and attraction strengths are not constants. There is, however, some special case where they are constants of motion. It is in two-dimensional models, when $n=2$. In fact, in a consequence of the fact that the orthogonal group $SO(2, {\mathbb R})$ is Abelian (because it is one-dimensional), in this special case we have:
\begin{equation}
{\hat{\rho}}^{a}\!_{b}= \rho^{a}\!_{b}, \qquad {\hat{\tau}}^{a}\!_{b}= \tau^{a}\!_{b}, \label{eq:361}
\end{equation}
and because of this, $M^{a}\!_{b}, N^{a}\!_{b}$ are constants of motion (for the doubly-isotropic models), just as $\rho^{a}\!_{b}=S^{a}\!_{b}$, $\tau^{a}\!_{b}=-V^{a}\!_{b}$ are. Because of this, the effective strengths $\left(M^{a}\!_{b}\right)^{2}$, $\left(N^{a}\!_{b}\right)^{2}$ in (\ref{eq:352}) (\ref{eq:358}) are constant in time. The matrices $L, R, D$ become 
\begin{eqnarray}
L&=&\left[ 
\begin{array}{cc}
\cos \alpha & -\sin \alpha \\ 
\sin \alpha & \cos \alpha  	
\end{array}\right], \quad
R= \left[
\begin{array}{cc}
\cos \beta & -\sin \beta \\
\sin \beta & \cos \beta	
\end{array}\right],\nonumber \\ 
D&=& 
\left[\begin{array}{cc}
Q^{1} & 0 \\
0 & Q^{2}	
\end{array}\right]= \left[\begin{array}{cc}
\exp q^{1} & 0 \\
0 & \exp q^{2}	
\end{array}\right]. \label{eq:362}
\end{eqnarray}

It is convenient to represent $GL^{+}(2, {\mathbb R})$ as ${\mathbb R}^{+}SL(2, {\mathbb R})$ by introducing the ``center of mass'' and the displacement between $q^{1}$ and $q^{2}$,
\begin{equation}
q=\frac{1}{2}\left(q^{1}+q^{2}\right), \qquad x=q^{2}-q^{1}, \label{eq:363}
\end{equation}
and their conjugate momenta
\begin{equation}
p=p_{1}+p_{2}, \qquad p_{x}=\frac{1}{2}\left(p_{2}-p_{1}\right). \label{eq:364}
\end{equation}

The angular velocities $\hat{\chi}$, $\hat{\vartheta}$ have the obvious form:
\begin{equation}
\hat{\chi}=\chi = \frac{d \alpha}{dt}\left[\begin{array}{cc}
0&-1\\
1&0
\end{array}\right], \qquad \hat{\vartheta}=\vartheta = \frac{d \beta}{dt}\left[\begin{array}{cc}
0&-1\\
1&0
\end{array}\right]. \label{eq:365}
\end{equation}

The quantities $M$, $N$ become
\begin{equation}
M = m \left[\begin{array}{cc}
0&1\\
-1&0
\end{array}\right], \qquad N = n \left[\begin{array}{cc}
0&1\\
-1&0
\end{array}\right], \label{eq:366}
\end{equation}
where, obviously,
\begin{equation}
m=p_{\beta}-p_{\alpha}, \qquad n=p_{\beta}+p_{\alpha}. \label{eq:367}
\end{equation}

Therefore, for $B=0$, the doubly affine-invariant kinetic energy of internal degrees of freedom becomes
\begin{equation}
\mathcal{T}= \frac{1}{2A}\left( p_{1}\!^{2} + p_{1}\!^{2} \right)+ \frac{1}{16A} \: \frac{m^{2}}{\sinh^{2}\frac{q^{2}-q^{1}}{2}} - \frac{1}{16A}\: \frac{n^{2}}{\cosh^{2}\frac{q^{2}-q^{1}}{2}}. \label{eq:368}
\end{equation}

Using the new variables (\ref{eq:363}) (\ref{eq:364}) (\ref{eq:367}), one obtains:
\begin{equation}
\mathcal{T}= \frac{p^{2}}{4A} + \frac{p_{x}\!^{2}}{A} + \frac{m^{2}}{16A \sinh^{2}\frac{x}{2}} - \frac{n^{2}}{16A \cosh^{2}\frac{x}{2}}. \label{eq:369}
\end{equation}

For a bit more general models (\ref{eq:270}) (\ref{eq:271}) (\ref{eq:310}) (\ref{eq:311}) one obtains essentially the same, up to certain modification of constants and correction terms proportional to $\left\|V\right\|^{2}$, $\left\|S\right\|^{2}$. The essence of the problem remains the same, because $\left\|V\right\|^{2}$, $\left\|S\right\|^{2}$ are constants of motion.

It is seen that in the geodetic problems like (\ref{eq:369}), the shear-part, i.e., $SL(2, {\mathbb R})$-component of dynamics admits a continuous family of bounded elastic vibrations. This is the part characterized by $\left|n\right|>\left|m\right|$, where at large $\left|x\right|$-distances the ``centrifugal attraction'' prevails the centrifugal repulsion. This is in spite of the non-compactness of the $x$-variable and of $SL(2, {\mathbb R})$ at all. The family of bounded motions has the dimension six, i.e., twice the dimension of $SL(2, {\mathbb R})$. Above the threshold, i.e., for $\left|m\right|>\left|n\right|$, one deals with the six-dimensional family of unbounded motion. This is the typical threshold behavior, which does exist on the purely geodetic level due to the curved geometry of $SL(2, {\mathbb R})$. In this way, the isochoric (incompressible) elastic motions may be described in purely geodetic terms, without any use of the potential. The only failure of this threshold behaviors on the level of the total $GL(2, {\mathbb R})$ are just dilatations. According to (\ref{eq:369}), the variable $p$ occurs in a quadratic way with constant coefficient, and its conjugate configuration variable $q$ does not occur at all. Therefore, except the solutions $q=const$, $p=0$, all solutions are singular wit respect to $q$. They are either infinitely expanding, or contracting to the singularity $q=0$. This is a consequence of the fact that $GL^{+}(2, {\mathbb R})$ is not semisimple, being isomorphic to ${\mathbb R}^{+}SL(2, {\mathbb R})= \exp ({\mathbb R})SL(2, {\mathbb R})$. Of course, this is not a dangerous failure for our philosophy. The $q$-motion may be stabilized by introducing any bounding potential, e.g., oscillator, thin potential well, etc.

Exactly the same holds for any $n>2$. The difference is only that for $n>2$, the spin quantities $M^{a}\!_{b}$, $N^{a}\!_{b}$ fail to be constants of motion, they also vibrate. But the $SL(2, {\mathbb R})$-threshold behavior is a consequence of the commutation rules is $SL(2, {\mathbb R})$, $GL(2, {\mathbb R})$. More or less, the same holds also for other models with the partial affine invariance of kinetic energy, like (\ref{eq:270}) (\ref{eq:271}), i.e., (\ref{eq:313}) (\ref{eq:314}).

Let us mention also about some other models more or less related to the above ones. First of all, let us remind the traditional d'Alembert model (\ref{eq:107a}) (\ref{eq:248}) with the materially isotropic co-moving inertial tensor $J^{AB} = {\rm I} \eta^{AB}$. Then the kinetic energy is given in the two-polar representation by
\begin{equation}
\mathcal{T}= \frac{1}{2{\rm I}}\sum_{a}P_{a}\!^{2} + \frac{1}{8{\rm I}} \sum_{a,b} \frac{\left(M^{a}\!_{b}\right)^{2}}{\left(Q^{a}-Q^{b}\right)^{2}} + \frac{1}{8{\rm I}} \sum_{a,b} \frac{\left(N^{a}\!_{b}\right)^{2}}{\left(Q^{a}+Q^{b}\right)^{2}}. \label{eq:370}
\end{equation}

Let us notice that now it is not logarithmic invariant $q^{a}$, but just $Q^{a}$ that is convenient. Obviously, this expression contains only centrifugal repulsion of invariants. Therefore, (\ref{eq:370}) is completely useless as a model of nonlinear elastic vibrations. To obtain a viable model, one has to modify it by adding some potential energy. In the doubly isotropic models, it will be a function ${\mathfrak V}\left(Q^{1}, \cdots, Q^{n}\right)$ of deformation invariants $Q^{a}$ alone. Obviously, the first two terms of (\ref{eq:370}) resemble the Calogero-Moser integrable lattice and indeed there is some relationship here.

It was mentioned also that (\ref{eq:368}) (\ref{eq:352}) (\ref{eq:358}) resemble the hyperbolic version of the Sutherland lattice. It turns out that some relationship with the usual, i.e., trigonometric Sutherland lattice does exist as well. To obtain it, one should compactify the deformation invariants by the formal substitution of
\begin{equation}
Q^{a}= \exp \left(i q^{a}\right) \label{eq:371}
\end{equation}
to the positively definite kinetic energy
\begin{equation}
T_{int}= - \frac{A}{2} \: Tr\left(\Omega^{2}\right)= \frac{A}{2} \: Tr\left(\Omega^{+} \Omega \right), A>0 \label{eq:372}
\end{equation}
Then $GL(2, {\mathbb R})$ is formally replaced by $U(n)$- the compact real form of $GL(n, {\mathbb C})$. Obviously,
\begin{equation}
\widehat{\Omega}= \varphi^{-1}\frac{d \varphi}{dt}, \qquad \varphi \in U(n). \label{eq:373}
\end{equation}
After this formal substitution one obtains for $\mathcal{T}_{int}$
\begin{equation}
\mathcal{T}_{int}= \frac{1}{2A}\sum_{a} P_{a}\!^{2} + \frac{1}{32A} \sum_{a,b}\: \frac{\left(M^{a}\!_{b}\right)^{2}}{\sin^{2}\frac{q^{a}-q^{b}}{2}} + \frac{1}{32A} \sum_{a,b}\: \frac{\left(N^{a}\!_{b}\right)^{2}}{\cos^{2}\frac{q^{a}-q^{b}}{2}}. \label{eq:374}
\end{equation}

The relationship with the usual Sutherland lattice is obvious. Let us observe that now the deformation invariants run over the circle. Because of this there is no problem with the positive definiteness of the terms in (\ref{eq:374}). Because in the compact space with the circular topology it is practically impossible to distinguish between repulsion and attraction.

There are also other important problems concerning affinely rigid body and the problem of dynamical affine invariance. They concern mainly two fields. One of them is the theory of systems of affine bodies. And of course another one is quantum mechanics of such bodies and their systems. There is no place in this paper for studying them. Some primary discussion was delivered in papers quoted in the references below, some other aspects are to be investigated in future.

\section*{Acknowledgements}

This paper contains results obtained within the framework of the research projects 501 018 32/1992 and N N501 049 540 financed from the Scientific Research Support Fund in 2007-2010 and 2011-2014. The authors are greatly indebted to the Polish Ministry of Science and Higher Education for this financial support.

\end{document}